\documentclass[fleqn,usenatbib]{mnras}

\usepackage[T1]{fontenc}
\usepackage{deluxetable}
\DeclareRobustCommand{\VAN}[3]{#2}
\let\VANthebibliography\thebibliography
\def\thebibliography{\DeclareRobustCommand{\VAN}[3]{##3}\VANthebibliography}

\def\code#1{\texttt{#1}}
\usepackage{graphicx}	
\usepackage{amsmath}	
\usepackage{amssymb}	
\usepackage{newtxtext,newtxmath}

\title[3D Radiative Transfer Modelling and Virial Analysis]{3D Radiative Transfer Modelling and Virial Analysis of Starless  Cores in the B10 Region of the Taurus Molecular Cloud}

\author[S. Scibelli et al.]{
Samantha Scibelli,$^{1}$\thanks{E-mail: sscibelli@arizona.edu}
Yancy Shirley,$^{1}$ Anika Schmiedeke,$^{2}$ Brian Svoboda,$^{3}$ Ayushi Singh,$^{4}$ \newauthor James Lilly,$^{5}$ Paola Caselli$^{6}$
\\
$^{1}$ Steward Observatory,  University of Arizona, 933 North Cherry Avenue,
Tucson, AZ 85721 \\
$^{2}$ Green Bank Observatory, 155 Observatory Rd P.O. Box 2. Green Bank, WV 24944, USA\\
$^{3}$ National Radio Astronomy Observatory, PO Box O, Socorro, NM 87801, USA\\
$^{4}$ Department of Astronomy and Astrophysics, University of Toronto, 50 St. George St., Toronto, Ontario, Canada, M5S 3H4\\
$^{5}$ Department of Physics \& Astronomy, University of Wyoming, Laramie, WY 82071, USA\\
$^{6}$ Max-Planck-Institut für extraterrestrische Physik, Giessenbachstrasse 1, 85748 Garching, Germany
}

\date{Accepted 2023 March 14. Received 2023 March 14; in original form 2023 January 10 }

\pubyear{2023}

\begin{document}
\label{firstpage}
\pagerange{\pageref{firstpage}--\pageref{lastpage}}
\maketitle

\begin{abstract}
Low-mass stars like our Sun begin their evolution within cold (10 K) and dense ($\sim 10^5 \mathrm{cm}^{-3}$) cores of gas and dust. The physical structure of starless cores is best probed by thermal emission of dust grains. We present a high resolution dust continuum study of the starless cores in the B10 region of the Taurus Molecular Cloud. New observations at 1.2mm and 2.0mm (12$^{''}$ and 18$^{''}$ resolution) with the NIKA2 instrument on the IRAM 30m have probed the inner regions of 14 low-mass starless cores. We perform sophisticated 3D radiative transfer modelling for each of these cores through the radiative transfer framework \textit{pandora}, which utilizes RADMC-3D. Model best-fits constrain each cores' central density, density slope, aspect ratio, opacity, and interstellar radiation field strength. These `typical' cores in B10 span central densities from 5\,$\times$\,10$^4$\,-\,1\,$\times$\,10$^6$\,cm$^{-3}$, with a mean value of 2.6\,$\times$\,10$^5$\,cm$^{-3}$. We find the dust opacity laws assumed in the 3D modelling, as well as the estimates from \textit{Herschel}, have dust emissivity indices, $\beta$'s, on the lower end of the distribution constrained directly from the NIKA2 maps, which averages to $\beta = 2.01\pm0.48$. 
From our 3D density structures and archival NH$_3$ data, we perform a self-consistent virial analysis to assess each core’s stability. Ignoring magnetic field contributions, we find 9 out of the 14 cores ($64\%$) are either in virial equilibrium or are bound by gravity and external pressure. To push the bounded cores back to equilibrium, an effective magnetic field difference of only $\sim 15 \mu$G is needed.

\end{abstract}




\begin{figure*}
\centering
\begin{center}$
\begin{array}{cc}
\includegraphics[width=84mm]{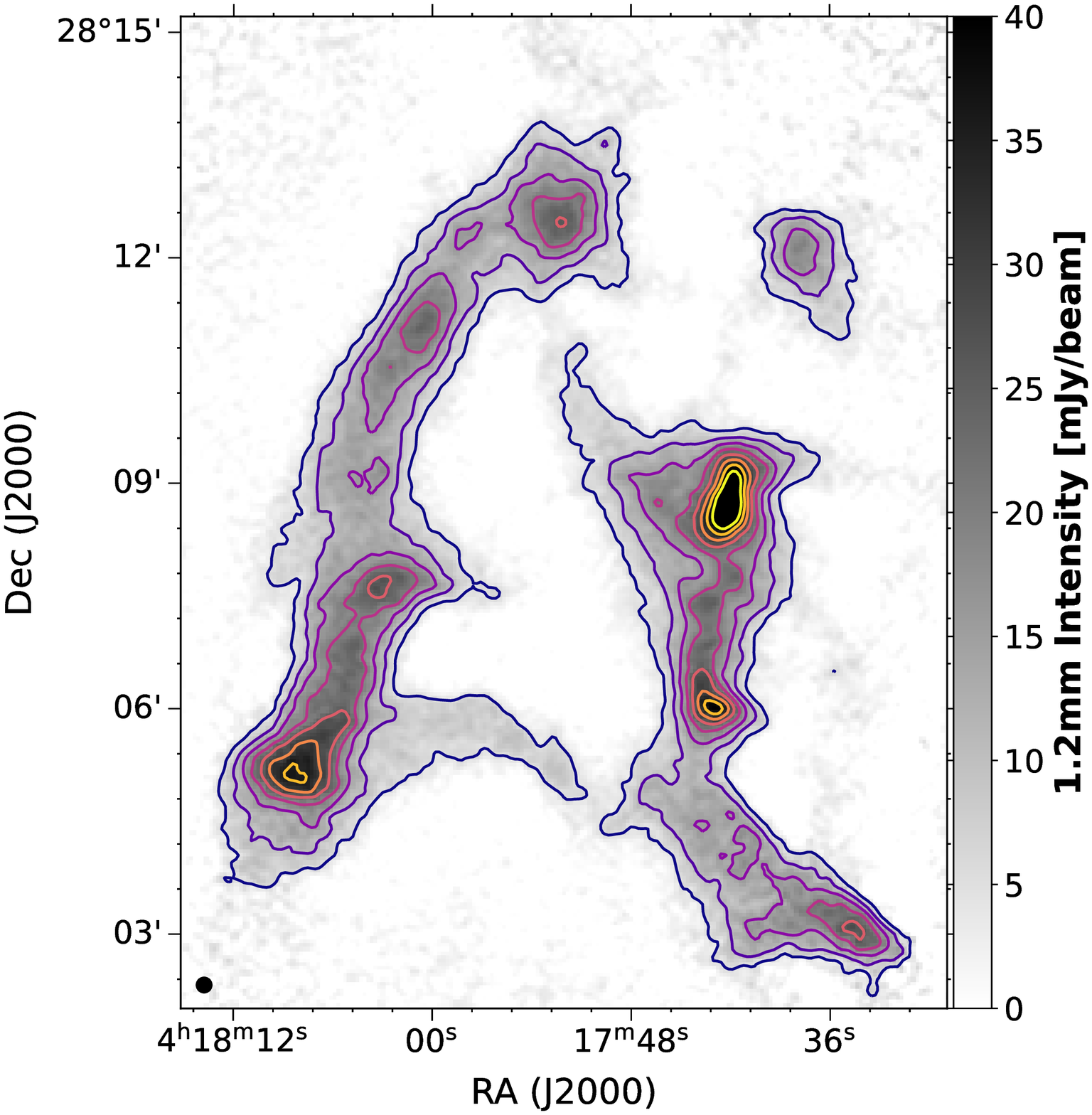}& 
\includegraphics[width=84mm]{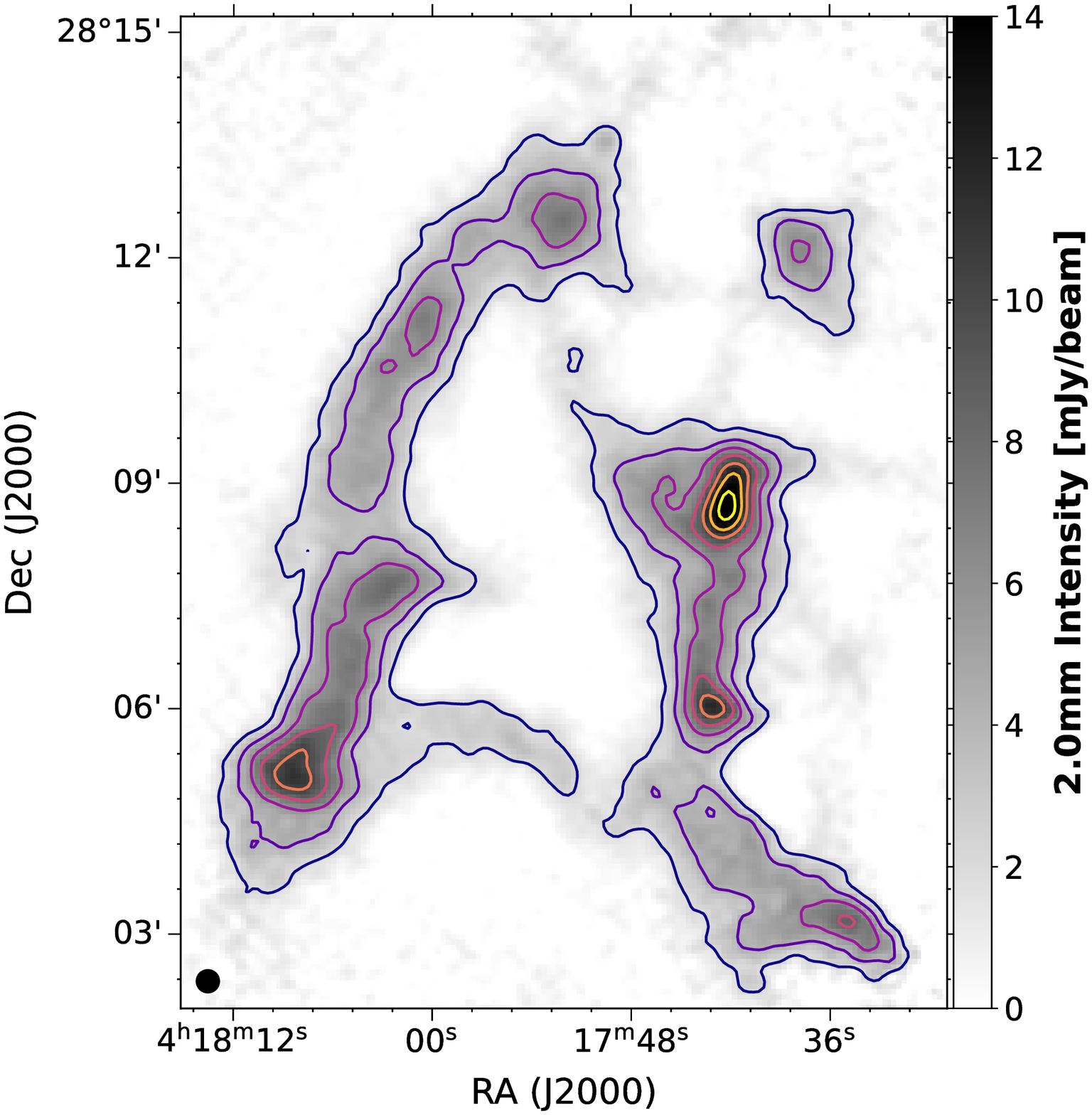} 
\end{array}$
\end{center}
\caption{\label{NIKA2map} Maps of the continuum emission from the NIKA2 instrument on the IRAM 30m telescope of the B10 region. (left) At 1.2mm the angular resolution is 12$^{''}$ (beam size in black in the bottom right corner). Contour levels from 5 -- 40 mJy/beam, increasing by 5 mJy/beam each level ($\sim6\sigma$). (right) At 2.0mm the angular resolution is 18$^{''}$ (beam size in black in the bottom right corner). Contour levels from 2 -- 14 mJy/beam, increasing by 2 mJy/beam each level ($\sim9\sigma$). }
\end{figure*}

\section{Introduction}

Starless cores, as well as bounded and collapsing prestellar cores, give us a unique glimpse at the physical and chemical conditions prior to stellar birth when the initial conditions of star formation are set \citep{2007ARA&A..45..339B}. These objects are ideal probes of the early beginnings of star formation because they are relatively undisturbed, meaning that they have shallow temperature gradients, the absence of an internal heat source, and the absence of shocks or outflows \citep{1994MNRAS.268..276W, 2001ApJ...557..193E, 2014prpl.conf...27A, 2016IAUS..315...95T}. Even so, the internal structure, formation, evolution and manner in which starless cores embedded within filaments collapse under gravity remains enigmatic. A significant problem for low-mass star formation is that initial conditions in the starless cores from which stars form are still not known to sufficient  accuracy. 

The central density of a core is a crucial (but not the sole) evolutionary parameter used to constrain starless core evolution. The best estimates typically come from radiative transfer models of optically thin dust continuum emission at far-infrared through millimeter wavelengths (e.g., \citealt{2001ApJ...557..193E, 2001A&A...376..650Z,  2005ApJ...632..982S, 2016A&A...592A..61L}). The classical picture of isolated star formation is illustrated by a spherically symmetric, non-magnetic, non-rotating, pressure confined core in equilibrium (Bonner Ebert Sphere, BES; \citealt{1955ZA.....37..217E, 1956MNRAS.116..351B}), whose collapse happens inside a thermally-supported, centrally condensed isothermal sphere that have central ``flat" density plateaus ($r_f$) inversely proportional to the square root of the central density, $r_f$ $\propto$ 1/$\sqrt{n_c}$ \citep{1977ApJ...214..488S}. It is imperative to resolve the size of $r_f$ if one wants to constrain the physical shape and structure of starless and prestellar cores. Unfortunately, ALMA and other interferometers resolve out $r_f$ \citep{2016ApJ...823..160D, 2017ApJ...838..114K} for all but the most extremely condensed cores with $n_c > \mathrm{few} \times 10^7 \mathrm{cm}^{-3}$ (L1544; \citealt{2019ApJ...874...89C}). High resolution ($\sim$ 10$''$) sub-millimeter, single-dish dust continuum data, that probe $r_f$ with enough sensitivity to constrain radial profiles are needed.

Because of potential non-ideal magneto-hydrodynamic (MHD) effects that impact core evolution, a `true' core profile most likely strays from the classical BES profile \citep{2014ApJ...785...69C}, and more generalized density profiles should be considered, such as Plummer spheres. MHD effects break spherical symmetry, therefore 3D modelling is needed to better constrain physical parameters such as the aspect ratio of the core in multiple dimensions. In this paper, we obtain high resolution 1.2mm observations of the dust continuum emission in the B10 region of the Taurus Molecular Cloud and we model the physical structure of the starless cores using 3D radiative transfer modelling. 

Perhaps the most widely accepted way to compare the stability of starless cores is via a virial analysis (e.g., \citealt{2020A&A...635A..34K, 2020A&A...638A..74L, 2021A&A...645A..55P}). The virial parameter calculates the balance between internal kinetic and gravitational energies \citep{1992ApJ...395..140B}, where $\alpha_{vir} = 5\sigma_v^2R/aGM$ for a core of mass $M$, radius $R$, a velocity dispersion $\sigma_{v}$ and correction factor $a$ (i.e., $a$ = 1 for a uniform density sphere). Other virial terms for external pressure and internal magnetic fields contribute to this stability analysis, but can often be challenging to include in a consistent way observationally \citep{2017ApJ...838..114K, 2019ApJ...877...93C}. Recently, it has been found that given the various discrepancies that can occur in observational measurements (such as how the core itself is defined and the background subtracted) it is relatively easy to underestimate the virial parameter \citep{2021ApJ...922...87S}. By modelling cores, the inherent density distribution in 3D can be used to calculate virial parameters and robustly compare evolutionary states without relying on line-of-sight observations.

We have targeted the B10 region in the Taurus Molecular Cloud, at a distance of $\sim$135 pc \citep{2014ApJ...786...29S}, to conduct this modelling study because it is less evolved compared to other regions in Taurus, containing only dense starless cores (no Class 0 or I protostars), whose evolution has not been significantly disturbed by the feedback from star formation \citep{2013A&A...554A..55H}. Ammonia, NH$_3$, mapping from \cite{2015ApJ...805..185S} show 10 dense cores in this region (i.e., within the mapping area discussed in this paper) with kinetic temperatures, T$_k$, from $9.47-10.9$\,K. \cite{2019ApJ...871..134S} find from low resolution (66$''$) line observations of HCN ($1-0$) and HCO$^+$ ($1-0$) that the cores in the B10 region do not show signs collapse or infall, suggesting that theses cores are not dynamically evolved and thus we will continue to refer to them as starless cores. Additional maps of CCS and HC$_7$N reveal that neither of these molecules are detected in B10, suggesting perhaps the initial C/O ratio in this region was low \citep{2019ApJ...871..134S}. 

Within B10 there is interesting chemistry, as the interstellar complex organic molecule (COM) acetaldehyde, CH$_3$CHO, was detected in 5 out of the 10 starless cores in this region \citep{2020ApJ...891...73S}. Additionally, methanol (CH$_3$OH) was mapped and readily seen throughout the region \citep{2020ApJ...891...73S, 2022ApJ...927..213P}, where 8 out of the 10 cores had detections of deuterated methanol, CH$_2$DOH \citep{2021MNRAS.501..347A}. To explain the presence of COMs in starless and prestellar cores, various chemical models usually set initial physical conditions based on one prestellar core (e.g., the evolved core L1544; \citealt{2017ApJ...842...33V, 2020ApJS..249...26J}), due to its well modeled physical structure. Thus, in order to better assist in the chemical modelling of more `typical' cores, full temperature and density distribution profiles for a representative sample of cores, like those in B10, is needed.

A young, dense region with interesting chemistry, B10 is the perfect test-bed for modelling the physical and evolutionary properties of one of the earliest stages of low-mass star formation. Below, we discuss the observational data of B10 in section\,\ref{sec:obs}. A description of the substructure, picked out by a dendrogram analysis, is laid out in section\,\ref{substructure}. In section\,\ref{pandora} we describe the procedure and results from our 3D radiative transfer modelling. We then use our best-fit models for each core to calculate virial parameters in a self-consistent way and discuss constraints on dust opacity (see section\,\ref{discussion}). Our conclusions are summarized in section\,\ref{conclusions}.

\section{Observations} \label{sec:obs}
Multiwavelength and multiscale data from dust continuum observations at the sub-mm to the far-infrared wavelengths are crucial for modelling the physical structure of B10. Below is a description of newly obtained IRAM 30m NIKA2 dust continuum observations of B10 at 1.2mm and 2.0mm, as well as archival PACS/SPIRE \textit{Herschel} data that has been recently re-analyzed in \cite{2022ApJ...941..135S}.

\begin{figure*}
\includegraphics[width=175mm]{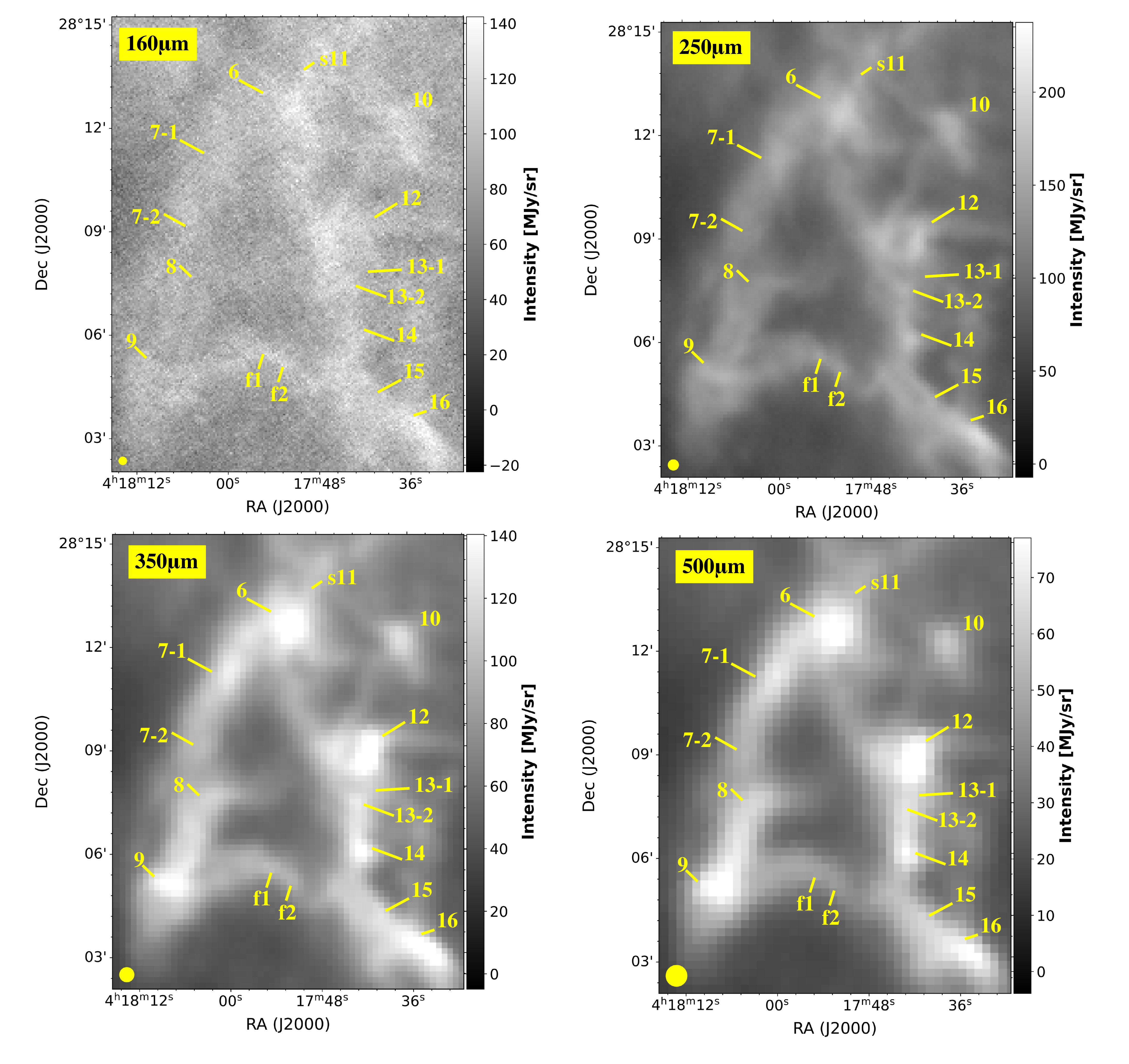}
\caption{\label{hershmaps} \textit{Herschel} intensity maps of the B10 region at 160, 250, 350 and 500$\mu$m in MJy/sr and at resolutions of 13.5$^{''}$, 18.1$^{''}$, 24.9$^{''}$ and 36.4$^{''}$, respectively. Beam size in yellow in the bottom right corner of each map. Overlaid in yellow are the labeled structures picked out by the high resolution 1.2mm NIKA2 data described in section\,\ref{substructure}. }
\end{figure*}

\begin{table*}
	\caption{Core Parameters 1.2mm Dendogram `Leaf' Structures}
	\label{Core_Parameters}
	\begin{tabular}{lllccccccc} 
        Core & RA$^{1}$ & DEC$^{1}$ & area$_\mathrm{ellipse}$ & area$_\mathrm{exact}$ &  radius$_\mathrm{ellipse_{major}}$  & radius$_\mathrm{ellipse_{minor}}$ &
        radius$_\mathrm{spherical_{exact}}$ & PA$^{2}$   \\ 
         &  J2000 & J2000 & arcsec$^2$ & arcsec$^2$   & arcsec & arcsec  & arcsec & deg  \\
        \hline
        6   &  4:17:52.2209 & +28:12:28.520  &  2281 & 5832  & 30 & 17 & 43 & -163 \\ 
        7-1 & 4:18:00.5836 & +28:11:06.373   &  1731 & 5013  &  37 & 11 & 40&  60  \\ 
        7-2 & 4:18:03.2918 & +28:09:06.813 & 484  & 1143  &  11 & 10 & 19 &  52   \\ 
        8 & 4:18:03.3442 &  +28:07:36.745   &  560 & 1575  &  16 & 8&  22 & -147   \\ 
        9 & 4:18:08.2906 & +28:05:08.226  & 2367  & 5895 &    39 & 14 & 43 &  63  \\ 
        10 &4:17:37.9308 & +28:12:07.410&  1638 & 5481 & 25 & 15 & 42 & 118  \\ 
        11s &  4:17:49.4966 &  +28:13:28.730 &  92 &  261 & 5  &  4 &  9 &  77  \\ 
        12 & 4:17:42.2077 & +28:08:43.138   &  1223 & 4032  &  22 & 13 & 36 & 68 \\ 
        13-1 & 4:17:41.8229 & +28:07:43.598 &  100 &  243  &  6 & 4 & 9 & 60  \\ 
        13-2 & 4:17:43.4190 & +28:07:23.105   & 120  & 342 &  6 & 5 & 10 & -160 \\ 
        14 & 4:17:43.0169 & +28:06:00.570    &  726 & 1773 &   20 & 8 & 23 & 100 \\ 
        15 & 4:17:41.2439  & +28:03:56.117   & 402  & 873  & 12 & 7 &  17 & 68  \\ 
        16 & 4:17:34.8089  & +28:03:05.023   &  1772 & 4149  &  31 & 13 & 36 & 161 \\ 
        f1 & 4:17:55.2360  & +28:05:34.579  &  204 & 558  &   8 & 6 &  13& 147 \\ 
        f2 & 4:17:52.5708  & +28:05:10.033   &  201 & 603  &  9 & 5 & 14 & 116 \\  
		\hline
	\end{tabular}
      \begin{description} 
      \item $^{1}$These are the peak dust positions we use in our modelling and not necessarily the `central' values outputted from the dendrogram analysis. $^{2}$ Position angle is calculated directly from the dendrogram analysis. Note: we include here the dimensions for the smaller `ellipse' approximations that \code{astrodendro} calculates, but the spherical symmetric `exact' radius value (derived from area$_\mathrm{exact}$ as in Figure\,\ref{dendofig}) should be used for comparison to our models.
      \end{description}
\end{table*}

\subsection{IRAM 30m}\label{iram30}
Observations of the B10 region, located within the Taurus Molecular Cloud, with the NIKA2 instrument on the IRAM 30m telescope in Pico Veleta, Spain were taken during the winter 2019 pool season (October and November). The observations cover a region of $15^{'} \times 15^{'}$ centered at $\alpha$(J2000)\,=\, 04$^\mathrm{h}$17$^\mathrm{m}$53$^\mathrm{s}$ and $\delta$(J2000)\,=\,+28$^{\circ}$08$^{'}$42$^{''}$. Dual-band capability of NIKA2 allowed us to simultaneously observe at 1.2mm and 2.0mm. The detector array illuminated by the 150 GHz (2mm) beam has been named Array 2 and the 260 GHz (1.2mm) channel, which has both a horizontal component and vertical component, is referred to as Array 1 and Array 3, respectively. The effective frequencies ($\nu_\mathrm{eff}$) for these bands (at 2mm precipitable water vapor) are 254.2, 257.1 for Array 1 and 3 (the 1.2mm band) and 150.6 GHz for Array 2 \citep{2020A&A...637A..71P}. We refer to the combined Array 1 and 3 as our `1.2mm map' and Array 2 as our `2.0mm map.' The effective beam FWHM are measured to be 12$^{''}$ and 18$^{''}$ for the 1.2mm and 2.0mm maps, respectively. Observing conditions required sky stability with a median opacity value of 0.35. Beam maps were done once a day on average, where Uranus was used as the primary calibrator.

The NIKA2 data was reduced using the Pointing and Imaging In Continuum (PIIC) software, which is part of the GILDAS\,\footnote{\url{http://iram.fr/IRAMFR/GILDAS/}} family of packages  (\citealt{2005sf2a.conf..721P, 2013ascl.soft05010G}). Our modified version of the provided NIKA2 pipeline was run to perform flat field correction, correlation correction, sky subtraction, baseline subtraction, calibration, geometry association, and re-gridding on the final map. Because B10 is an extended object with complex geometry, we do not define any a priori source definition and use an iterative mode of reduction (10 iterations for each map). For each iteration, all data reduction operations are repeated, using a new source definition. The rms noise levels for our final maps are $\sigma_{1.2mm}$\,=\,0.75\,mJy/beam and $\sigma_{2mm}$\,=\,0.22\,mJy/beam at 1.2mm and 2.0mm, respectively. Brightness maps of the two NIKA2 maps are shown in Figure\,\ref{NIKA2map} in units of mJy/beam. 

To validate our reduction process, we calculate if the relative flux between our two reduced maps is reasonable. We construct a ratio map, $R_{1,2} = \frac{I_{1.2mm}}{I_{2.0mm}}$, of the two NIKA2 intensity maps after re-gridding, convolving to a common 18$^{''}$ resolution, and converting from units of mJy/beam to MJy/sr. Note that only pixels $>$ 6$\sigma_{1.2mm}$ were used to create the ratio map (see section\,\ref{betasection} for more description on the beam convolution). We find reasonable estimates for the B10 region, with a median $R_{1,2}$ of 6.6. Our analysis follows that of \cite{2017A&A...604A..52B}, who find for B213, another nearby region of Taurus, their $R_{12}$ histogram peaks at $\sim$\,7, using NIKA data. 

\subsection{Herschel}
We make use of dust continuum intensity maps of the Taurus Molecular Cloud, from the \textit{Herschel Space Observatory} Gould Belt Legacy Survey \citep{2010A&A...518L.102A}. The intensity maps at 160$\mu$m, 250$\mu$m, 350$\mu$m, and 500$\mu$m (Figure\,\ref{hershmaps}) have been run through an optimized Spectral Energy Distribution (SED) pipeline (see \citealt{2022ApJ...941..135S}) that was first corrected for zero-point effect by using the \textit{Planck} dust models (see \citealt{2014A&A...571A..11P}). The maps, in units of MJy/sr, are at resolutions of 13.5$^{''}$, 18.1$^{''}$, 24.9$^{''}$ and 36.4$^{''}$, respectively. We have also made use of the corresponding `Herschel Optimized Tau and Temperature' or `HOTT' maps from \citealt{2022ApJ...941..135S}\footnote{\url{https://www.cita.utoronto.ca/HOTT}}. The column density, $N$($\mathrm{H}_2$ [cm$^{-2}$]), dust temperature, T$_{d}$ [K], and dust emissivity, $\beta$, maps we use are all set to the 500$\mu$m intensity map resolution, i.e., 36.4$^{''}$.

\section{Core and Filament Substructure}\label{substructure}

The high resolution (12$^{''}$) 1.2mm dust map of B10 allows us to separate out dust substructure into cores and filaments. The Python code \code{astrodendro} was implemented to deconstruct the B10 region's substructure \citep{2008ApJ...679.1338R}. The algorithm works by starting from the brightest pixels in the data and adding fainter and fainter pixels to ``leaves,'' ``branches,'' and ``trees,'' similar to that of other dendogram-like procedures (i.e., \code{getsources}; \citealt{ 2012A&A...542A..81M}).

A minimum value, or \code{min$\_$value}, in the code sets the noise level where pixels below this value are not included in the analysis. In our analysis we choose this minimum height to be 5\,mJy/beam ($\sim\,7\sigma_{1.2mm}$), to avoid including noise around the edges of the B10 filament (see Figure\,\ref{NIKA2map}). This value defines the outer edge and area within where over-densities, i.e., cores or ``leaves'', will be found. 

The minimum number of pixels necessary for a structure  to be considered independent within the dendrogram (denoted \code{min$\_$npix} in the code) needs to be calculated and defined. To determine this value we first find the number of pixels contained within the beam's solid angle of the observations, $N$, assuming a Gaussian beam with a FWHM, denoted $\theta_0$,

\begin{equation}
    \Omega_A = \frac{\pi \theta_0}{4 \ln(2)},
\end{equation} which gets divided by the area per pixel within the image, $CDELT$. Thus, the final value for \code{min$\_$npix}, or $N$, becomes; 

\begin{equation}
    N = \frac{\Omega_A}{CDELT}. 
\end{equation}

\begin{figure}
\centering
\begin{center}$
\begin{array}{c}
\includegraphics[width=82mm]{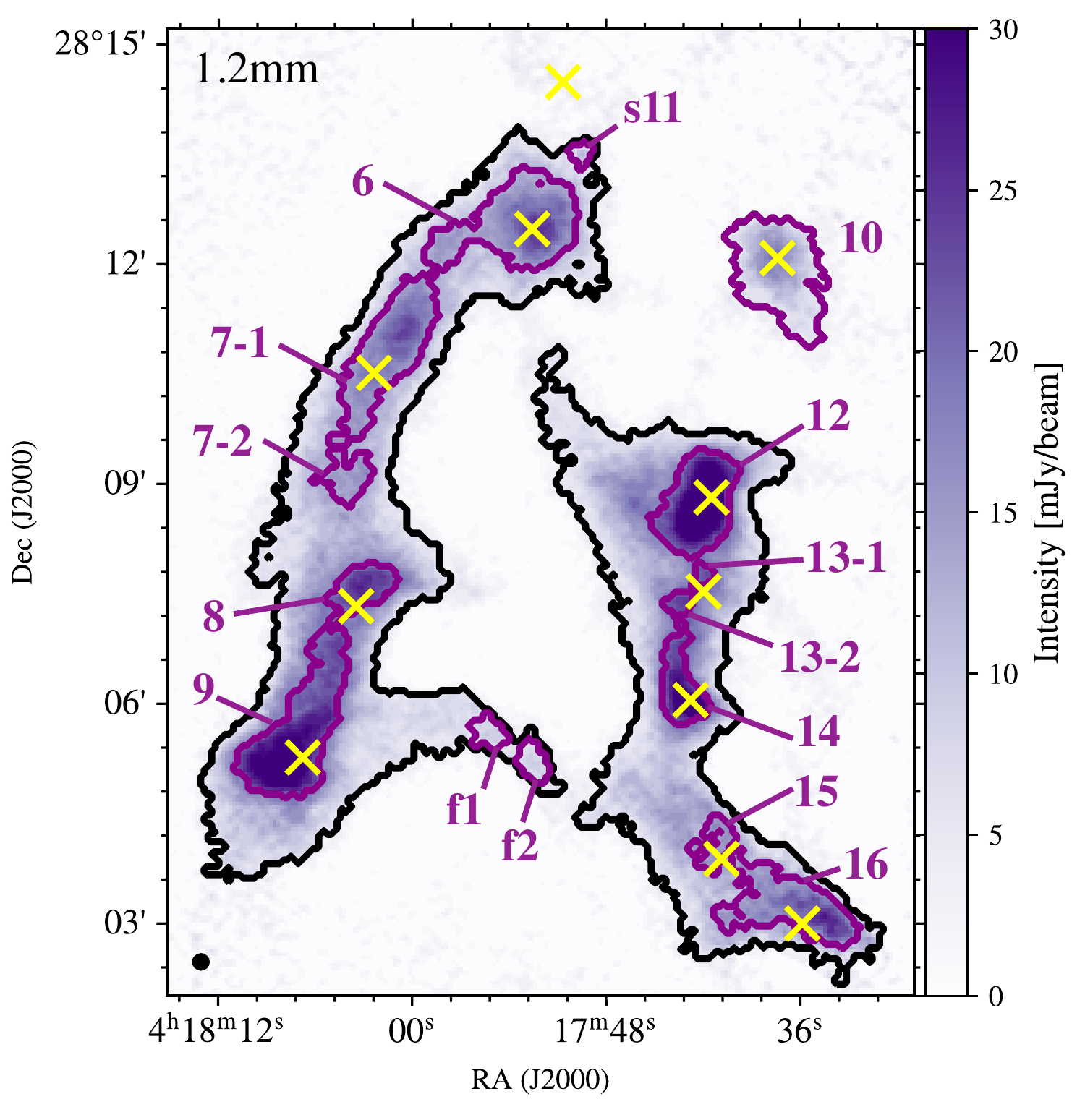}
\end{array}$
\end{center}
\caption{\label{dendofig} Continuum emission of the B10 region at 1.2mm at an angular resolution is 12$^{''}$ (beam size in black in the bottom right corner). Outlined in black are the filamentary structures, or ``branches'' picked out by the dendrogram analysis code \code{astrodendro}. Purple contours are the core structures or ``leaves'' picked out by \code{astrodendro}. A total of 15 ``leaves'' are found. Yellow crosses mark where the peaks of the NH$_3$ cores lie \citep{2015ApJ...805..185S}.   }
\end{figure}

For the 1.2mm  map $N = 4.5$. While this parameter marks the minimum pixel value for an independent structure, the code can still find sub-structures that do not obey this constraint. Therefore, a minimum difference, or \code{min$\_$delta} in the code, is needed to determine the number of `leaves' or substructure within the filaments. This is essentially the minimum difference between local maxima required to consider them as part of different dendrogram components. 
We set this parameter to the lowest value possible without picking out leaves smaller than our pixel size in arcseconds ($<$ 3$^{''}$), which leads to a value of 2.6\,mJy/beam ($\sim4\sigma_{1.2mm}$). 
A total of 15 cores or `leaves' were found within the filamentary structures or `branches' (Figure\,\ref{dendofig}). Only one core (10) is considered both a `branch' and `leaf' as it is isolated in the top right of the region.

We find similar core structures to that determined in the NH$_3$ mapping analysis done in \cite{2015ApJ...805..185S} that found 10 cores within this region. We stick with a naming convention that correlates with the NH$_3$ cores (Figure\,\ref{dendofig}, Table\,\ref{Core_Parameters}). In our analysis, core 7 and 13 are broken into two separate substructures, and we do not see core 11 most likely due to low sensitivity at the edges of our NIKA2 map. We do identify a substructure we call `s11' that correlates  spatially with the Class II object within this region \citep{2010ApJS..186..259R} and thus do not include it in any future analysis. From our map we also identify two completely new leaves, f1 and f2, that lie along a filament to the east of core 9 that were not identified in the NH$_3$ map. The core and filament structures defined here by the dendrogram analysis are used as starting points for our radiative transfer analysis in the proceeding sections, where we further constrain the morphology and physical structure of B10, aided by the incorporation of the other intensity maps at different wavelengths.

\section{3D Radiative Transfer modelling} \label{pandora}

In addition to picking out new structures, our NIKA2 maps allow us to probe the inner region (12'' or 1620 AU at 1.2mm and 18'' or 2430 AU at 2.0mm) of the starless cores embedded in the B10 filament. The goal now is to use this data in conjunction with radiative transfer modelling techniques to better constrain the evolutionary parameters for each core. We use the framework \textit{pandora} as introduced by \cite{2016A&A...588A.143S} to construct 3D radiative transfer models for the cores in the B10 region. To do this, within the framework, the publicly available radiative transfer program RADMC-3D is used (\citealt{2012ascl.soft02015D} , version 2.0) as well as the python package \code{astropy} for post-processing \citep{astropy:2013, astropy:2018, astropy:2022}. We discuss below 1) the \textit{pandora} set-up, 2) our general method for fitting cores, 3) a more detailed description of our fitting prescription for core 12, and, lastly, 4) we discuss the results for all cores and provide comparisons. 

\subsection{Set-up for the \textit{pandora} Framework}

The model grid is laid out on a positively right-handed Cartesian coordinate system, where the $x$-axis points west on the sky, the $y$-axis points to the north, and the $z$-axis points towards the observer. The origin of the system is denotad by a \code{refPosition}, which is chosen to be the center of B10, i.e., $\alpha$(J2000)\,=\, 04$^\mathrm{h}$17$^\mathrm{m}$53$^\mathrm{s}$ and $\delta$(J2000)\,=\,+28$^{\circ}$08$^{'}$42$^{''}$. An adaptive mesh refinement technique within RADMC is used to improve the spatial resolution of the numerical radiative transfer simulation \citep{1984JCoPh..53..484B, 1989JCoPh..82...64B, 1998JCoPh.143..519K}. The global parameters for this AMR grid are listed in Table\,\ref{tab:global_params}. 

Dust temperature is calculated self-consistently within RADMC-3D using the Monte Carlo method of \cite{2001ApJ...554..615B}. There is also a Modified Random Walk method that is implemented in RADMC-3D and described in \cite{2010A&A...520A..70R}. By including a random walk method, computation time is saved since RADMC-3D can make one single large step of the photon package. For a more detailed description on how the dust temperature is calculated, see section 3.6 in \cite{2016A&A...588A.143S}. 

Important to reproducing a 3D core structure are the density structures within our model. An overall density structure is obtained by the superposition of the density profiles of all dust cores. The density is summed up, where in each cell \textit{j} is determined by, 

\begin{equation}
n_j = \sum_{i=1}^N n_{i,j}(r),
\end{equation} where $i$ is the index of the dust cores and $N$ is the number of cores. A modified Plummer-like profile (see \citealt{2011A&A...530L...9Q}) is used to define the cores. The distribution is defined as follows,

\begin{equation}\label{nr_pandora}
n_i(\vec{\textbf{r}}) = \frac{n_c}{ (1 + \left(\vec{\textbf{r}})^2\right)^{\eta/2}},
\end{equation} where $n_c$ is the central number density given in H$_2$ cm$^{-3}$ and the magnitude of $\textbf{r}$ is given by the Euclidean norm, including scaling factors, 

\begin{equation}\label{r_def}
|\vec{\textbf{r}}| = \sqrt{\left(\frac{r_x}{r_{0,x}}\right)^2 + \left(\frac{r_y}{r_{0,y}}\right)^2 + \left(\frac{r_z}{r_{0,z}}\right)^2},
\end{equation} 
where $r_{x,y,z}$ are the components of $r$ and $r_{0,x}$, $r_{0,y}$ and $r_{0,z}$ set the size of the flat density plateau in each of the axes. For full 3D treatment $r_{0,z}=r_{0,x}$ in our models.  We can reach a spherical symmetric distribution by setting $r_{0} = r_{0,x} = r_{0,y} = r_{0,z}$. Inside $r_{0}$ is our `flat' region. When $r >> r_0$ the profile reaches a power-law distribution with an exponent $\eta$. 

\begin{table}
	\caption{Global \textit{pandora} parameters }
	\label{tab:global_params}
	\begin{tabular}{lc} 
        \textbf{Distance to source} & \\
        $d$ & 135 [pc]  \\
        \hline
        \textbf{Model Center} &  \\
        \code{refPosition} &  04:17:53 +28:08:42 [J2000] \\
        \hline
        \textbf{AMR Gridding} &  \\
         \code{nbCells}  & 11 \\
          \code{cubeHalfSize}  & 0.5 \\
          \code{amr\_minCellSize}  & 10 \\
          \code{amr\_maxTotalNBCells}$^{a}$  & 1e8 \\
          \code{amr\_flag\_dustDensGradient}$^{b}$  & 1 \\
          \code{amr\_dustDensGradient}  & 20 \\
          \code{numberPhotons}  & 1e6 \\
		\hline
	\end{tabular}
      \small
      \begin{description}
      \item $^{a}$The maximum number of total cells allowed. $^{b}$This parameter will refine the grid based on the gradient in dust density.
      \end{description}
\end{table}

\begin{table}
	\caption{Core Grid \textit{pandora}}
	\label{tab:pandora_grid_params_current}
	\begin{tabular}{ll} 
    $n_0$ (cm$^{-3}$) & [1.0e4, 5.0e4, 1.0e5, 2.0e5, 3.0e5, 4.0e5, 5.0e5,\\ & 6.0e5 7.0e5, 8.0e5, 9.0e5, 1.0e6, 2.0e6] \\
    \hline
    $\eta$ & [1.5, 2.0, 2.5, 3.0, 3.5, 4.0, 4.5, 5.0, 5.5] \\
    \hline
    $r_{0,x}$ (AU)& [1620, 1890, 2160, 2430, 2700, 2970, 1890, 2160, 2430,\\ & 2700, 2970, 3240, 3510, 3780, 4050, 4320, 4590, 4860, \\ &5130, 5400, 5670, 5940, 6210, 6480, 6750]  \\
    \hline 
    $r_{0,y}$ (AU)& [1620, 1890, 2160, 2430, 2700, 2970, 1890, 2160, 2430,\\ & 2700, 2970, 3240, 3510, 3780, 4050, 4320, 4590, 4860, \\ &5130, 5400, 5670, 5940, 6210, 6480, 6750]  \\
    \hline
    $s_{isrf}$ & 0.3, 0.6, 1.0, 2.0, 3.0 \\
    \hline 
    O${\&}$H94 & 0, 1, 10, 11 \\
    \hline 
	\end{tabular}
      \small
      \begin{description}
      \item For O${\&}$H94: 0 = bare0.tab (MRN distribution of grains, no initial gas density, with no ice mantles), 1 = bare5.tab (MRN distribution of grains, coagulated at 10$^{5}$ cm$^{-3}$, with no ice mantles),
      10 = thin0.tab (MRN distribution of grains, no initial gas density, with thin ice mantles), 11 = thin5.tab (MRN distribution of grains, coagulated at 10$^{5}$ cm$^{-3}$, with thin ice mantles). 
      \end{description}
\end{table}

\begin{figure}
\centering
\begin{center}$
\begin{array}{c}
\includegraphics[width=81mm]{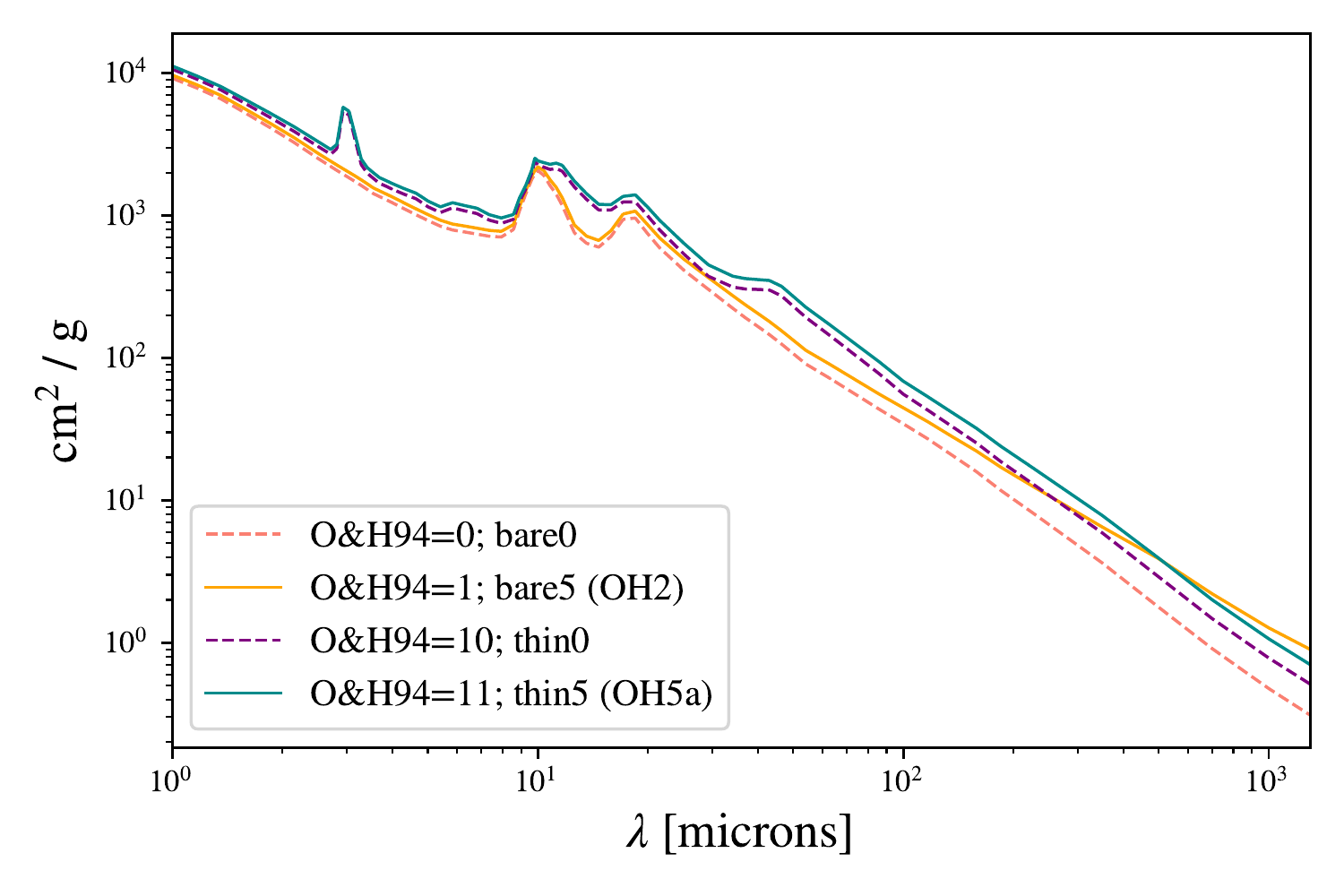} 
\end{array}$
\end{center}
\caption{\label{opacityfig} Dust opacity laws (extinction/absorption coefficient per gram of dust) used in our \textit{pandora} grid from \citealt{1994A&A...291..943O}. A gas-to-dust ratio of 100 has been assumed. Laws either have a thin ice mantle (`thin') or no ice mantle (`bare'), and have either a gas density of 10$^{5}$ cm$^{-3}$ (`5') or no initial gas density (`0'). }
\end{figure}

\begin{figure}
\centering
\begin{center}$
\begin{array}{c}
\includegraphics[width=81mm]{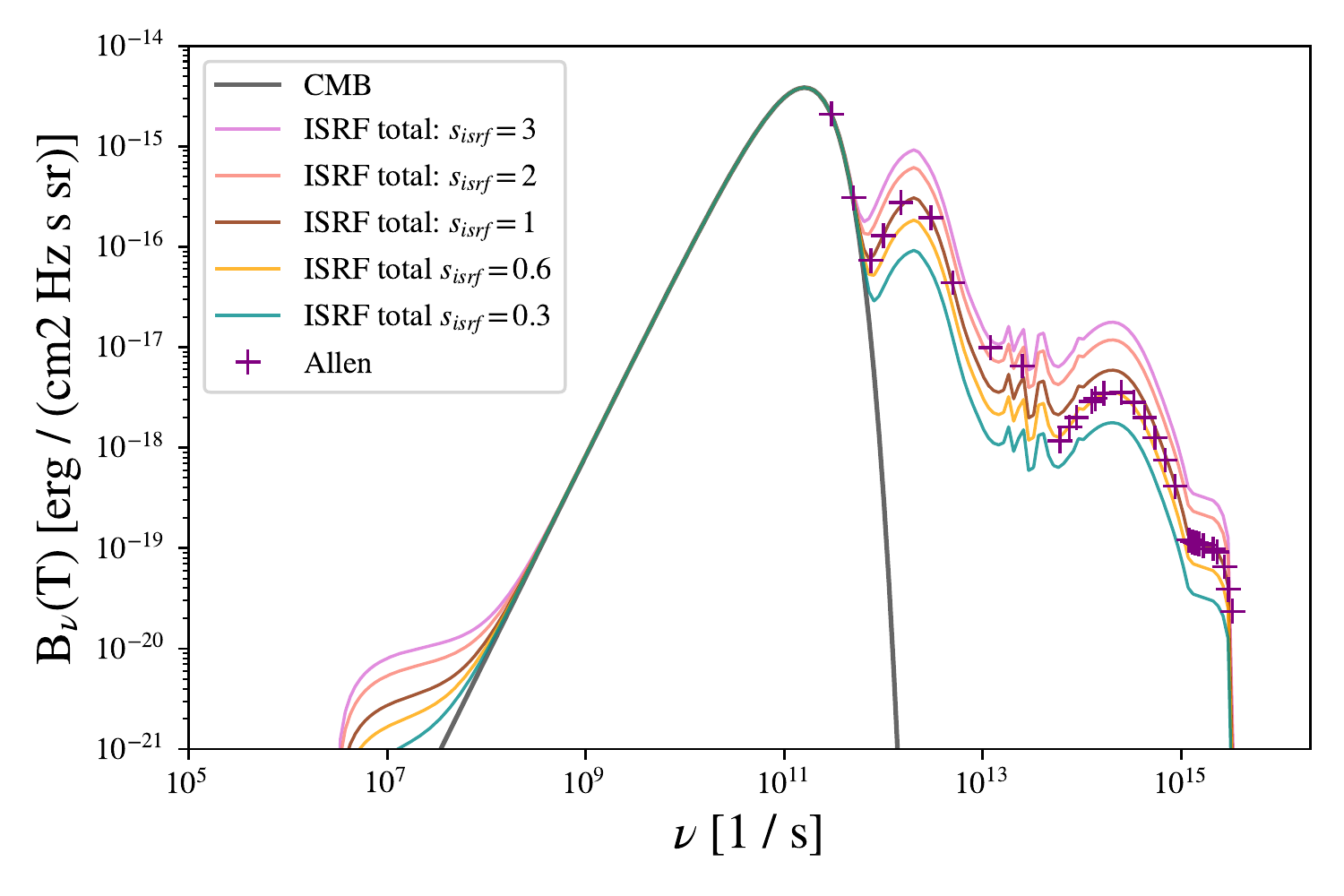} 
\end{array}$
\end{center}
\caption{\label{sisrf} The Interstellar Radiation Field (ISRF) used in the \textit{pandora} models \citep{1983A&A...128..212M, 2007ApJ...657..810D}. We scale the stellar and dust contributions of ISRF by a factor, $s_{isrf}$, in our models while allowing the CMB contribution (in black) to remain as is. }
\end{figure}

\begin{figure*}
\begin{center}$
\begin{array}{c}
\includegraphics[width=170mm]{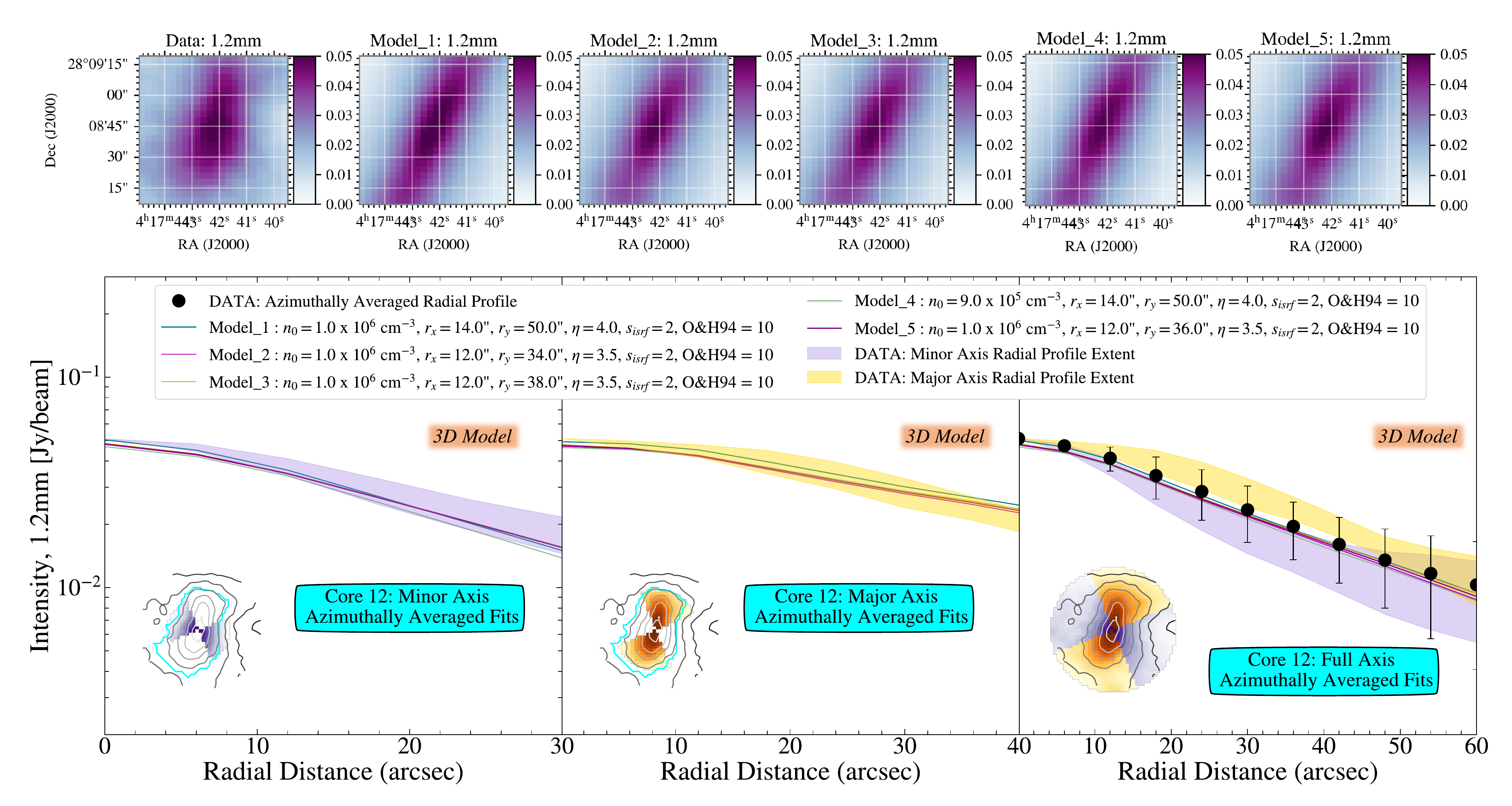}
\end{array}$
\end{center}
\caption{\label{2D_core12} Best fit \textit{pandora} model results for core 12. (top) 2D intensity maps for 1.2mm emission show the NIKA2 data and the five best-fit models (intensity units in Jy/beam). (bottom) In order for these best-fit models to qualify, the average radial profile around the model minor axis sectors (left panel; purple fill) and the major axis sectors (middle panel; yellow fill) needs to fit within the extent of the data plus its standard deviation error. By following this prescription, these models fit the full azimuthally average data (right panel; black points) and thus models need to have azimuthally averaged profiles located in between the sectored profiles or in the overlapping regions in order to qualify as best-fit. For the remaining cores we show only the full azimuthally averaged profiles (Figure\,\ref{sector_plot}). 
}
\end{figure*}

The initial morphology, positions, radii, and central densities of the cores in the modeled B10 dust maps were aided by the dendogram analysis described in section\,\ref{substructure} as well as preliminary 1D isothermal modelling that we do not present here. Once parameters are set, the \textit{pandora} framework creates the adaptive mesh grid and calculates the dust temperature, as well as the dust continuum distribution. 

The code then outputs continuum fits files, convolved with a single Gaussian of same FWHM as the effective telescope beam size (pixel scale), i.e., 12.0$''$(3.0$''$), 18.0$''$(4.0$''$), 13.5$''$(3.0$''$), 18.1$''$(6.0$''$), 24.9$''$(10.0$''$), and 36.4$''$(14.0$''$), for the 1.2mm and 2.0mm NIKA2 maps as well as the 160, 250, 350 and 500$\mu$m \textit{Herschel} emission maps, respectively.
It should be noted that while the NIKA2 beam profiles measured at the telescope are best fit by a three-Gaussian function (see Table 5 and Figure 6 in \citealt{2020A&A...637A..71P}), we find that the deviations introduced by using a single Gaussian FWHM beam are $< 3\%$ on the modeled profile compared to using the three-Gaussian beam profile, and thus do not affect our results. \cite{2017A&A...604A..52B} also find in their NIKA data that the error due to assuming a Gaussian beam is negligible.
The procedure described is reproduced at all wavelengths and done simultaneously. In the following sections we describe in detail the individual core parameters, as well as our methods for determining our best-fit 3D models. 
Again, for more specifics on the \textit{pandora} framework itself see \cite{2016A&A...588A.143S}.

\subsection{Fitting Cores} \label{fittingcores}

Grids of \textit{pandora} core models were run with varying central density (n$_c$), flat radius in the x-direction ($r_{0,x}$), flat radius in the y-direction ($r_{0,y}$), and exponent in the Plummer profile ($\eta$) from equation\,\ref{nr_pandora}. We also adjusted the dust opacity law, from tables in \cite{1994A&A...291..943O} (O\&H94)\footnote{\url{https://hera.ph1.uni-koeln.de/~ossk/Jena/tables.html}}, for grains coagulated at 10$^5$ years either with no initial gas density or $n = 10^5\mathrm{cm}^{-3}$ and an assumed gas-to-dust ratio of 100 (see Table\,\ref{tab:pandora_grid_params_current} and Figure\,\ref{opacityfig}).
Our opacity model `11' is the same referenced to as `OH5a' in \cite{2016A&A...592A..61L}, who found it closest to the observed properties of their starless core sample where $\kappa_{1.2mm}=0.007$\,cm$^2$/g. Additionally, our opacity models `11' and `1' are very similar to `OH5' and `OH2', respectively, in \cite{2001ApJ...557..193E}, the difference being they pick the slightly higher 10$^6$\,cm$^{-3}$ gas density (see Figure\,\ref{opacityfig}). For O\&H94\,=\,10, $\kappa_{1.2mm}=0.005$\,cm$^2$/g (same as used in \citealt{2002ApJ...569..815T}). The intensity of the interstellar radiation field is varied as well, scaled by a set factor, $s_{isrf}$, based on the model from \cite{1983A&A...128..212M} and \cite{2007ApJ...657..810D}, and plotted for reference in Figure\,\ref{sisrf}.

\begin{figure*}
\begin{center}$
\begin{array}{c}
\includegraphics[width=170mm]{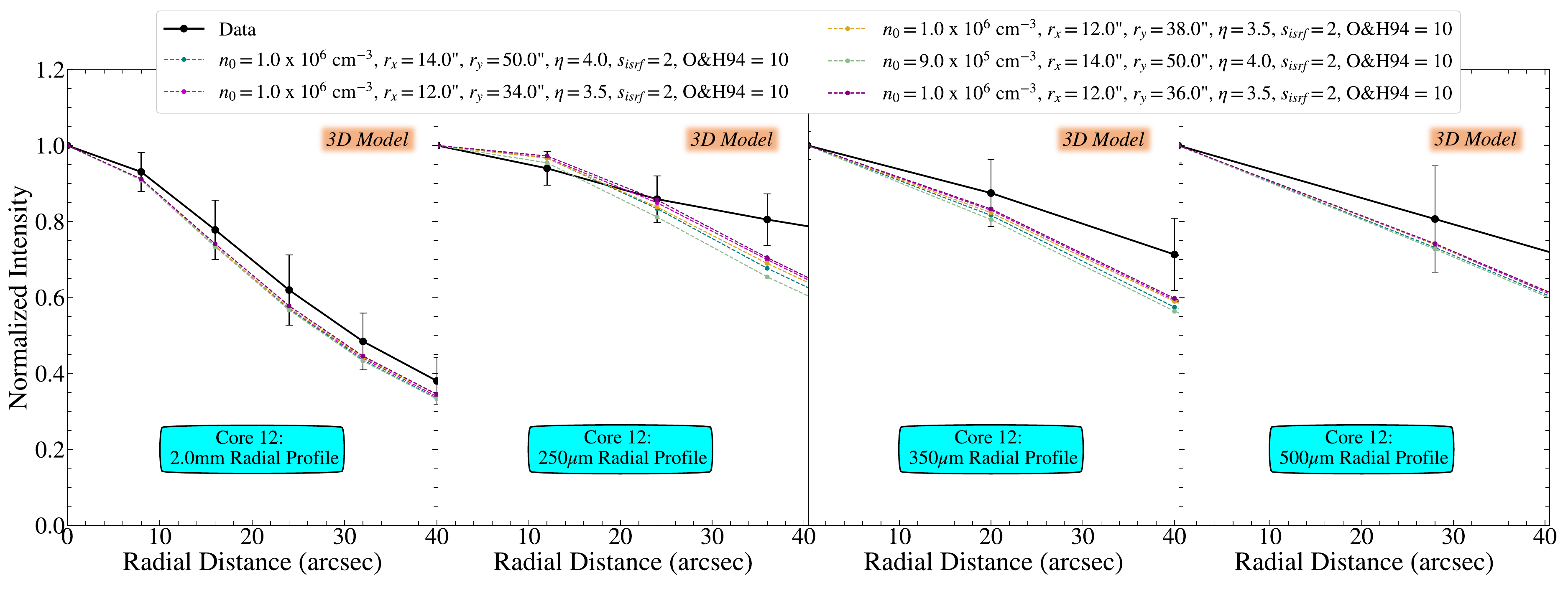}
\end{array}$
\end{center}
\caption{\label{core12_rad_2mm_250_350_500} Best fit \textit{pandora} model results for core 12  out to the major axis limit of 40". Radial profiles are plotted, from increasing (poorer) resolution, for the 2.0mm (18'' resolution), 250$\mu$m (18.1'' resolution), 350$\mu$m (24.9'' resolution) and 500$\mu$m (36.4'' resolution). Each azimuthally average radial profile was done within an annulus corresponding to two times the pixel scale of the map to properly sample the data. Note that we to not attempt to fit the 160$\mu$m emission because the cores are barely seen (see Figure\,\ref{hershmaps}). At the shorter wavelengths it is hard to fit only a few data points and thus we do not use \textit{Herschel} radial profiles to constrain our best fits. We do acknowledge that the models are slightly steeper than what has been observed at the \textit{Herschel} wavelengths, likely do to the extended emission not picked up in the NIKA2 data. 
}
\end{figure*}

In total 1,040,000 models were run for 6 different wavelengths and the parameters are listed in Table\,\ref{tab:pandora_grid_params_current}. 
Grids are limited due to each model taking about a couple of minutes of computing time. To speed up the process, we ran sub-grids in parallel using high performance computing (HPC) resources provided by the University of Arizona using thousands of hours of CPU time. 

The main diagnostics we use to find best-fits for our \textit{pandora} models are 1) normalized radial profile and 2) peak intensity comparisons. For the data itself, as well as for the \textit{pandora} models, we take the 1.2mm dust continuum emission maps and azimuthally average the emission within a 6$^{''}$ annulus (2 pixels) where the standard deviation within this annulus is the corresponding error. The extent of the radial distance for the fit, $r_d$, is 60$^{''}$, chosen as a standard value to encompass emission within the majority of the dendrogram structure for each core. A $\chi^2$ value is used as our diagnostic to find best-fit radial profiles, which we define here as, 

\begin{equation}\label{equ:chi}
    \chi^2 =  \sum_{i=1}^{n}   \left(\frac{I_i^\mathrm{obs} - I_i^\mathrm{mod}}{\sigma_i^\mathrm{obs}}\right)^2\;\;,
\end{equation} where $n$ is the number of radii points, $I_i^\mathrm{obs}$ is the observed intensity at each point, $I_i^\mathrm{mod}$ is the modeled intensity and $\sigma_i^\mathrm{obs}$ is the standard deviation associated with the observed intensity within the annulus used to calculate each point. 

When comparing the radial profile of the observed data to the \textit{pandora} models, we also make sure that the modeled best-fits lie within the azimuthally averaged major and minor axis radial profiles out to the \textit{dendrogram radius} to 1) better probe the 2D shape and 2) set constraints for the smaller ($<60^{''}$) cores. The position angle from the dendrogram analysis for each core, listed in Table\,\ref{Core_Parameters}, is also used to find the angle along which the minor and major axis lie and when comparing to the models this position angle offset sets the axis for which a `major' and `minor' sector is defined.

To help break additional degeneracy issues, we also compare modeled peak intensities (in Jy/beam) to the data at all available wavelengths, i.e., the NIKA2 1.2mm and 2.0mm data as well as the \textit{Herschel} maps at  160$\mu$m, 250$\mu$m, 350$\mu$m and 500$\mu$m. Because we are only modelling the cores and not the filaments they are embedded within, the intensity of the center pixel is used to calculate a flux in order to mitigate effects from background emission in a larger beam. Errors on the peak intensity for the 1.2mm and 2.0mm points are considered to be three times the rms noise level of the maps, i.e., $3\sigma_{1.2mm}$ and $3\sigma_{2mm}$, respectively. For the \textit{Herschel} data, the median rms value from three empty regions on the map was used to estimate an error. We note that comparisons of radial profiles at the other wavelengths was not performed due to poorer resolution and thus lack of any real constraint on physical properties that depend on resolution, such as the aspect ratio ($r_{0,x}$, $r_{0,y}$).

\subsubsection{Test Case: Modelling Results for Core 12 } \label{core12}

We first discuss the modelling results for core 12, located in the upper region of the right most filament in the B10 region (see Figure\,\ref{dendofig}). Because of this, each core in our model grid is centered on the position of core 12 ($\alpha$\,=\,04$^\mathrm{h}$17$^\mathrm{m}$42.2077$^\mathrm{s}$ and $\delta$(J2000)\,=\,+28$^{\circ}$08$^{'}$43.138$^{''}$). Rather than a conglomerate of all 14 cores positioned in a single B10 map, we found a series of singular core models was better to implement because it allowed us to vary the $s_{isrf}$ and the opacity law for each core.
Currently, the \textit{pandora} framework does not allow multiple cores in a single `B10' model to have varying $s_{isrf}$ or opacity laws.

Once the 1,040,000 core models were run at each wavelength, we then match best-fit models by comparing to the data. First, we restrict our grid to contain only models with peak 1.2mm intensities that fall within the error bar and that have peak 250$\mu$m intensities that fall within the error bar. The 250$\mu$m peak is chosen because it should be near the peak of the SED and thus sensitive to s$_{isrf}$. For core 12, there were 8,627 models left that fit this criteria. 

\begin{figure}
\begin{center}$
\begin{array}{c}
\includegraphics[width=82mm]{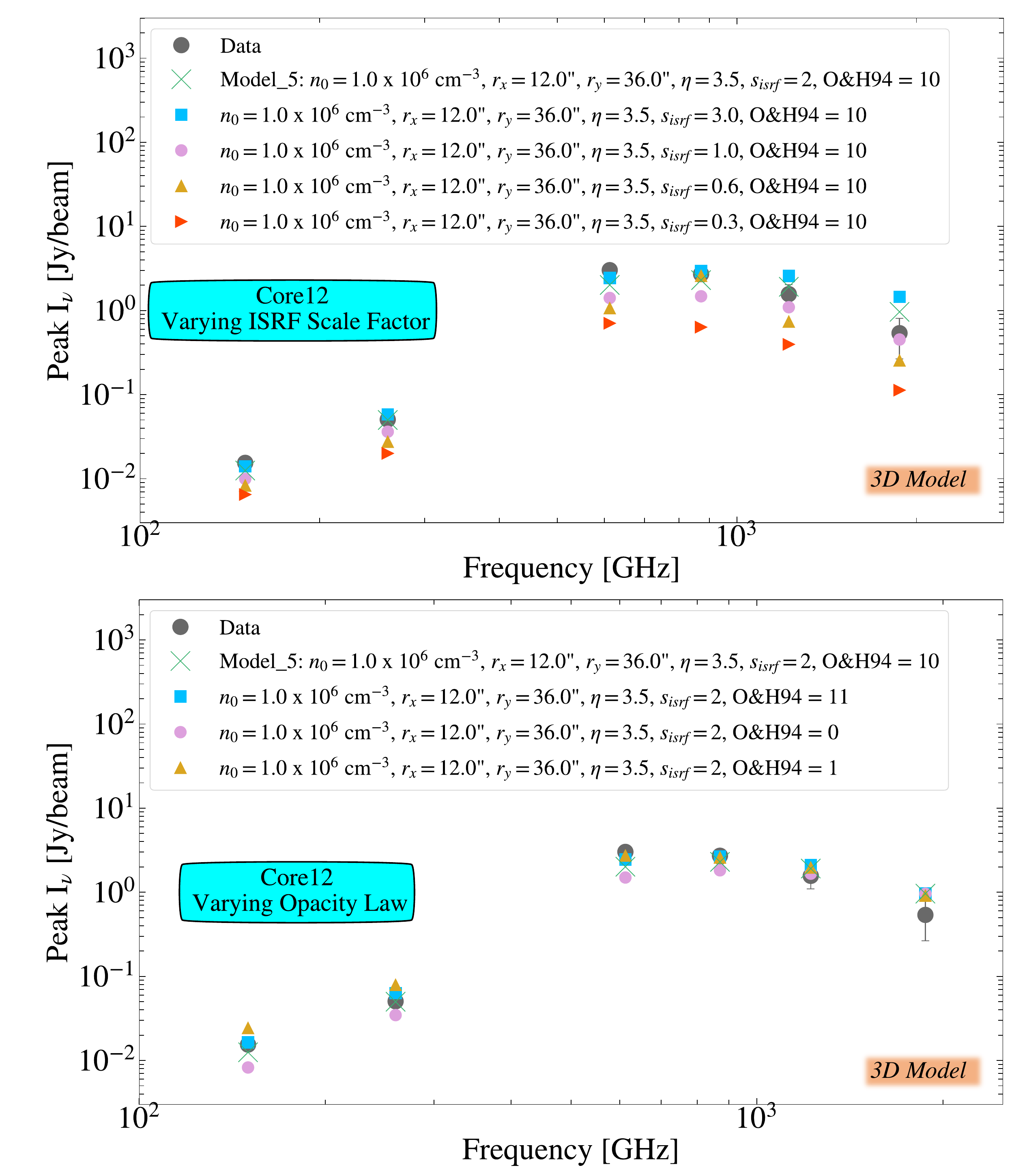} 
\end{array}$
\end{center}
\caption{\label{sed_opac_sisrf} Peak intensity versus frequency (i.e., SED plots) for core 12. We choose the best-fit Model\,\_5 to plot (as green `X' markers) against the data (grey circles) and then show how different opacity laws (top panel) and different strengths for the interstellar radiation field (bottom panel) can affect the SED. From left to right show the 2.0mm, 1.2mm, 500$\mu$m, 350$\mu$m, 250$\mu$m, and 160$\mu$m emission peaks.
}
\end{figure}

For our next cut-off, the only models considered are those that have normalized radial profiles with the lowest $\chi^2$ (equation\,\ref{equ:chi}) and fall within the normalized major and minor axis radial profiles. The normalized profiles are calculated similarly to the full radial profile, i.e., within 6$''$ annuli, but are done within the major and minor axis sectors (see Figure\,\ref{2D_core12}). The more spherical a core, the more these sector radial profiles overlap (e.g., core 6) and the more elongated a core, the more the sector radial profiles separate (e.g., core 16). For core 12, the elongated shape creates a well-defined major and minor axis. Only the models that fall in between and overlap with the major and minor sectors (and that have 1.2mm and 250$\mu$m peak intensities in agreement) are considered as best-fits (see Figure\,\ref{2D_core12}). There are five models that fit these criteria for core 12 (Table\,\ref{tab:pandora_best_fit_params} and Figure\,\ref{2D_core12}). 

Radial profiles for core 12 (out to the major axis limit of 40") at the other wavelengths (Figure\,\ref{core12_rad_2mm_250_350_500}) are also investigated. We find the best-fit models and the data for core 12 match that at 2.0mm emission (within errors), but have a harder time fitting the \textit{Herschel} profiles. The models drop off at a steeper slope than the data, perhaps not surprising due to the fact that we are not modelling the filamentary structure behind or surrounding the cores, which \textit{Herschel} better traces. We also point out that the resolution and pixel scale gets progressively worse in the 2.0mm, 250$\mu$m, 350$\mu$m, 500$\mu$m maps and thus we are not properly sampling the data at core scales ($\leq$ 4 data points per profile in the \textit{Herschel} data; Figure\,\ref{core12_rad_2mm_250_350_500}). Thus, as stated in section\,\ref{fittingcores}, we continue to rely solely on the highest resolution 1.2mm maps in the radial profile analysis. 

We find the central densities of our best-fits for core 12 range from $9 -10 \times 10^{5}$\,cm$^{-3}$, with an average value of $9.8\times 10^{5}$\,cm$^{-3}$ and a standard deviation of $0.4\times 10^{5}$\,cm$^{-3}$. This is consistent with results from ALMA ACA observations of the same source in \cite{2019PASJ...71...73T}, as they also conclude this core (labeled MC5-N in their paper) has a central density of $\sim 10^6$ cm$^{-3}$. The elongated nature of core 12 is picked up in our best-fit results, as for each model $r_x < r_y$ where $r_x$  ranges from $12 - 14^{''}$ and $r_y$  ranges from $34 - 50^{''}$. The slope of the Plummer profile $\eta$, as in equation\,\ref{nr_pandora}, is the best constrained as either $3.5$ or $4.0$ and when $\eta=4.0$, the core is more elongated in $r_x$ and $r_y$. In all the five best-fit models the opacity law found is for thin ice mantles and no initial gas density is best fit (O\&H94 = 10). A larger-than-unity scale factor for the interstellar radiation field is found with best fit values of s$_{isrf}$=2.0 (Table\,\ref{tab:pandora_best_fit_params}).

If the opacity is changed, we find the SED for core 12 is affected most at the 1.2mm and 2.0mm wavelengths (see top panel Figure\,\ref{sed_opac_sisrf}). The models for thin ice mantles (O\&K94 = 11 and 10) do not differ by much, but for the bare ice grains either the 1.2mm and 2.0mm emission is underestimated if no initial gas density is assumed (O\&K94 = 0), or overestimated if a gas density of 10$^5 \mathrm{cm}^{-3}$ is assumed (O\&K94 = 1). By changing the interstellar radiation field scale factor, $s_{isrf}$, a more noticeable trend is seen (bottom panel Figure\,\ref{sed_opac_sisrf}). If the $s_{isrf}$ is increased, we find the peak emission values also increase.

\begin{table}
	\scriptsize
	\caption{Best-fit \textit{pandora} Core Models}
	\label{tab:pandora_best_fit_params}
	\setlength{\tabcolsep}{5.0pt}
	\begin{tabular}{lccccccccc} 
	Core & \# &  $n_c$ & $r_x$ & $r_y$ & $|r_{2D}|$  & $\eta$ & $s_\mathrm{isrf}$ &  O\&H94 &  *$\chi_\mathrm{norm}^2$ \\ 
    &  & (cm$^{-3}$) &  ($''$) & ($''$) &  ($''$) &   &  \\
	\hline 
	\hline
       6     & Model\_1 & 5e4 & 44 & 42 & 61  & 4.0 & 2.0 & 10  & 0.1  \\
     & Model\_2 & 5e4 & 38 & 50 &   63 & 4.0 & 2.0 & 11  & 0.1  \\
     & Model\_3 & 1e5 & 50 & 34 &  60  & 4.0 & 1.0 & 0 & 0.1 \\
      & Model\_4 & 5e4 & 46 & 40 &  61  & 4.0 & 2.0 & 10  & 0.1  \\
    \hline
    7-1    & Model\_1  & 5e4  & 46 &  26 & 53  & 3.5 & 1.0 & 10  & 0.1 \\
            & Model\_2 & 5e4 & 46&   20&  50  & 3.0 & 1.0  & 10 & 0.1 \\
           & Model\_3  & 1e5  & 46 &  24&  52 & 3.5 & 0.6 &  0 &  0.1\\
           & Model\_4 & 5e4  &  34&  24 & 42 & 3.0 & 1.0 &  11 & 0.1 \\
           & Model\_5 & 5e4  &  38&  22 & 44 & 3.0 & 0.6 &  11 & 0.1 \\
           & Model\_6 & 5e4  &  38&  22 &  44 & 3.0 & 1.0 &  10 & 0.1 \\
    \hline
    7-2   &  Model\_1 &  1e5  & 20  & 12  &  23 &  2.0  & 0.6 &  0   &  2.0 \\
            & Model\_2 &  1e5  & 16  & 16  &  23 & 2.0 & 1.0  &  0 &   2.0 \\
            & Model\_3 &  1e5  & 18  &  12  & 22  &  2.0 &  0.6 &  0  &  2.0 \\
             & Model\_4  &  1e5  &  16 & 14  &  21  & 2.0 & 0.6 &  0 & 2.0 \\
            & Model\_5  & 1e5   & 14  & 16  & 21  &  2.0 & 0.6  & 10   &  2.0 \\
    \hline
    8           & Model\_1 &  1e6  &  12 & 14   & 18&  2.5 & 0.6  &   0 & 0.1  \\
             & Model\_2    &  7e5  &  14 & 14   &20 &  2.5 & 0.6  &   0 & 0.1  \\
            & Model\_3     &  9e5  &  12 & 14  & 18&  2.5  & 0.6  &   0 & 0.1  \\
    \hline
    9            & Model\_1 &  3e5  & 44  & 44  & 62 & 5.0 &   1.0 &  0 &  2.0  \\
                & Model\_2  &  3e5  &  44 &  42  & 61  & 5.0 & 1.0   & 0 &  2.0   \\
                & Model\_3 &  3e5  &  46 &  48  &  66 & 5.5 & 1.0   & 0 &   3.0  \\
                & Model\_4 &  3e5  &  42  & 44  &  61  & 5.0 &   1.0 & 0 &  3.0  \\
    \hline 
    10           & Model\_1 &   2e5 & 24  &  40  & 47& 5.0   & 1.0  &  0  &  1.0 \\
               & Model\_2 &  2e5  &  30 &  30 & 42 &  5.0 &  1.0  & 0    &  1.0 \\
               & Model\_3  &  2e5  &  34 &  28 & 44 &  5.0 & 0.6  &  0  &   1.0\\
    \hline
    12      & Model\_1 &  1e6  & 14  &  50  & 52 & 4.0  & 2.0   & 10   &  <0.1 \\
            & Model\_2 &  1e6  &  12 & 34   &36 & 3.5  & 2.0   & 10   &  <0.1 \\
             & Model\_3 & 1e6   &  12 &  36  & 38 & 3.5  & 2.0   &  10  &  <0.1 \\
             & Model\_4 & 9e5   &  14 & 50   & 52& 4.0  & 2.0  & 10   & 0.1  \\
            & Model\_5 &  1e6  &  12 &  38  & 40 &  3.5 & 2.0   & 10   &  <0.1 \\
    \hline 
    13-1     & Model\_1 &  3e5  & 14  &  20  &  24 & 2.0  & 0.6   & 0   &  2.0 \\
            & Model\_2 &  5e4  &  12 &  12  & 17  & 1.5  & 1.0   & 11   &  2.0 \\
            & Model\_3 &  3e5  & 14  &  18  & 23  & 2.0  & 0.6   &  0  &  2.0 \\
            & Model\_4 &  3e5  & 18  &  38  & 42  & 2.5  & 0.6   &  0  &  2.0 \\
            & Model\_5 &  5e4  & 14  &  12  & 18  & 1.5  & 1.0   &  11  &  1.0 \\
            & Model\_6 & 3e5   & 14  & 16   &  21 & 2.0   & 0.6  & 0   & 2.0  \\
    \hline 
    13-2    & Model\_1 & 2e5   & 16  & 12   & 20 & 2.0  & 0.6   &  0  & 2.0  \\
            & Model\_2 & 2e5   & 14  & 14  & 20 & 2.0  & 1.0  &  0  & 2.0  \\  
            & Model\_3 & 2e5   & 18  & 12   &22 &  2.0 &  0.6 &  0  &  2.0 \\
            & Model\_4 & 1e5   & 16  & 14  & 21 &  2.0 &  0.6 &  11 & 2.0  \\
            & Model\_5 & 2e5   & 16  & 14  & 21 &  2.0 & 1.0  &  0  &  2.0 \\
    \hline 
    14      & Model\_1 &  2e5  &  48 & 12   & 49 &  3.5 &  0.6 &  0  & <0.1  \\
            & Model\_2 &  2e5  & 46  &  12  & 48 &  3.5 & 0.6  &  0  & <0.1   \\
             & Model\_3 &  3e5  & 38  &  12  & 40 &  3.5 &  0.6 &  0  & <0.1  \\
             & Model\_4 &  1e5  & 30  &  12  & 32&  3.0 &  1.0 &  11  & <0.1   \\
           &  Model\_5 &  5e4  & 34  &  12  & 36 &  3.0 &  2.0 &  1  &  <0.1   \\
    \hline 
    15          &  Model\_1 & 1e5   & 44  &  28  &  52 &  3.0 &  0.3 & 0   & 1.0  \\
            &  Model\_2 &  2e5  & 18  &   32 & 37 & 2.5  &  0.3 &  10  &  1.0 \\
            &  Model\_3 &  1e5  & 34  &   32 & 47  &  3.0 & 0.3  &  0  &  1.0 \\
           &  Model\_4 &  2e5  & 18  &   26 &  32 &  2.5 & 0.3  &  10  &  1.0 \\ 
    \hline 
    16       & Model\_1 &  6e5  &  18 &  46  & 49& 5.0  &  2.0 & 0   &  0.1 \\
            & Model\_2 &  5e5  & 18  & 48  & 51&  5.0 &  2.0 & 0   & 0.1  \\
            & Model\_3 &  6e5  &  18 &  44  & 48& 5.0  &  2.0 &  0  & 0.1  \\
            & Model\_4 &   5e5 & 18  &   46 & 49& 5.0  & 2.0  & 0   & 0.1  \\
    \hline 
    f1      & Model\_1 &   2e5 & 12  &  46  & 48 & 4.0  & 2.0  &  0  & 0.2  \\
            & Model\_2 &   2e5 & 12  &  50  & 51 & 4.0 & 2.0  &  0  & 0.2  \\
            & Model\_3 &   1e5 & 12  &  50  & 51 & 4.0  & 2.0  &  10  & 0.2  \\
            &Model\_4&   1e5 & 12  & 34  &   36  & 3.5  &   3.0 & 0 & 0.2 \\
            & Model\_5 &  2e5  &  12 &  44  & 46 & 4.0  &  2.0 &  0  &  0.2 \\
            & Model\_6 &  2e5  & 12  &  48  & 49 &4.0  & 2.0  &  0  &  0.2 \\
            &Model\_7& 1e5  & 12  & 38 &  40  & 3.5 & 3.0 & 0 & 0.2 \\
            & Model\_8 &   1e5 & 12  &  48  & 49 & 4.0 & 2.0  &  10  & 0.2  \\
    \hline 
    f2     & Model\_1 &  5e4  & 16  &  12  &  20 & 3.0  & 2.0  &  10  &  0.2 \\ 
            & Model\_2  &   5e4 & 16   &  12 & 20  & 3.0  & 3.0  &  0  &  0.2 \\ 
            & Model\_3 &  5e4  & 14  &  14  & 20  &  3.0 &  2.0 &  10  &  0.2 \\
            & Model\_4 &  5e4  & 16  &  12  &  20 & 3.0  &  3.0 &  10  &  0.2 \\
             & Model\_5 &  5e4  & 14  &  14  &  20 &  3.0 &  2.0 &  11  &  0.2 \\
              & Model\_6 &  5e4  & 14  &  14  &  20 &  3.0 &  3.0 &  10  &  0.2 
	\end{tabular}
      \begin{description}
      \item *Calculated via equation\,\ref{equ:chi} for the full azimuthally averaged radial profile within 60" for each core.
      \end{description}
\end{table}

\begin{figure*}
\centering
\begin{center}$
\begin{array}{ccccccccccccc}
\includegraphics[width=55mm]{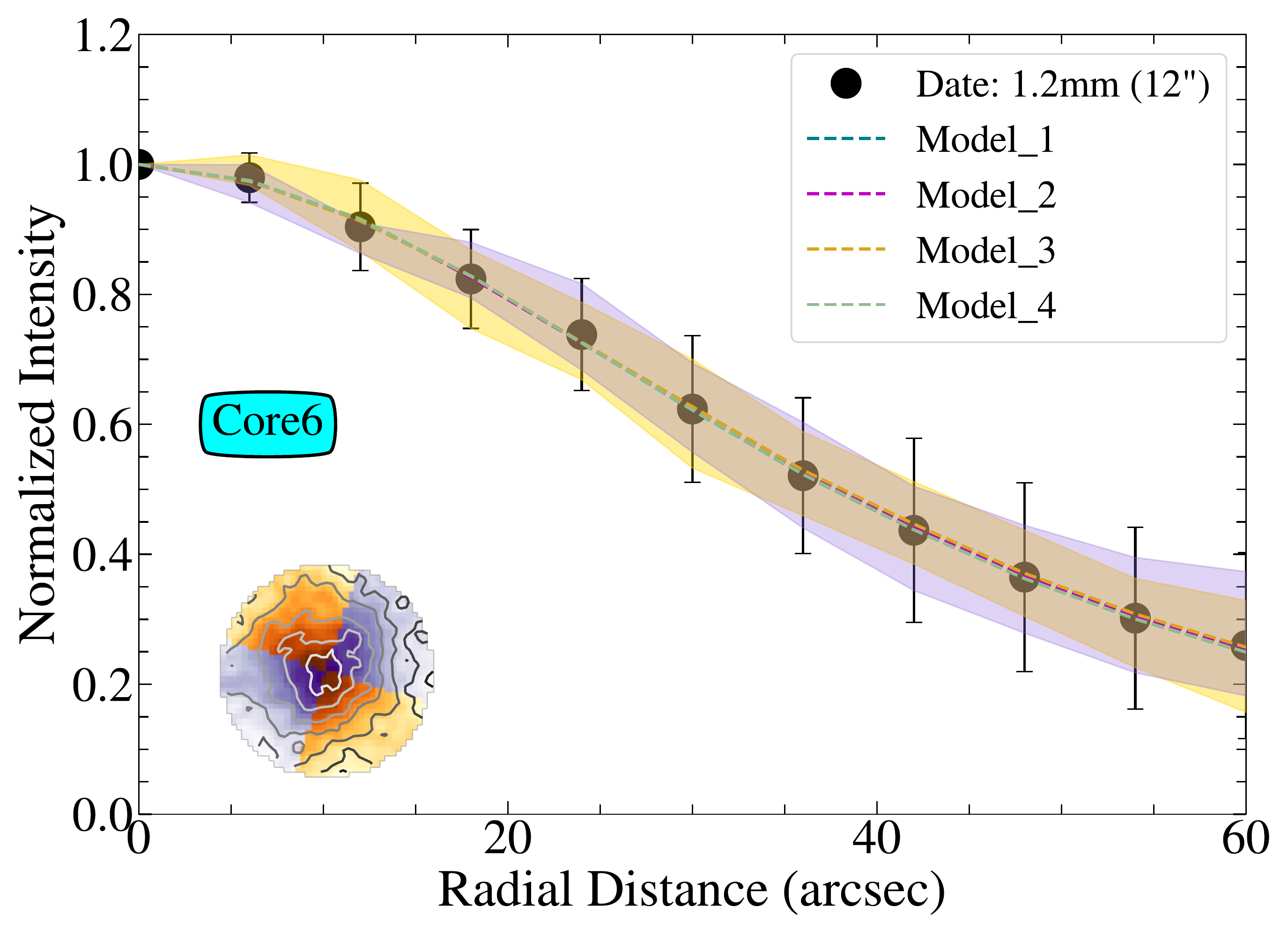} & 
\includegraphics[width=55mm]{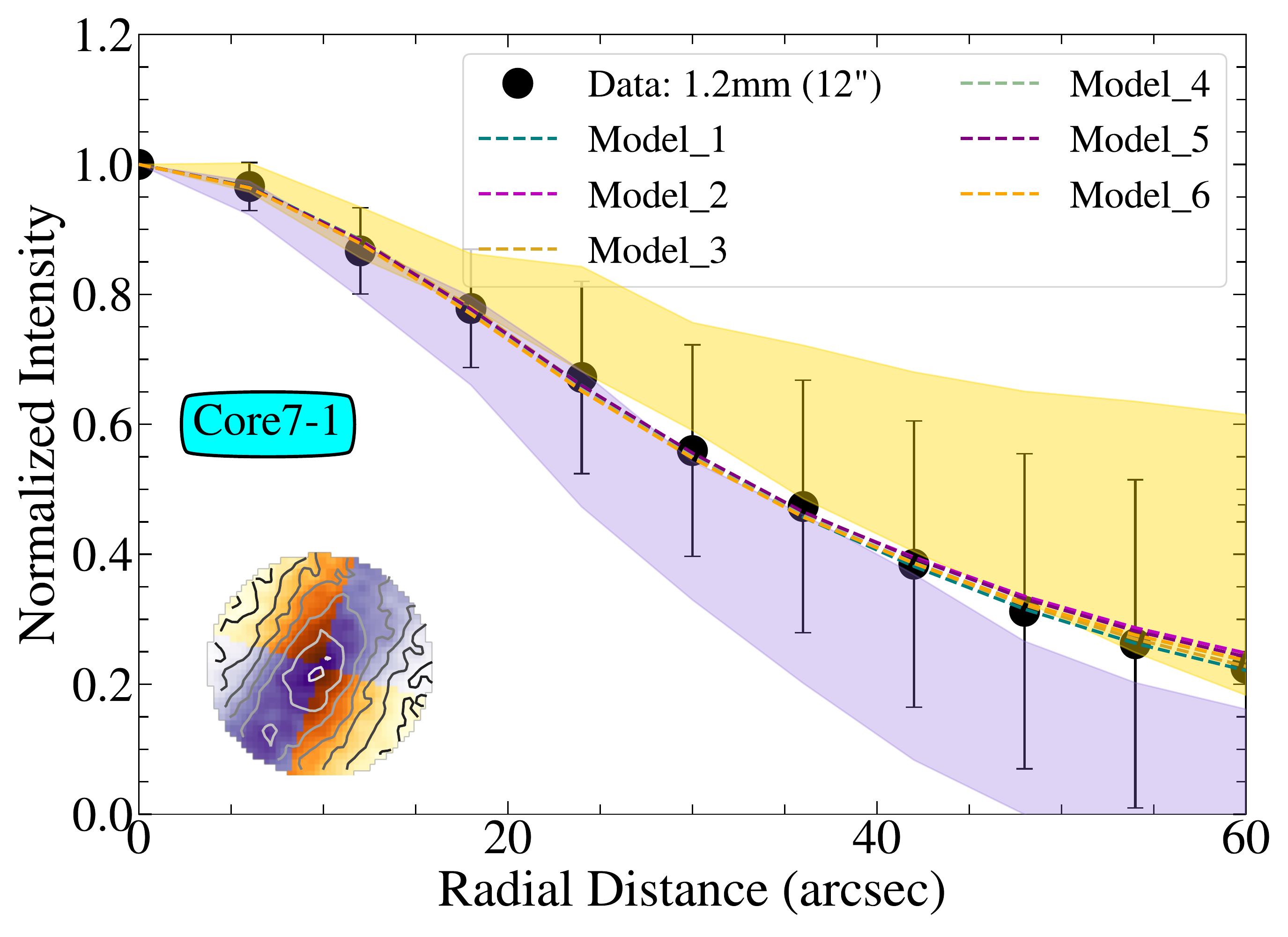} &
\includegraphics[width=55mm]{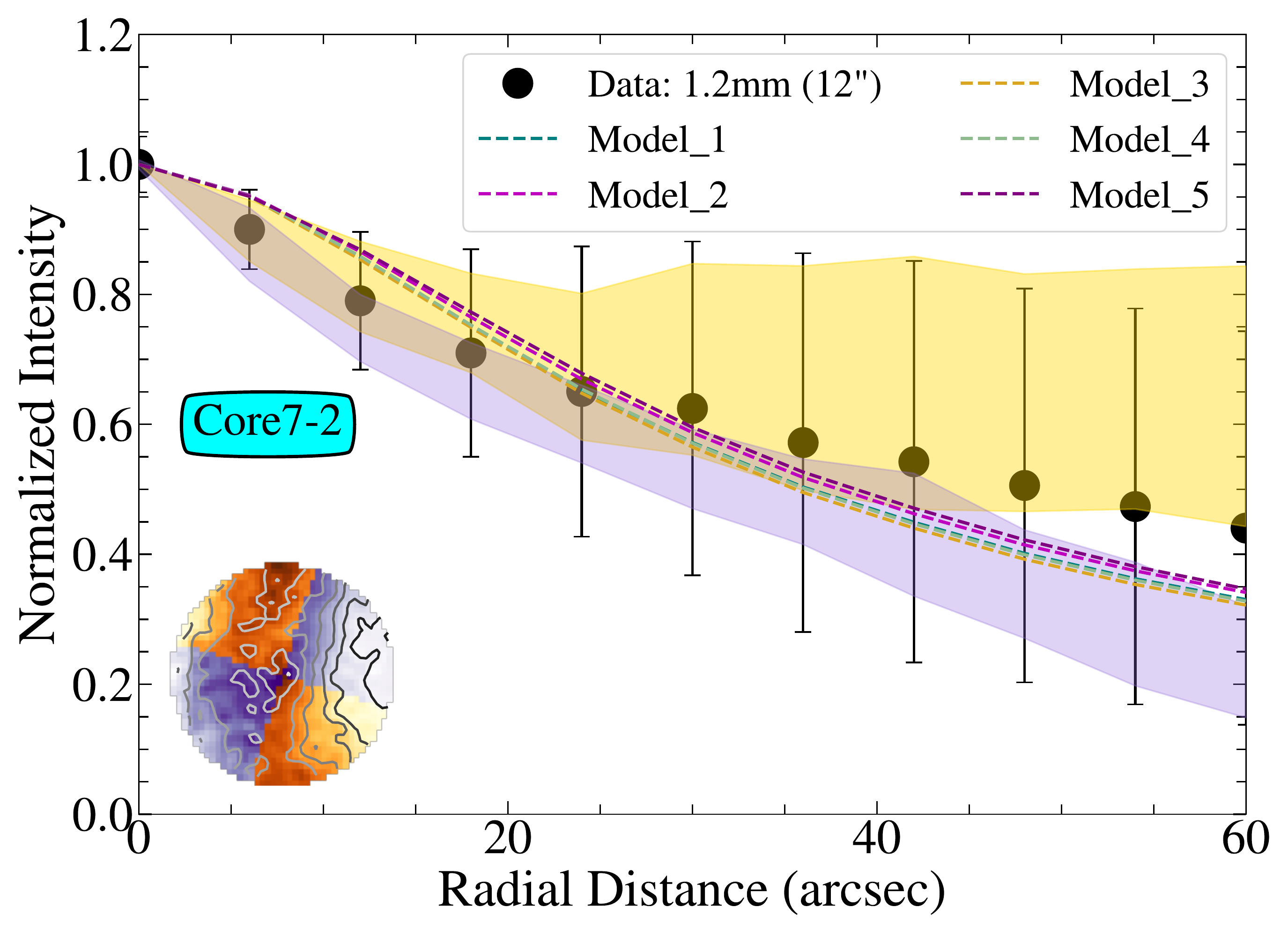} \\
\includegraphics[width=55mm]{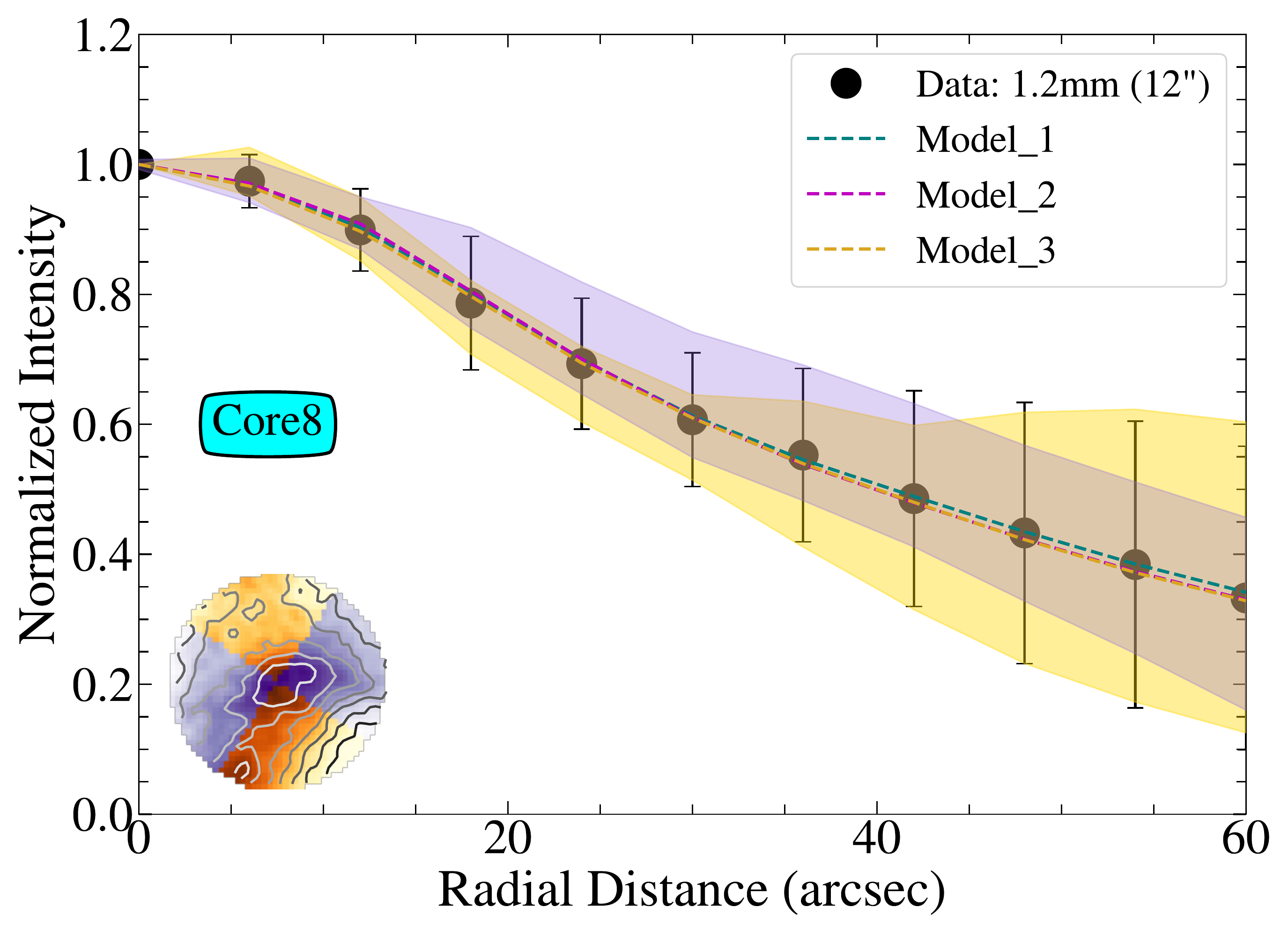} &
\includegraphics[width=55mm]{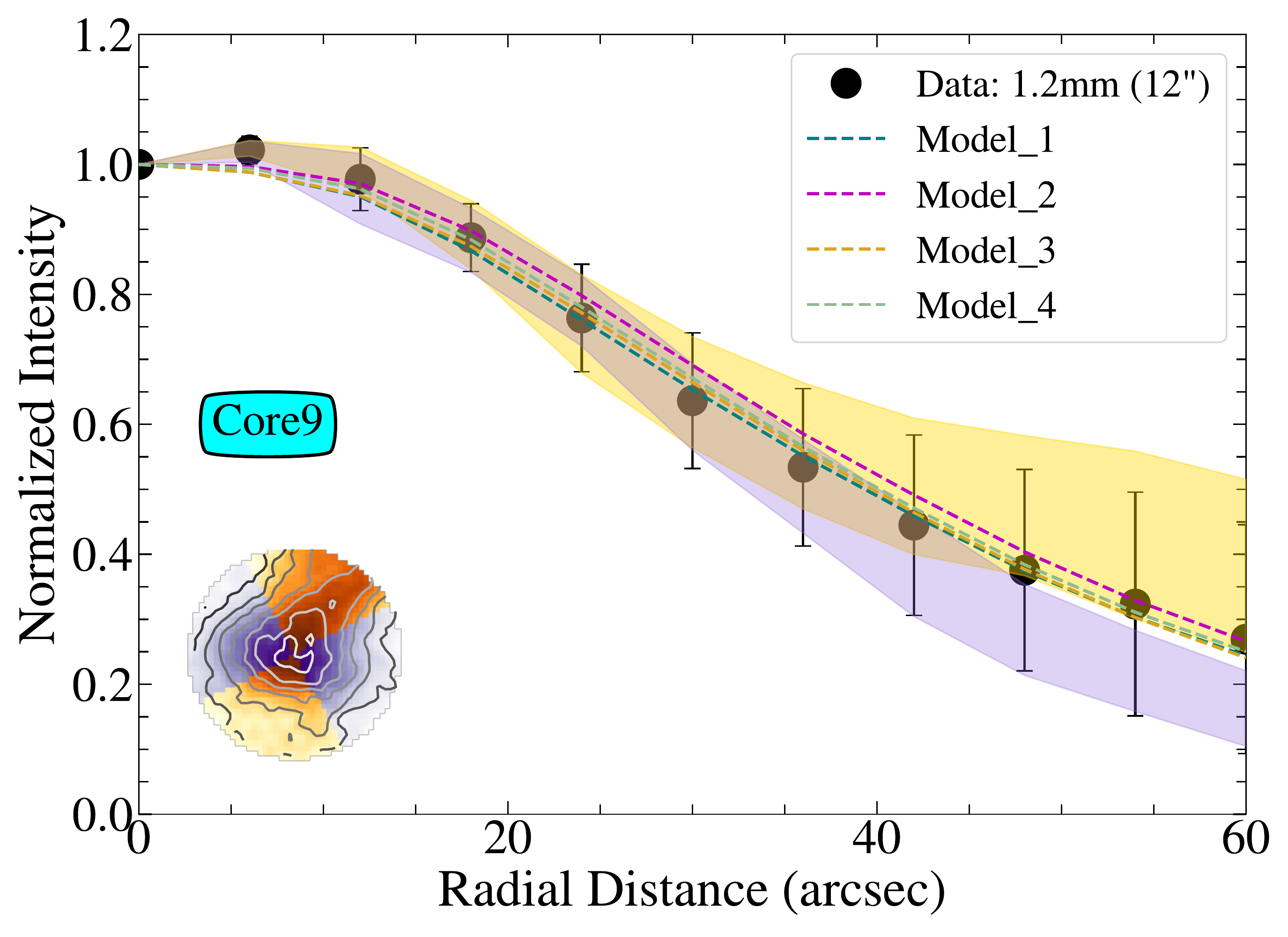} & 
\includegraphics[width=55mm]{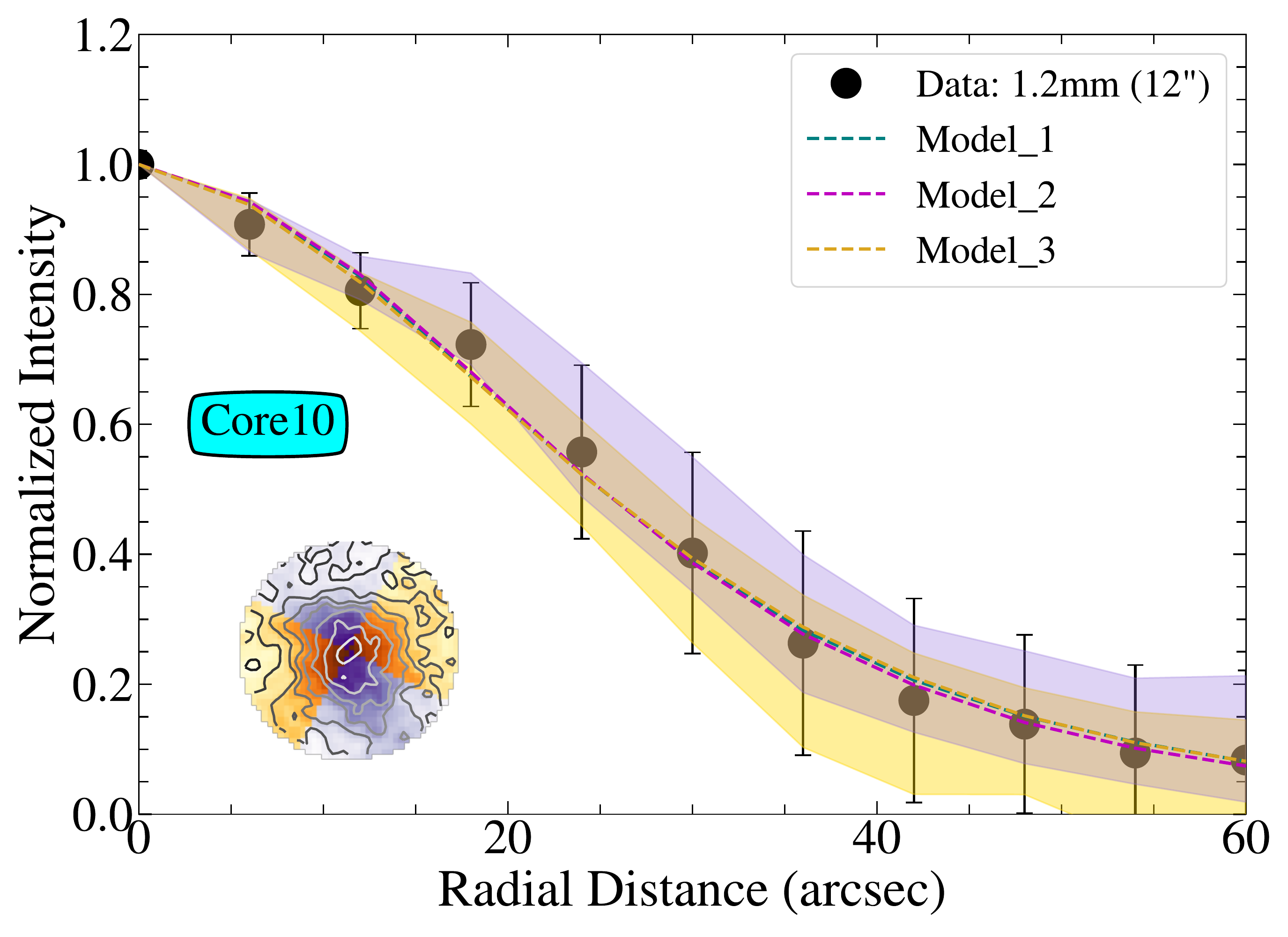} \\
\includegraphics[width=55mm]{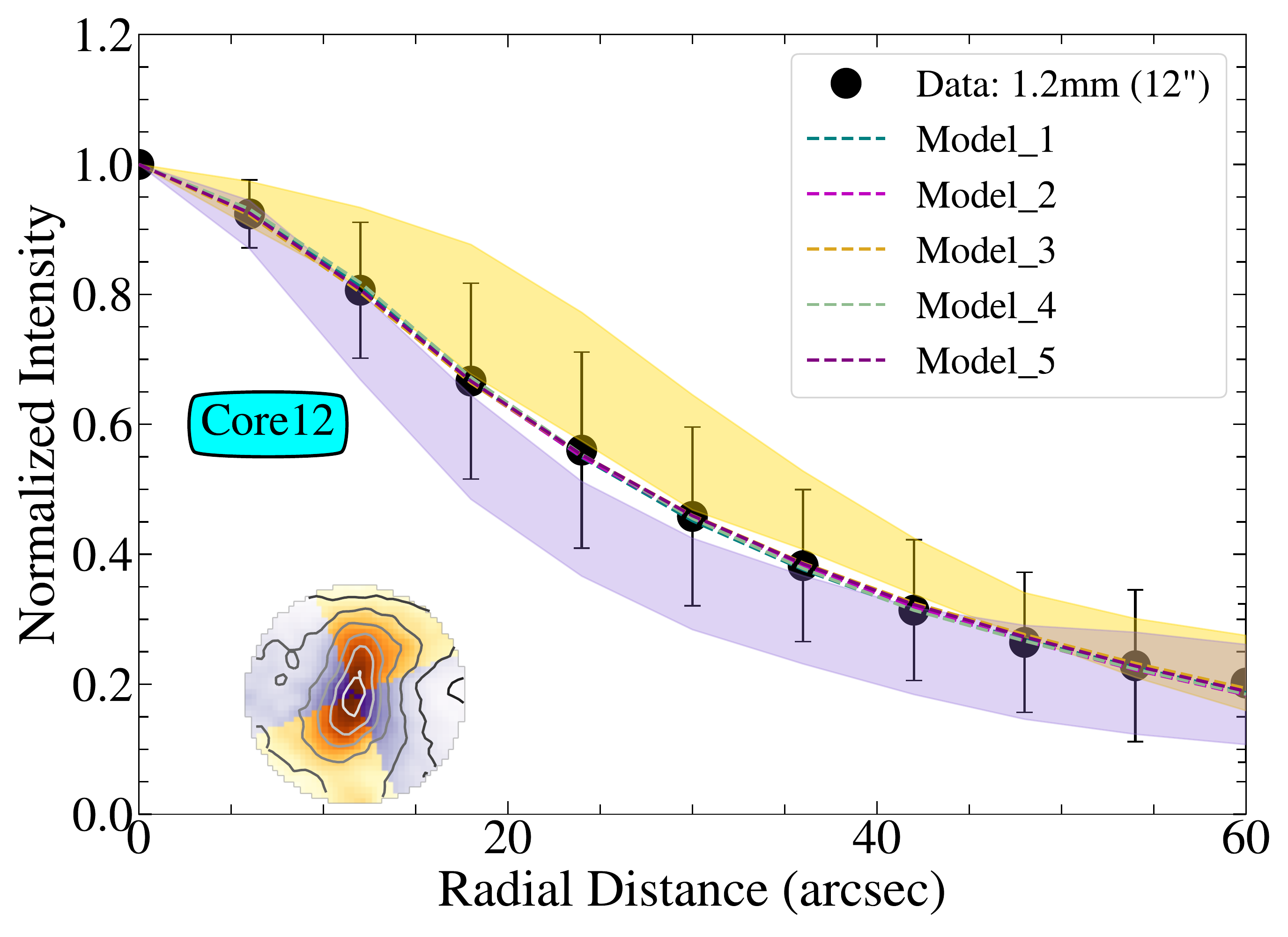} &
\includegraphics[width=55mm]{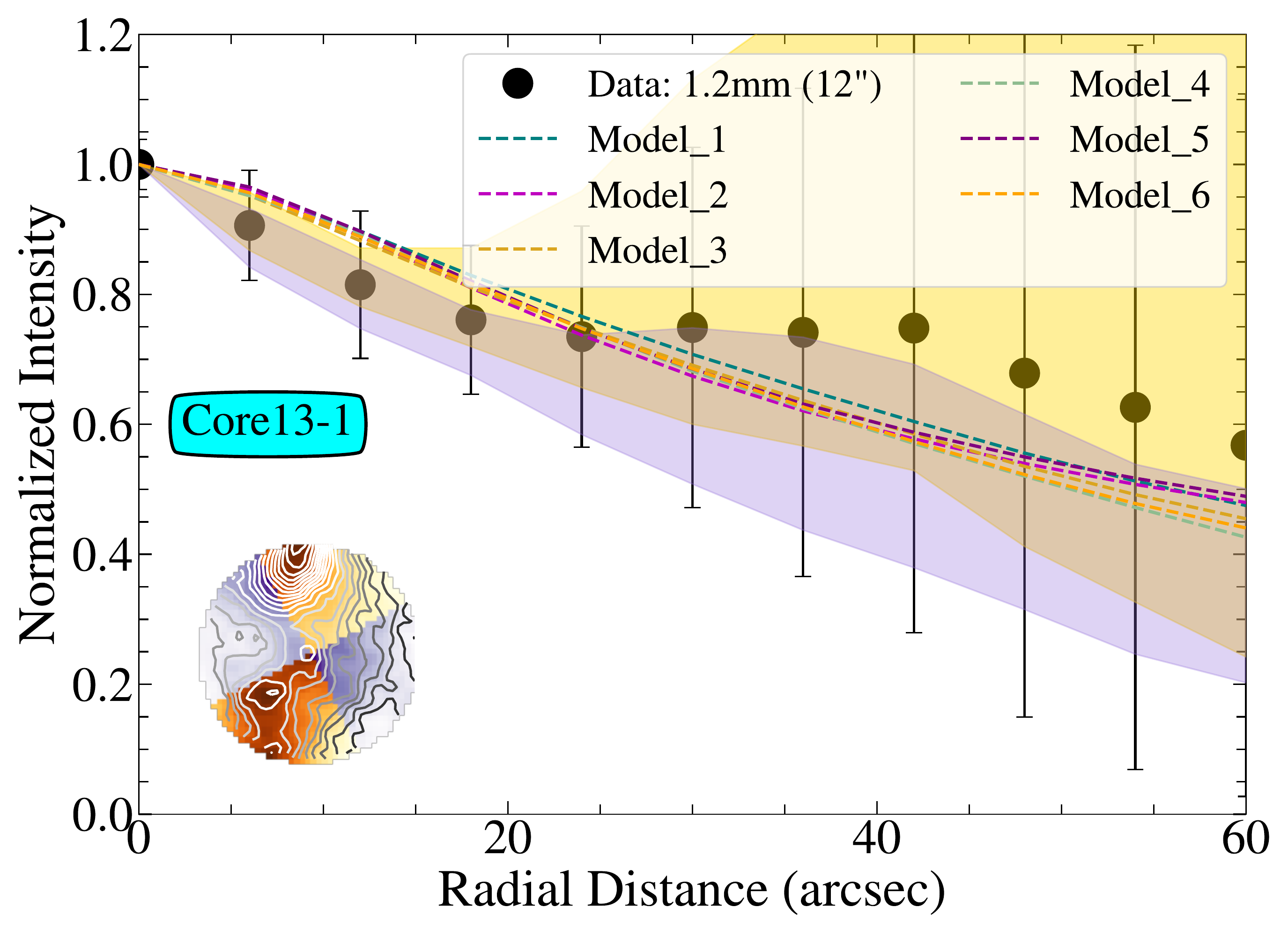} &
\includegraphics[width=55mm]{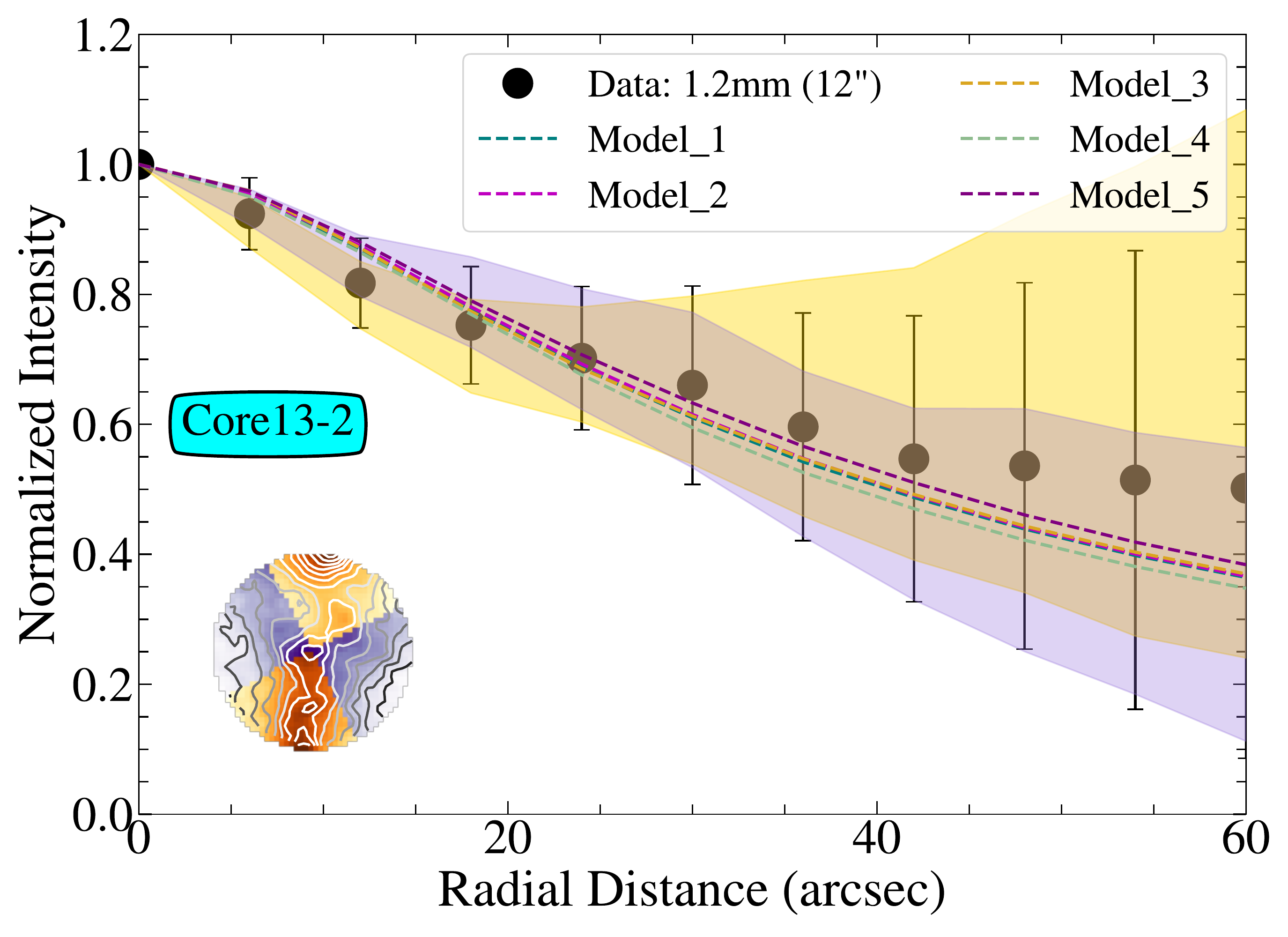} \\
\includegraphics[width=55mm]{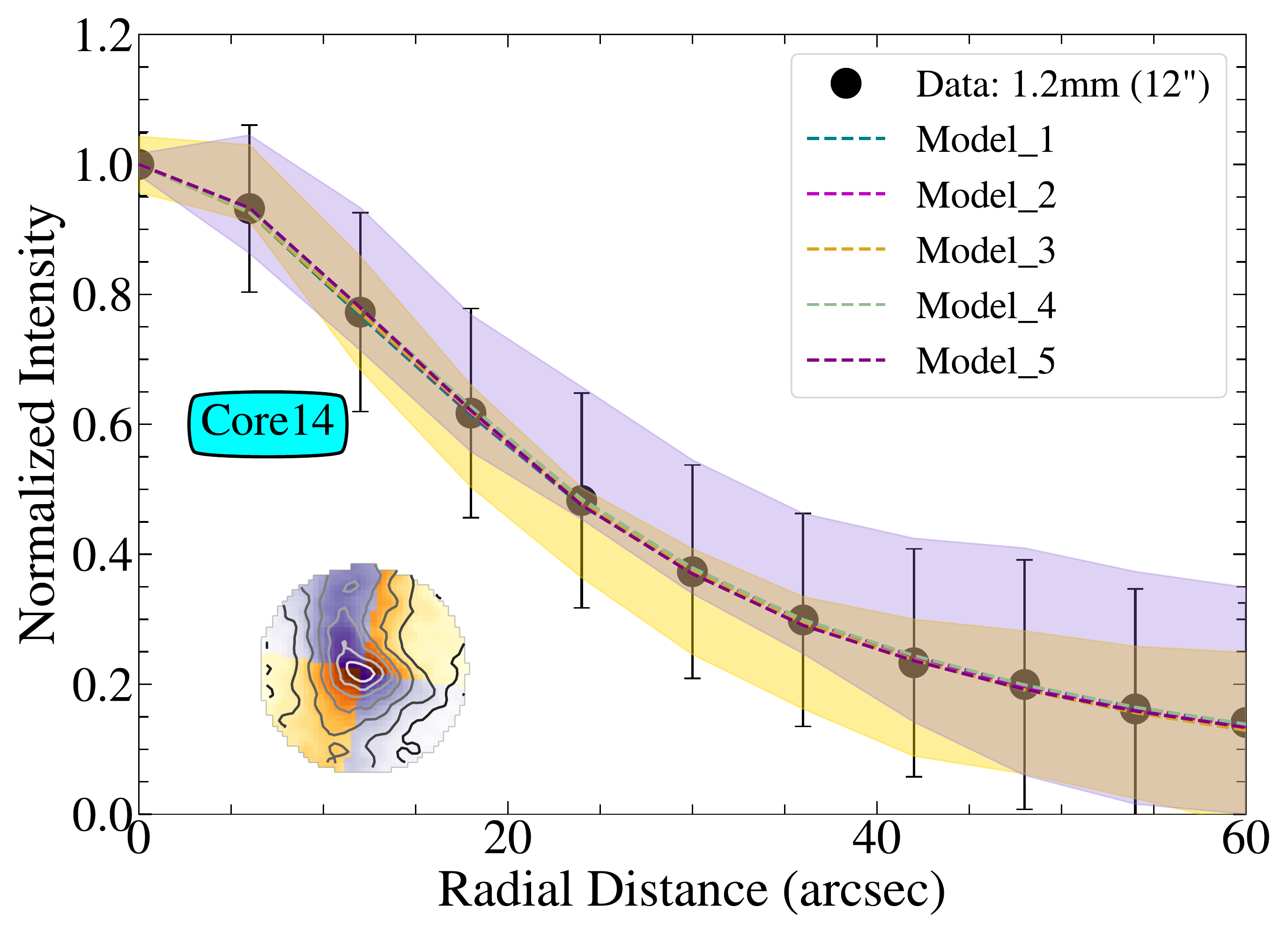} &
\includegraphics[width=55mm]{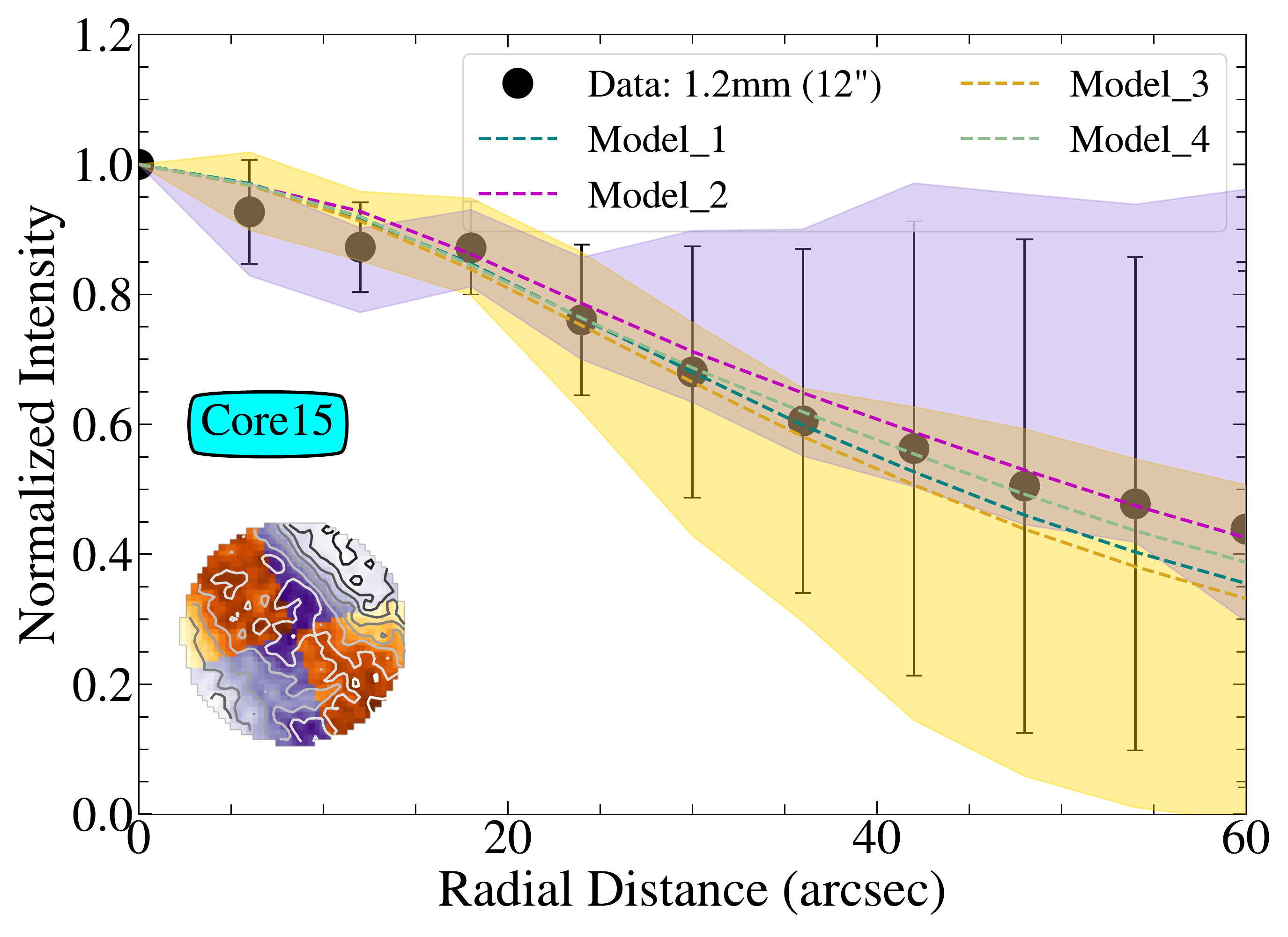} &
\includegraphics[width=55mm]{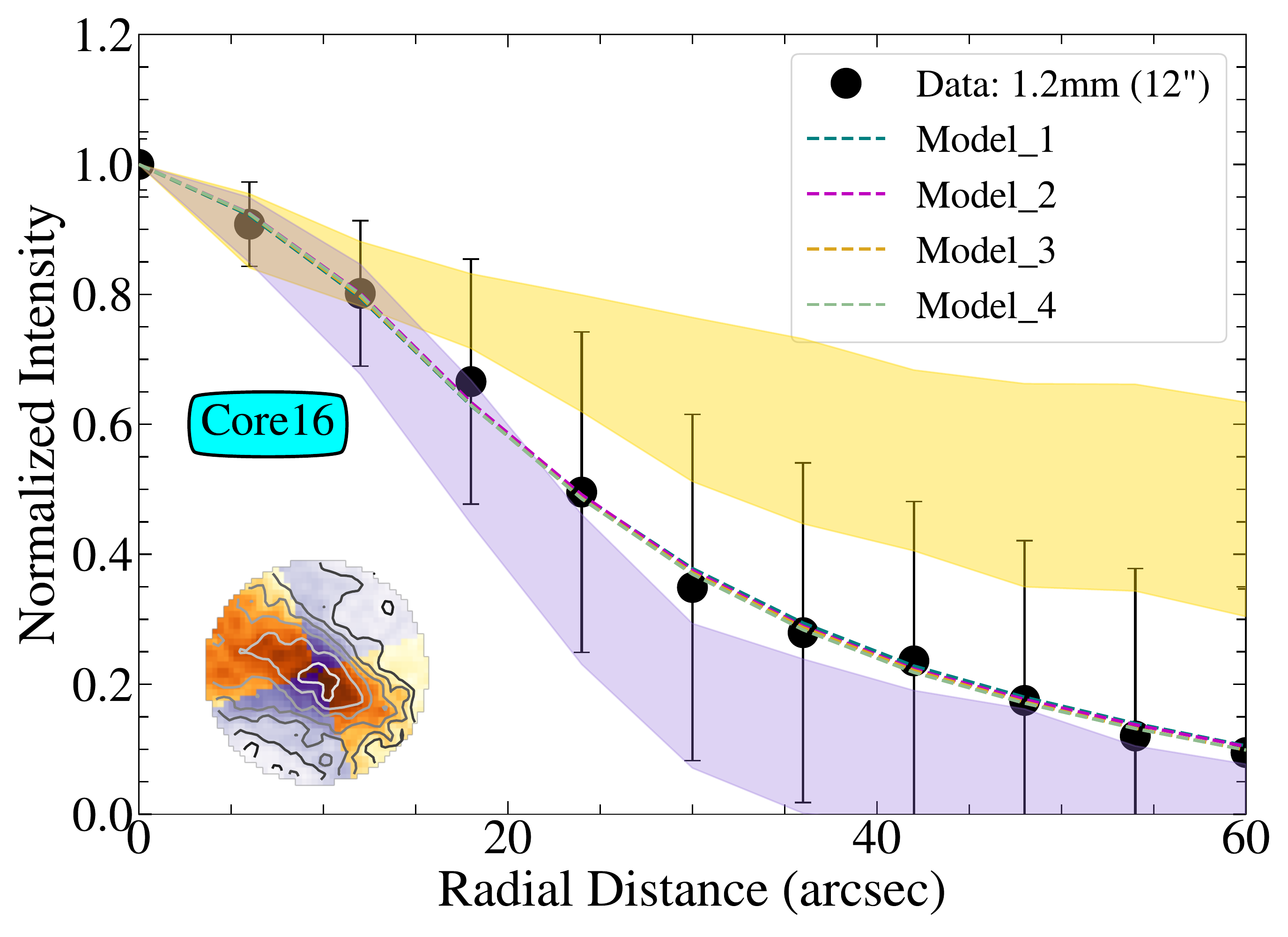} \\
\includegraphics[width=55mm]{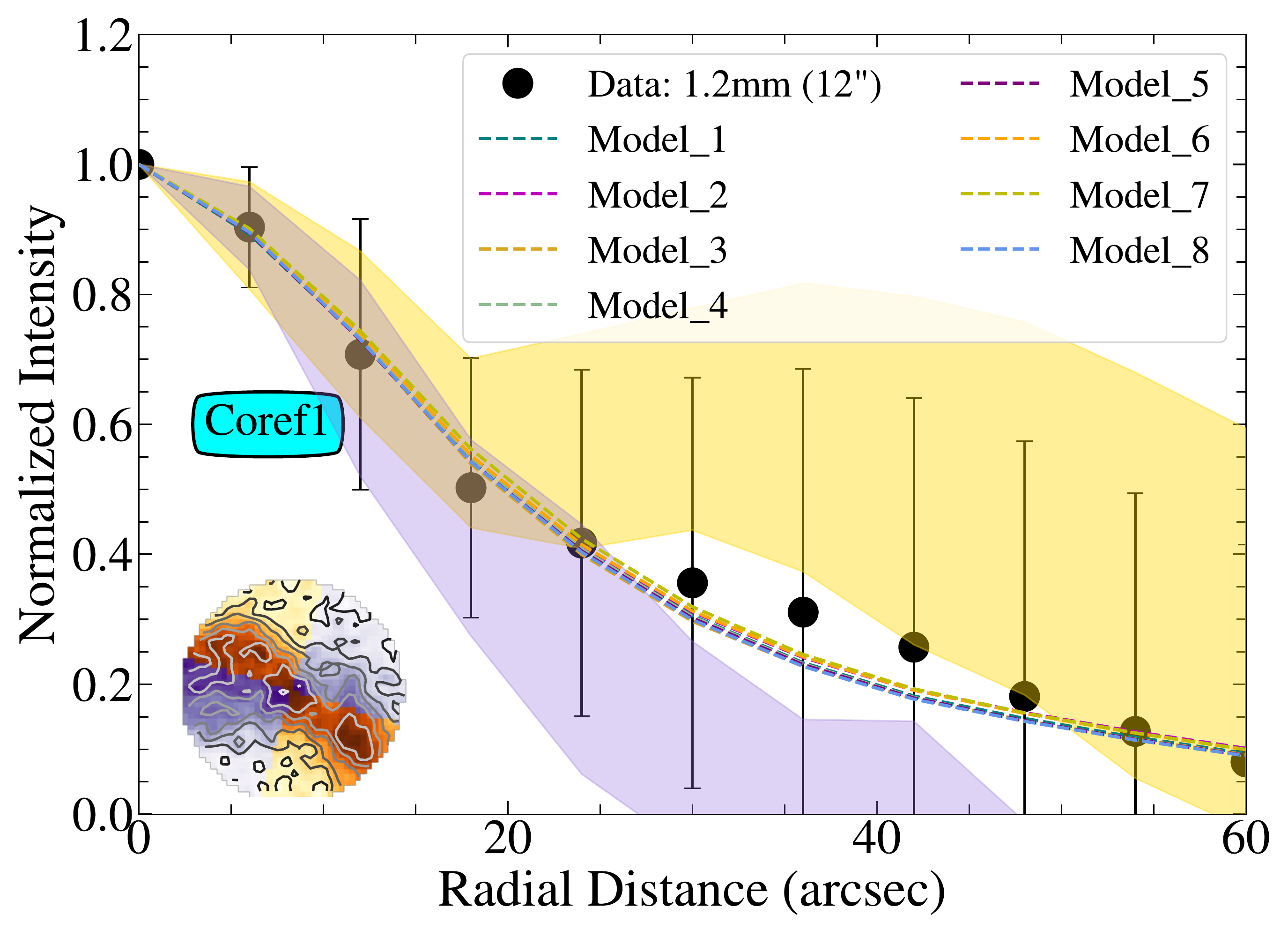} & 
\includegraphics[width=55mm]{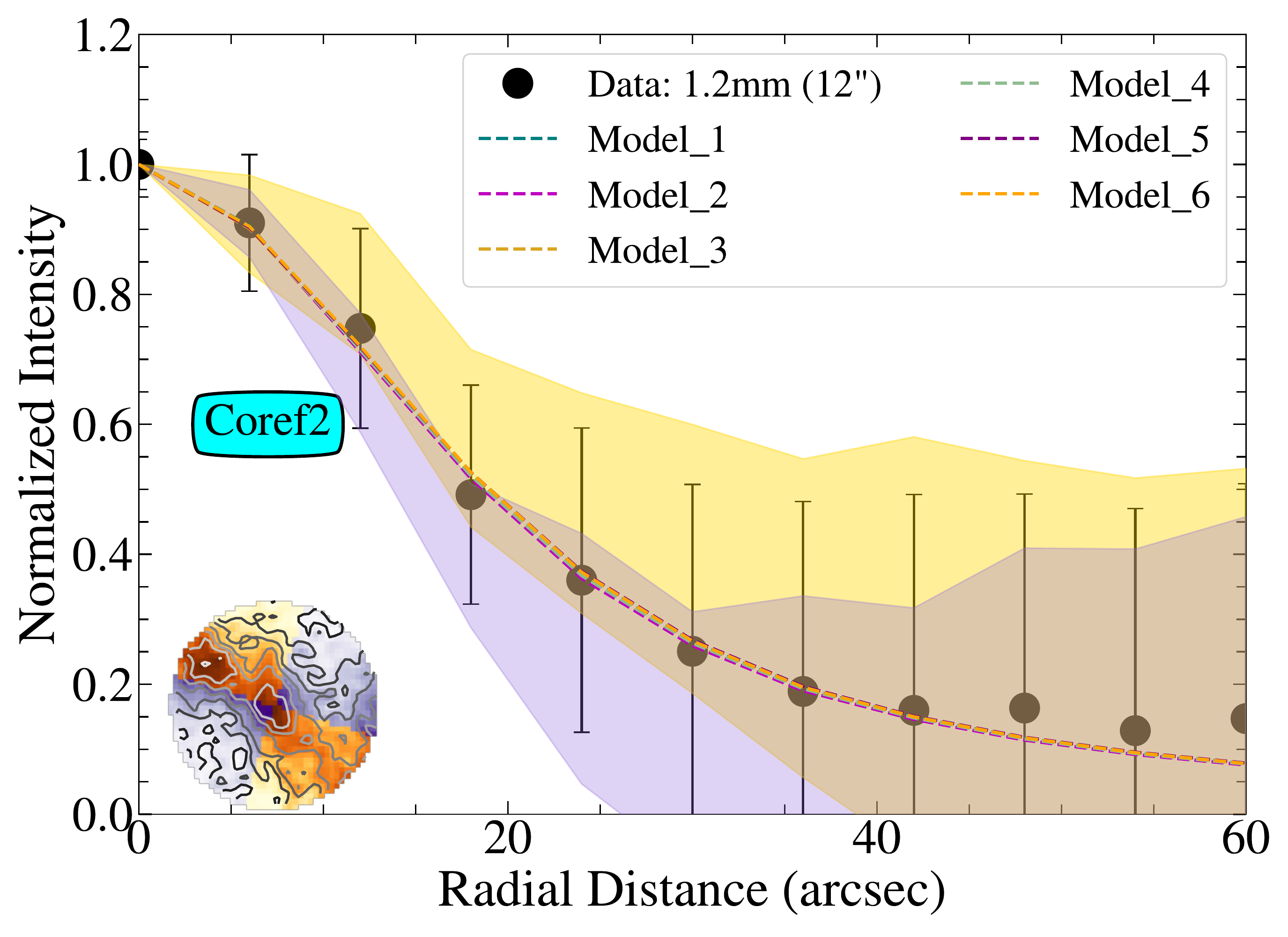} 
\end{array}$
\end{center}
\caption{\label{sector_plot} Each panel plots the normalized azimuthally averaged radial profiles for each individual `leaf' or core picked out by our dendrogram analysis. The azimuthally averaged and  normalized 1.2mm radial profiles are plotted in black, along with error bars derived as the standard deviation within the 6$''$ annulus. Similarly calculated are the normalized profiles in 90 degree sectors centered on the major and minor axes of the cores, based on their position angles. To represent these profiles in each panel, a purple and yellow filled in region represents the extent, including the standard deviation errors, that the major and minor axis radial profile cover. An inserted image of the each core (outlined in grey contour) out to 60$''$ shows with the same color-scheme where spatially each sector-ed radial profile is calculated on the core itself. The best-fit \textit{pandora} models are overlaid as dashed lines ordered starting from `Model\_1' for each core. 
}
\end{figure*}

\begin{figure*}
\centering
\begin{center}$
\begin{array}{cccccccccccccc}
\includegraphics[width=56mm]{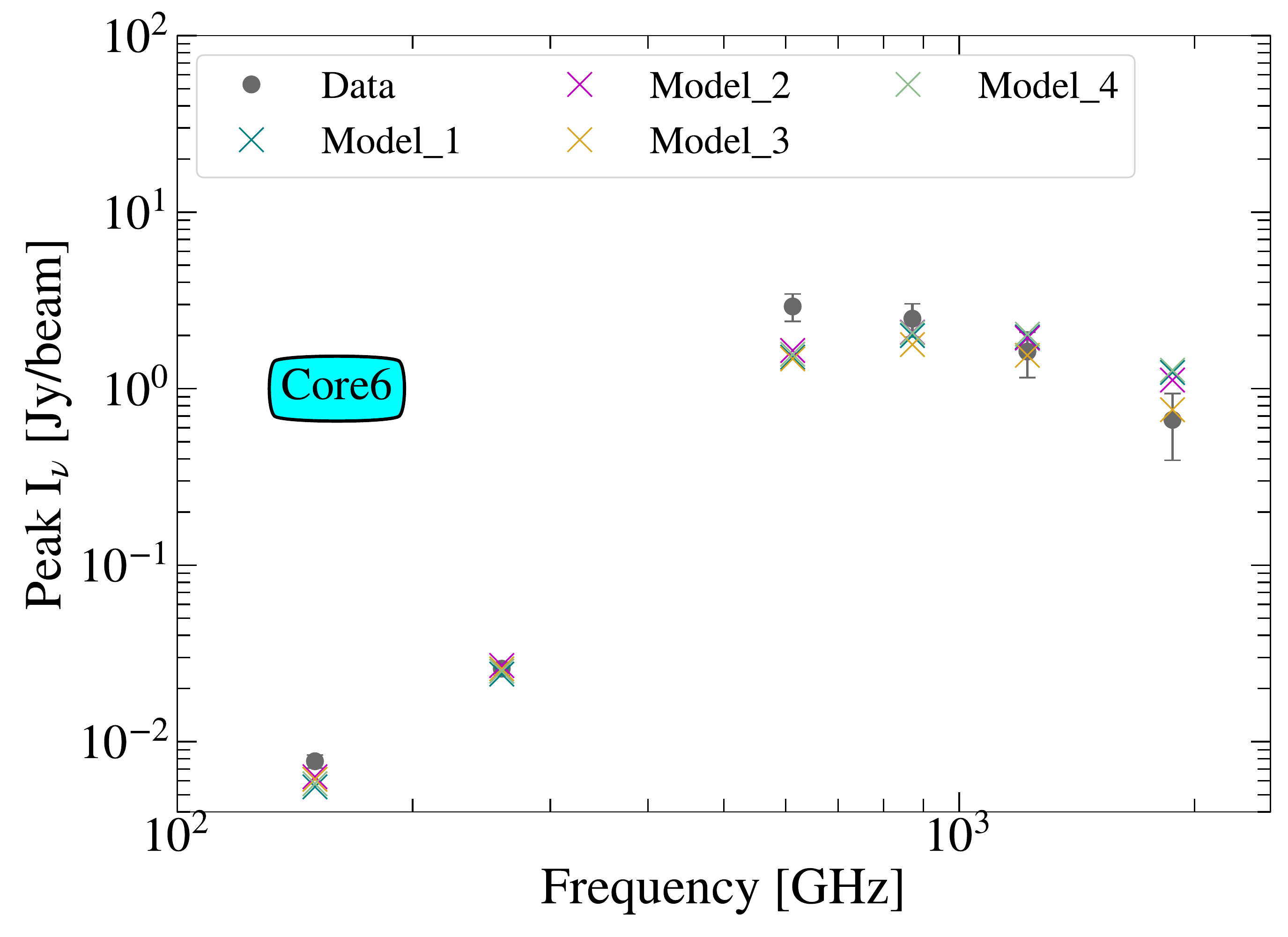} & \includegraphics[width=56mm]{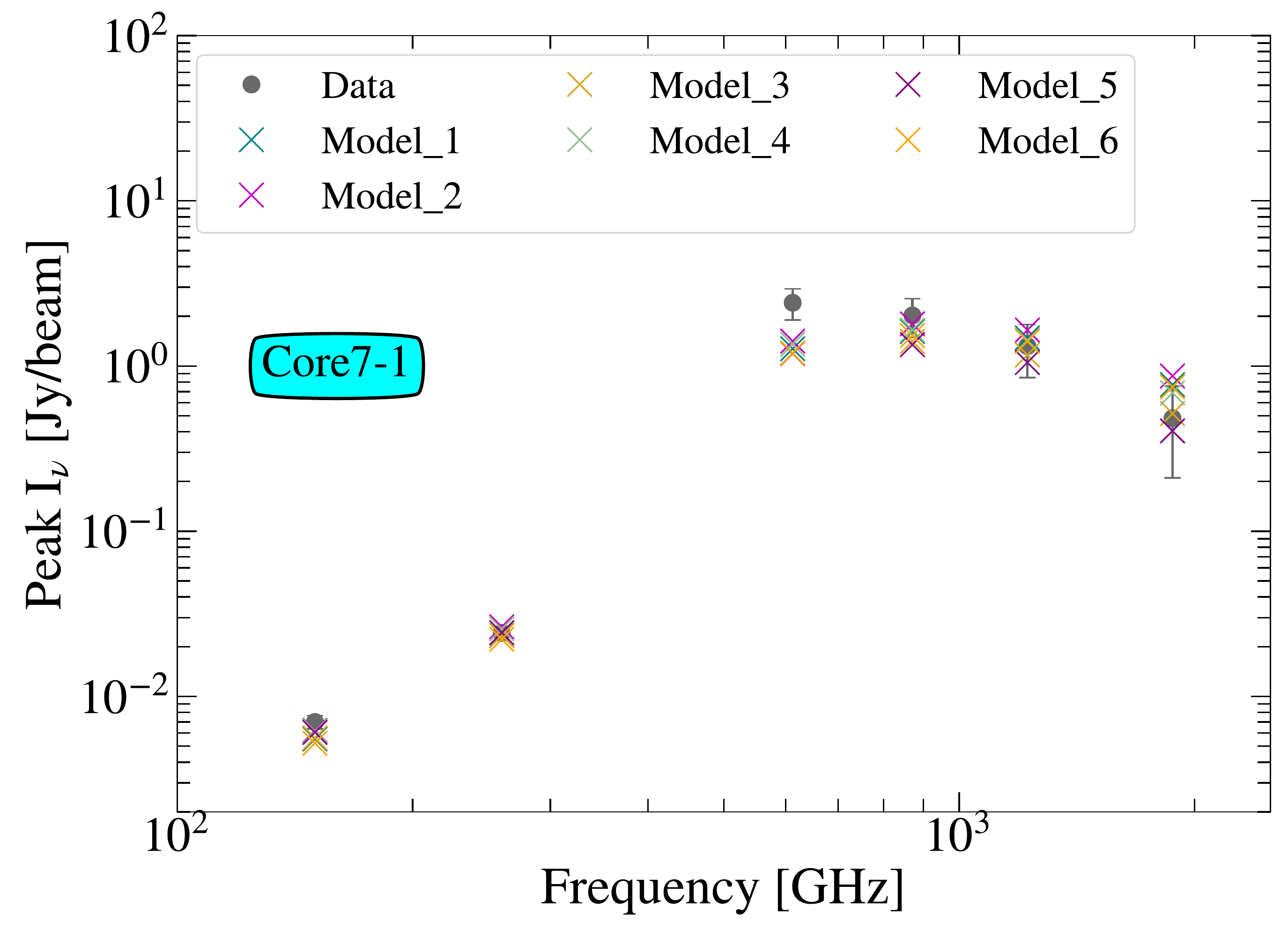} & \includegraphics[width=56mm]{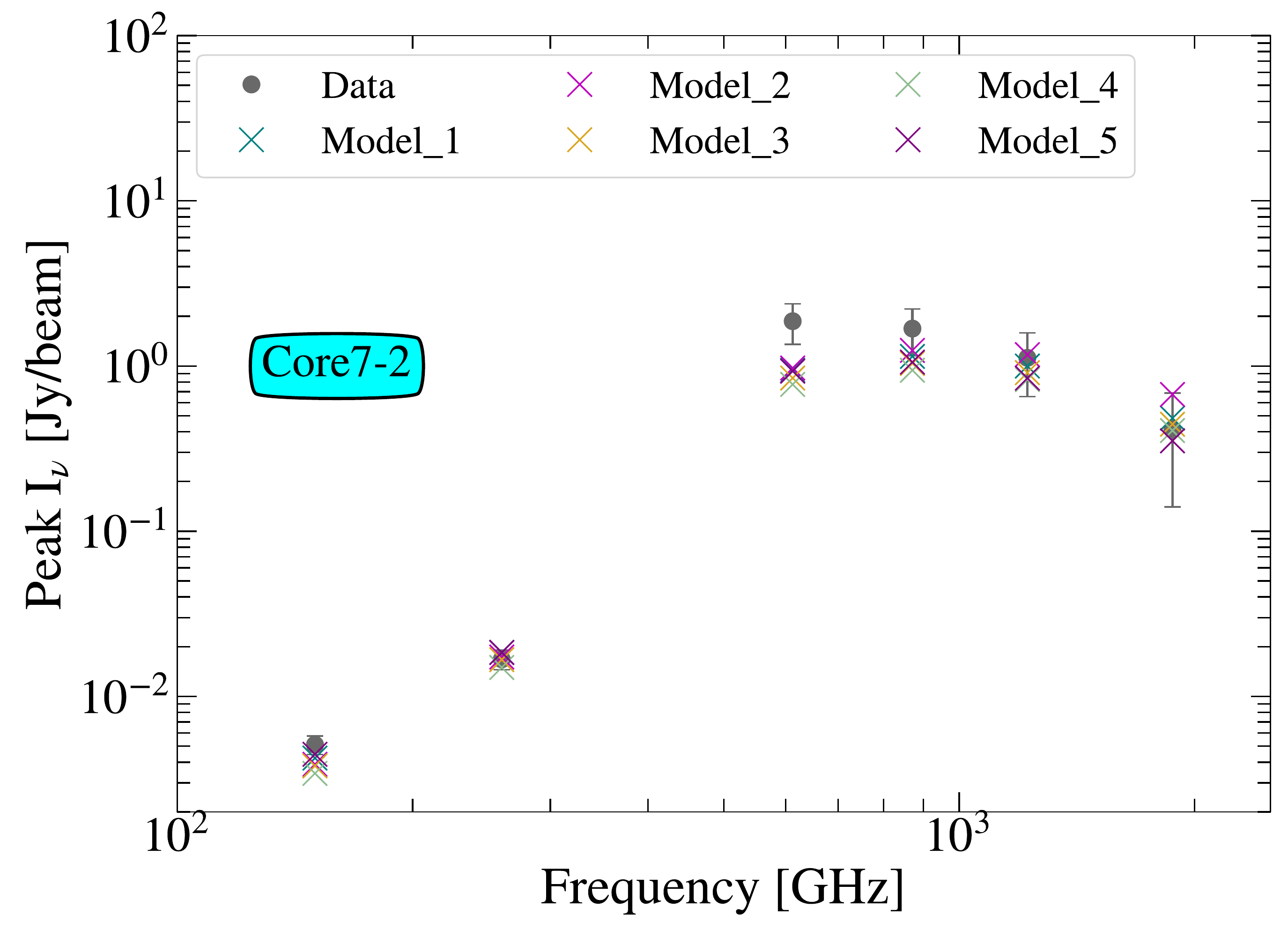} \\
\includegraphics[width=56mm]{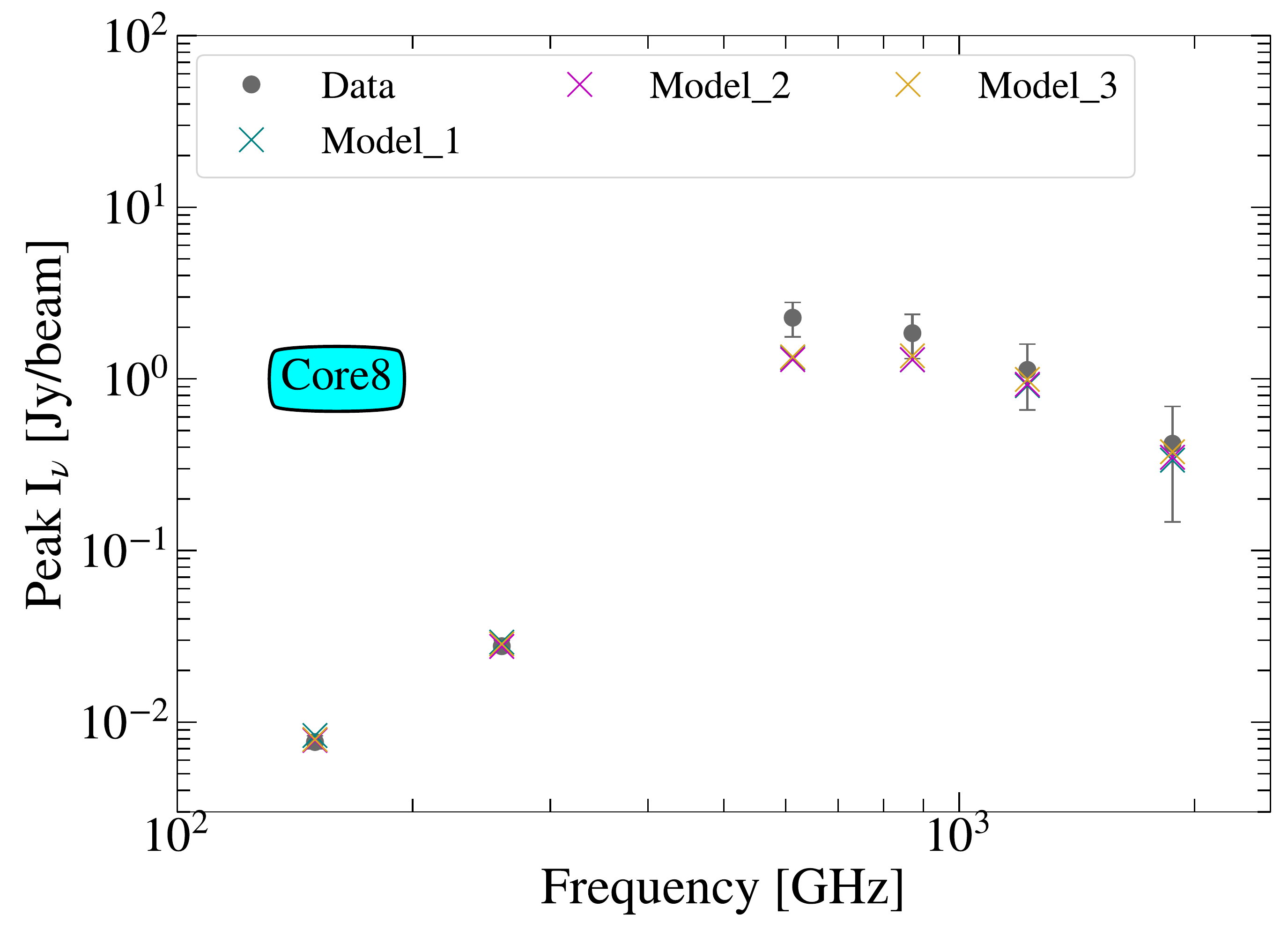} & \includegraphics[width=56mm]{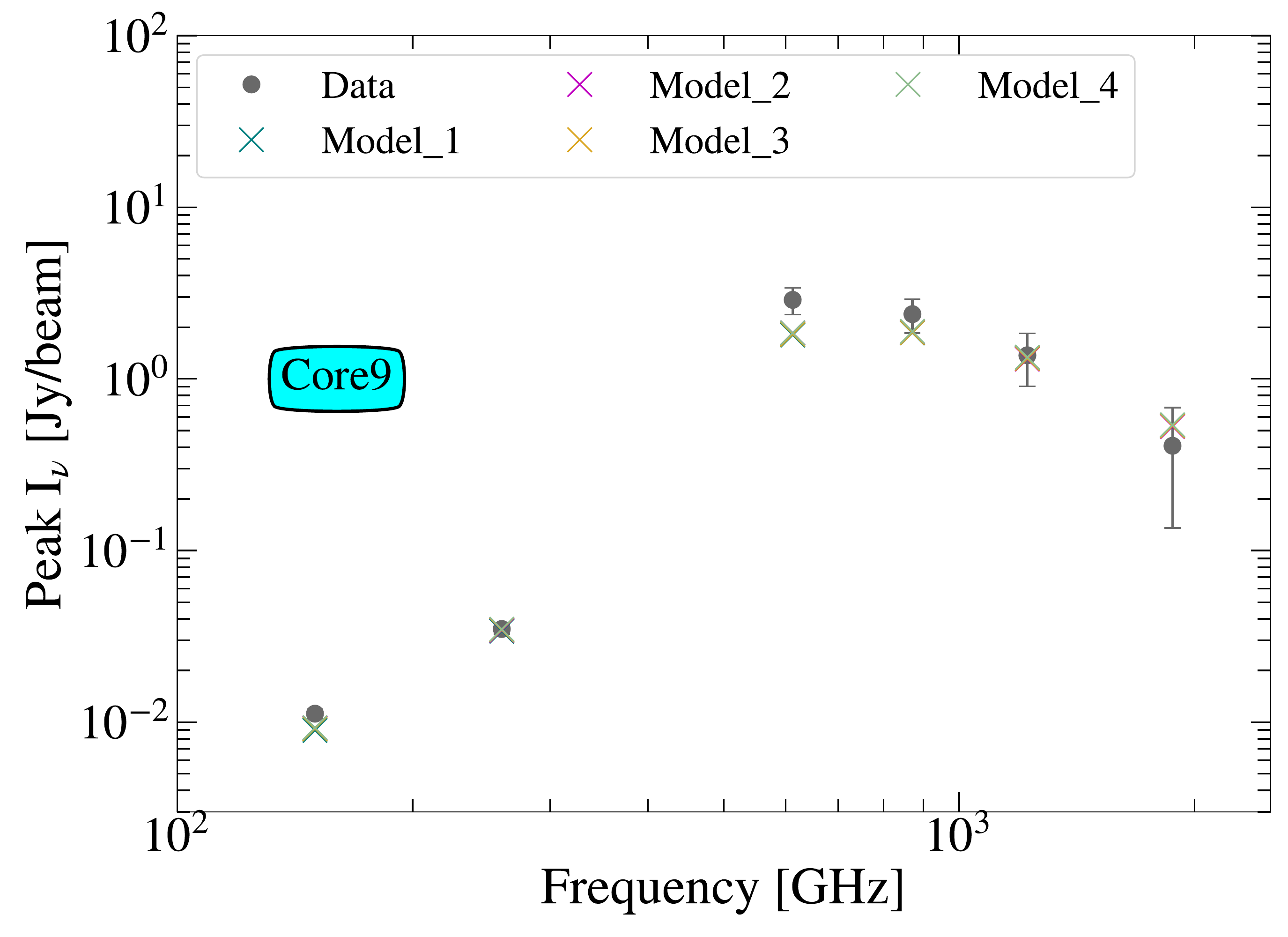} & \includegraphics[width=56mm]{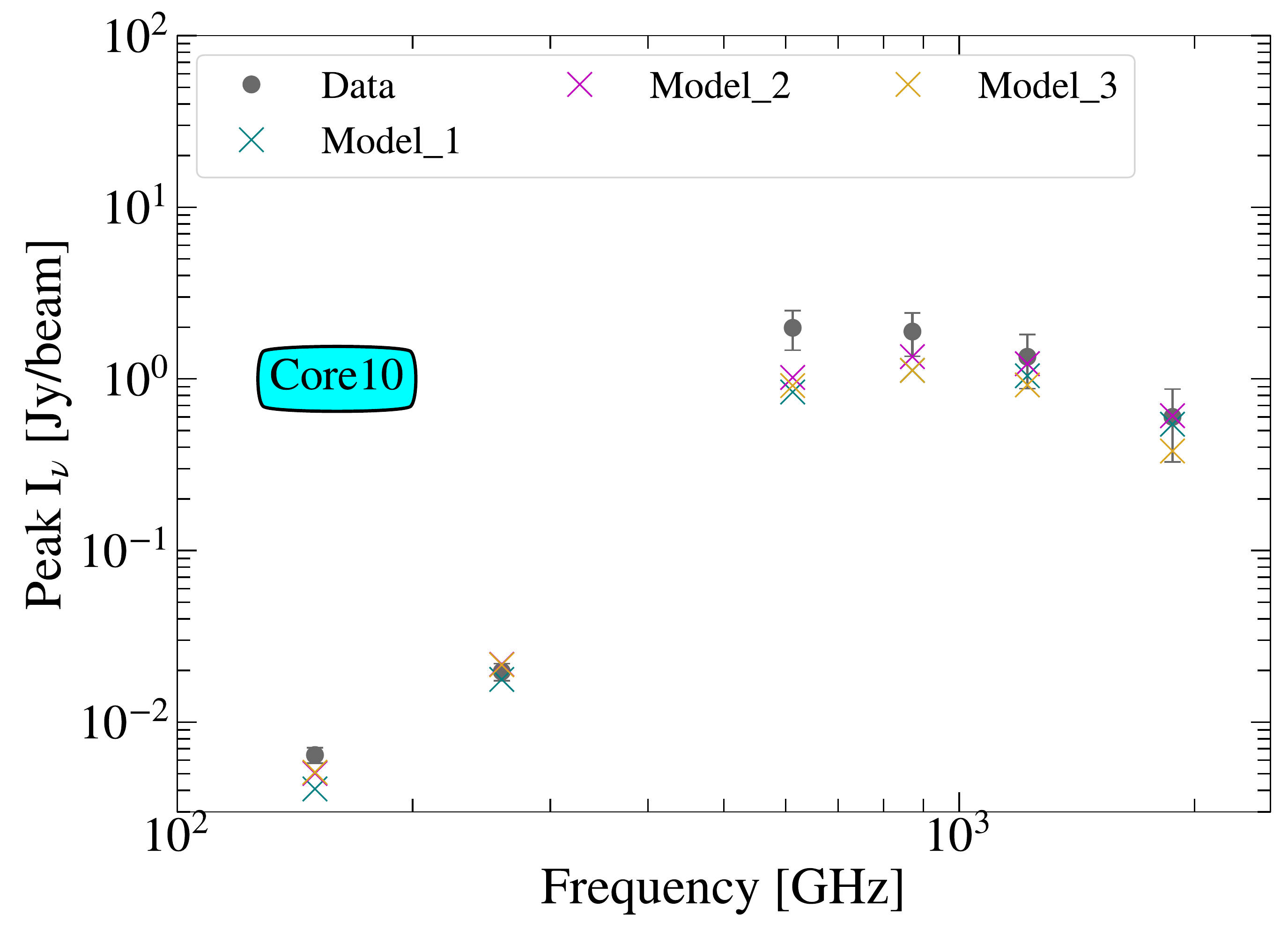} \\
\includegraphics[width=56mm]{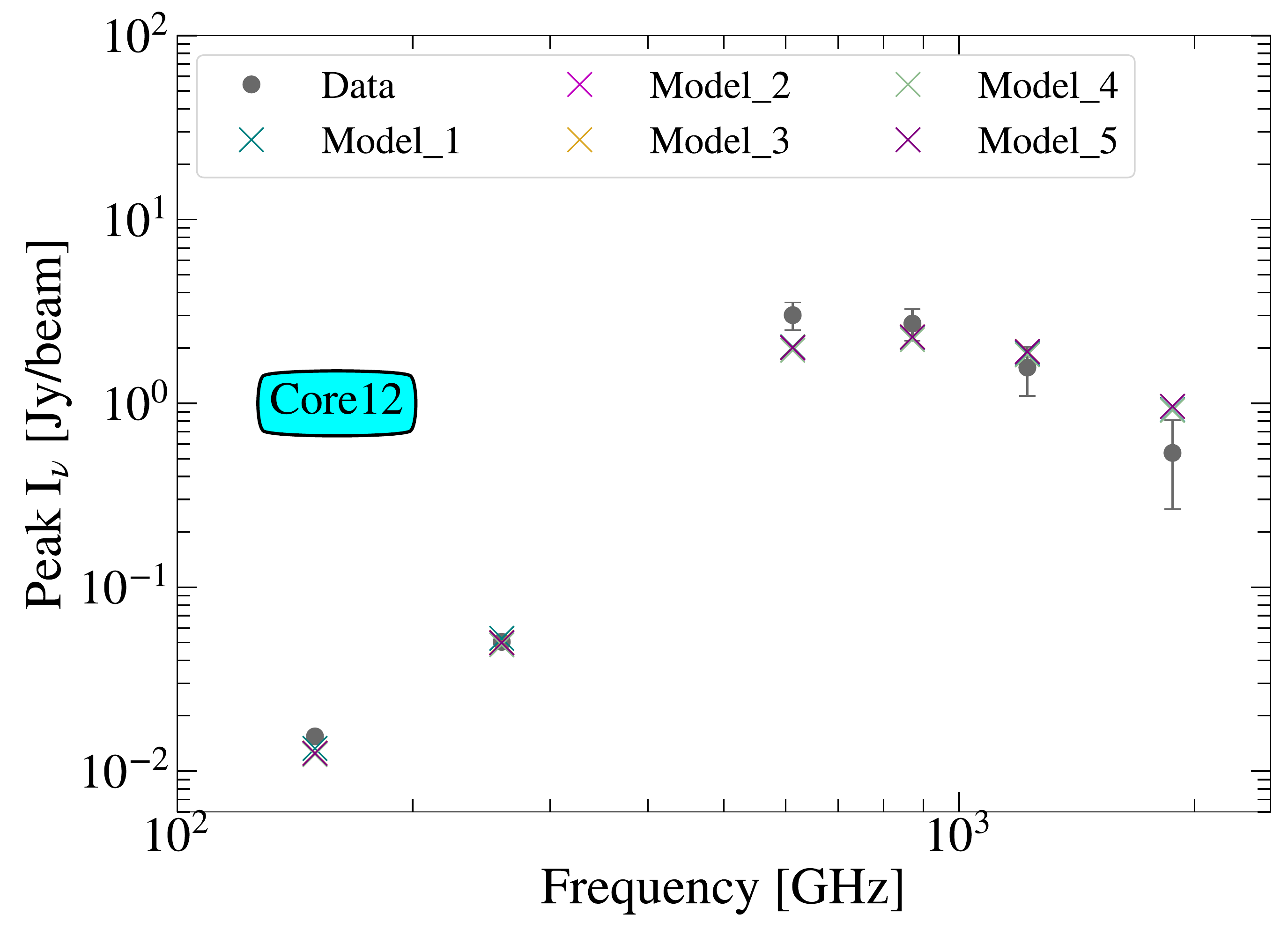} & \includegraphics[width=56mm]{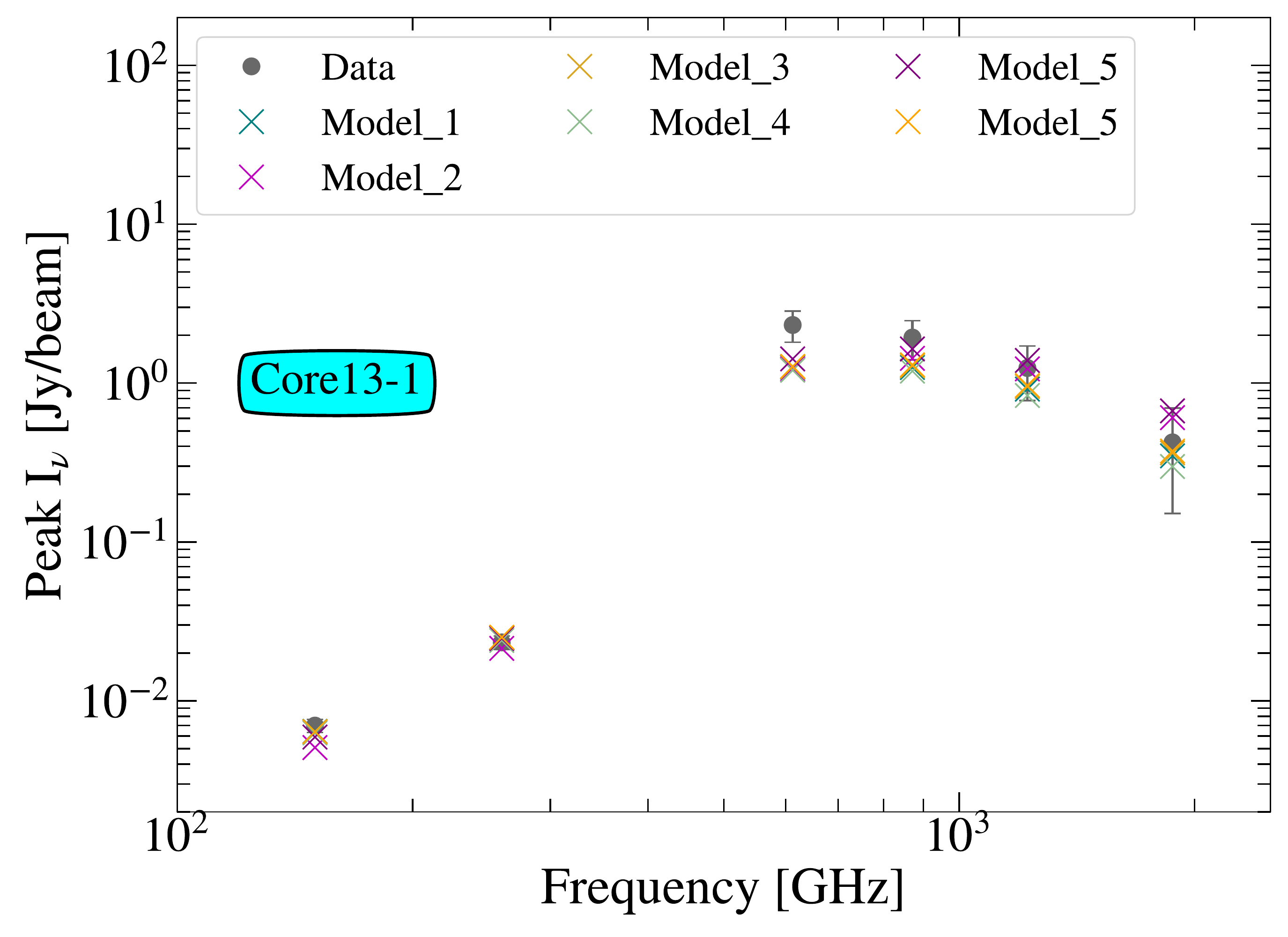} & \includegraphics[width=56mm]{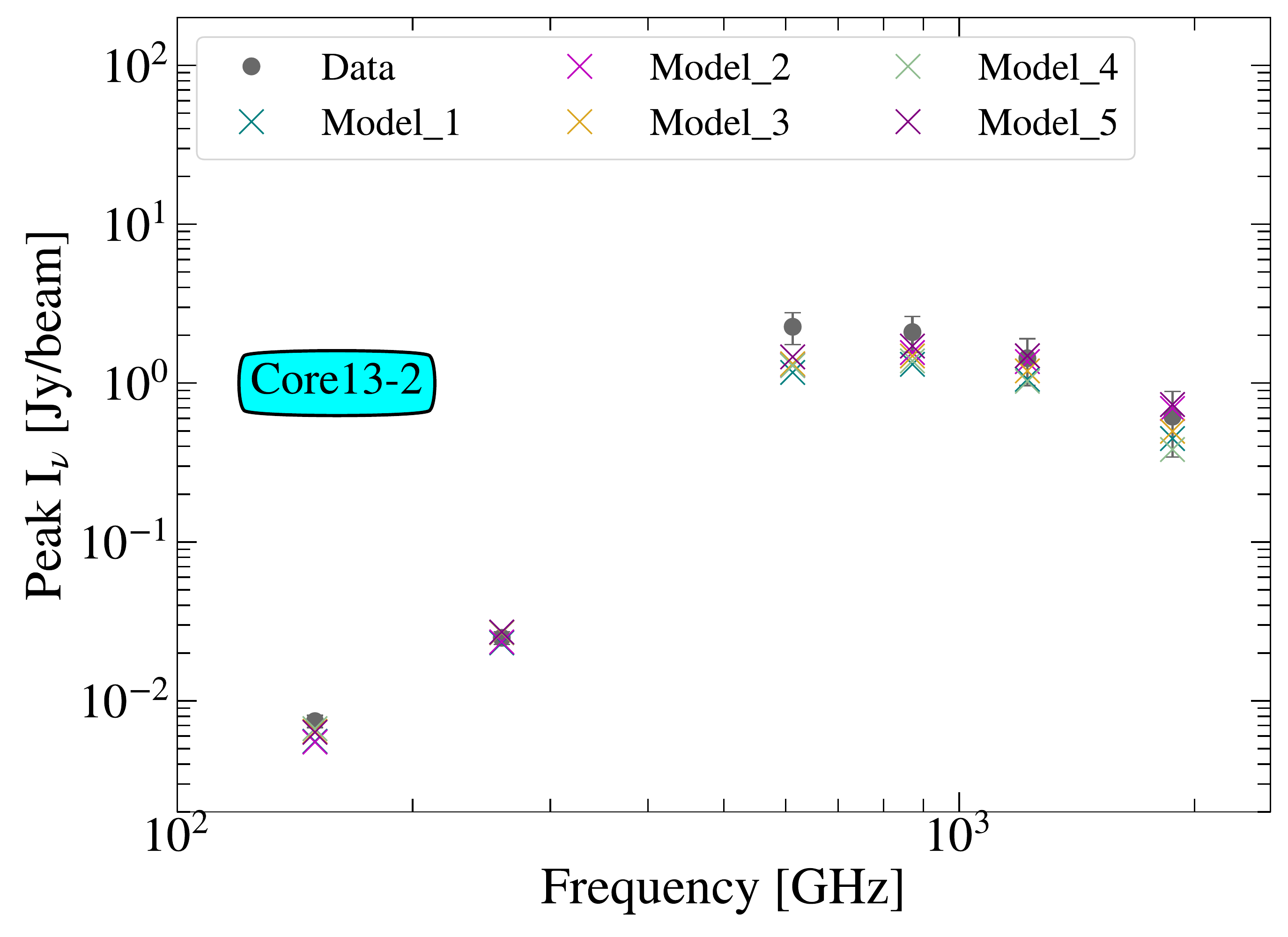} \\
\includegraphics[width=56mm]{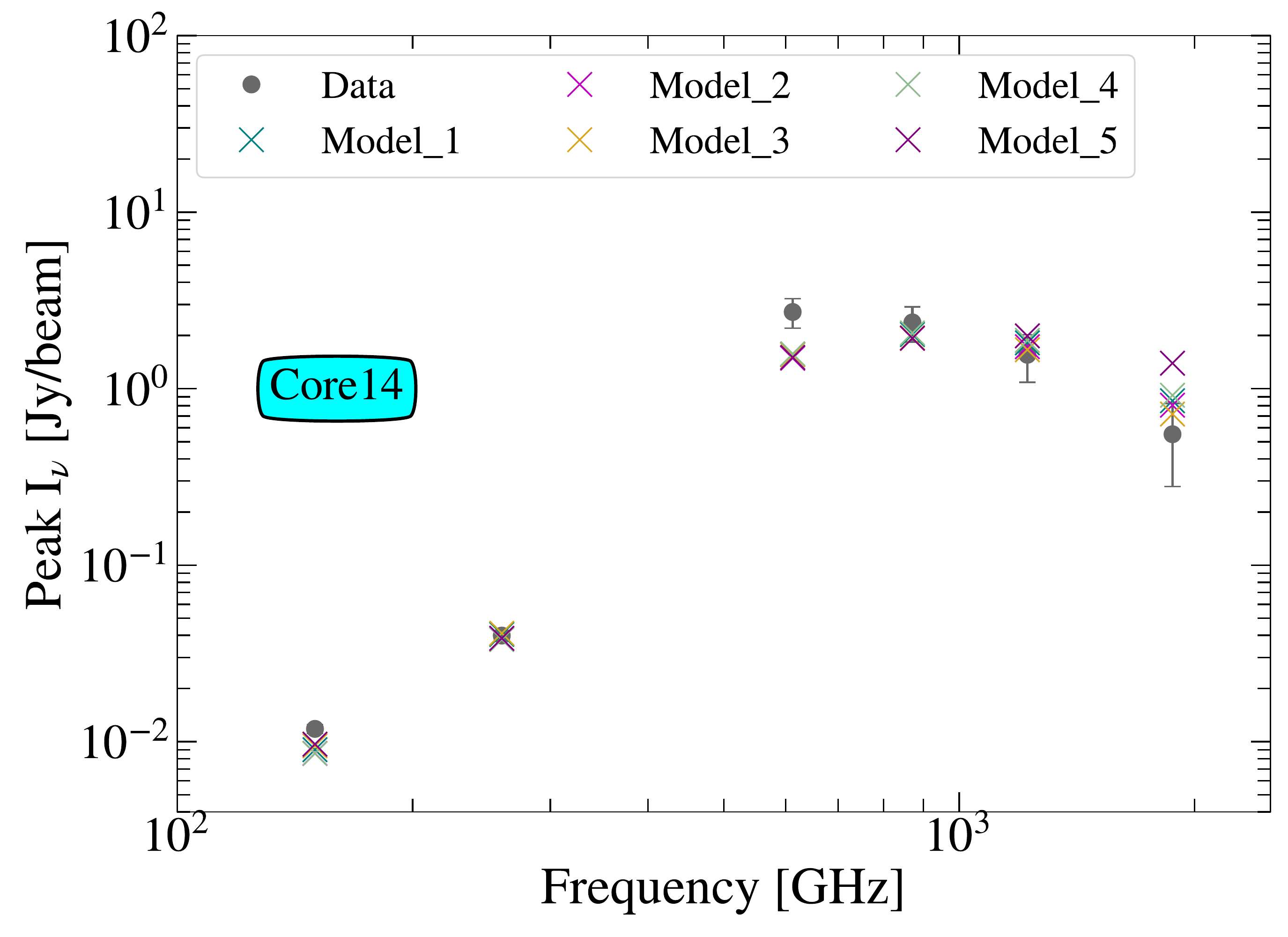} & \includegraphics[width=56mm]{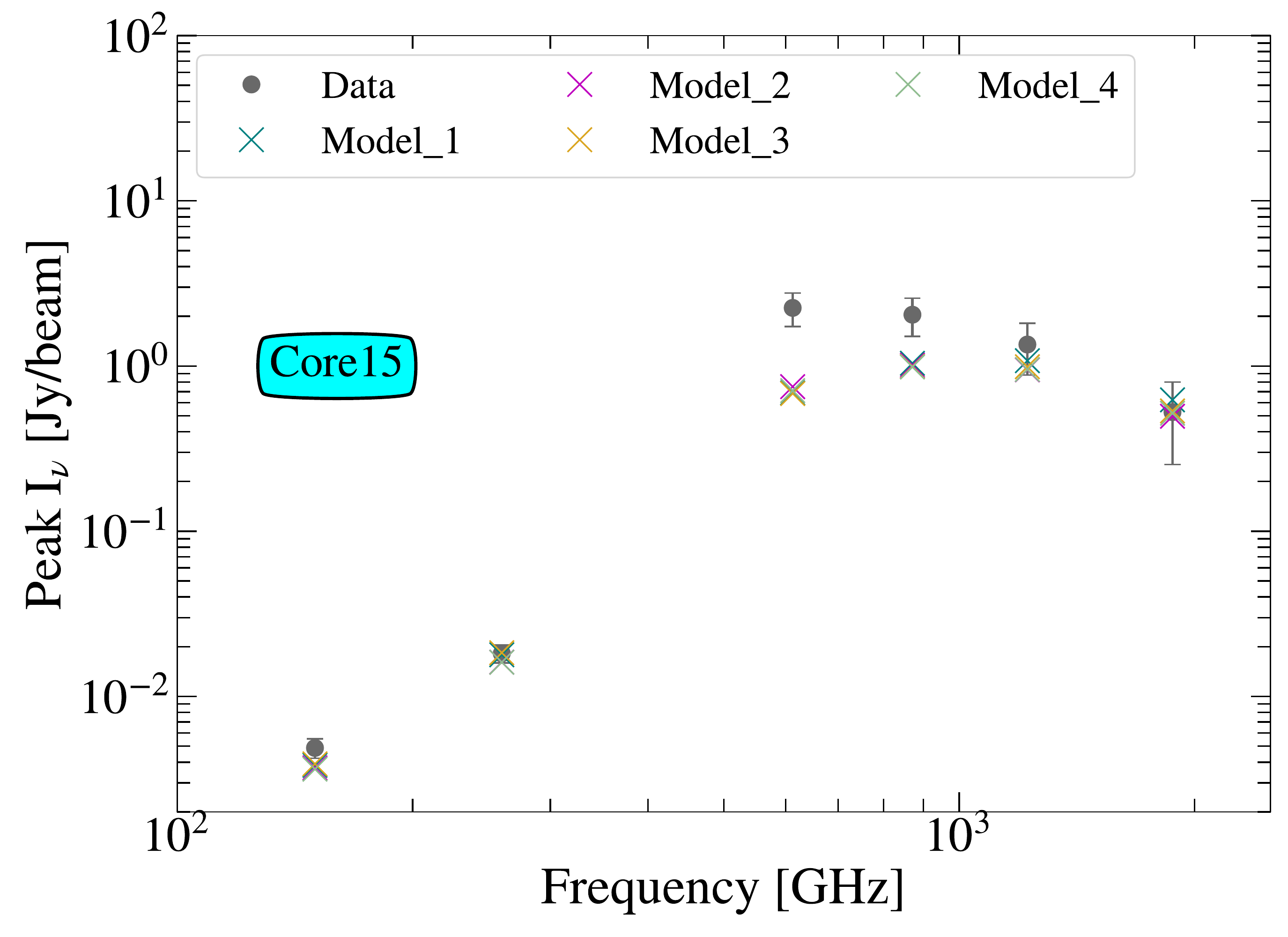} & \includegraphics[width=56mm]{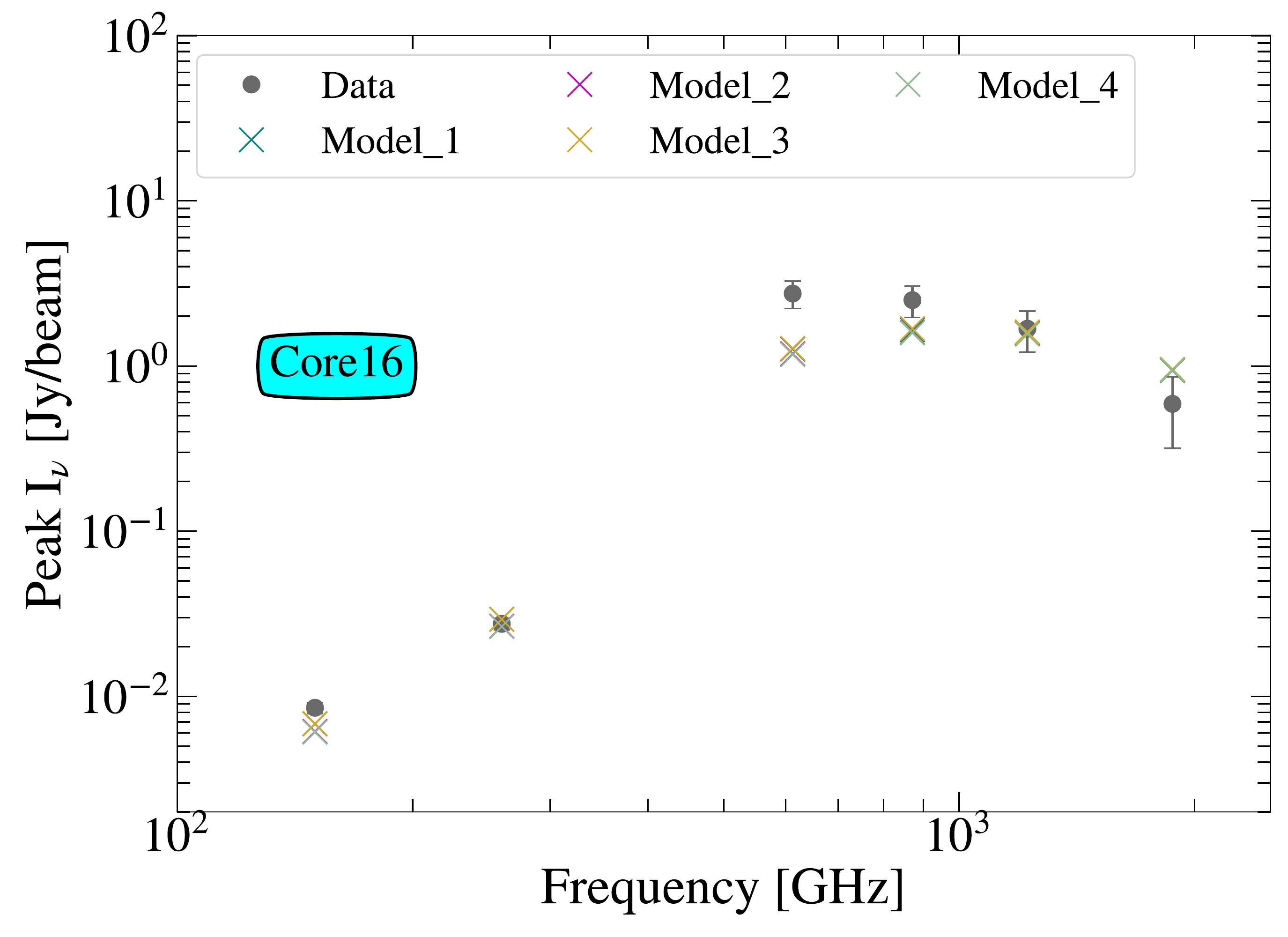} \\
\includegraphics[width=56mm]{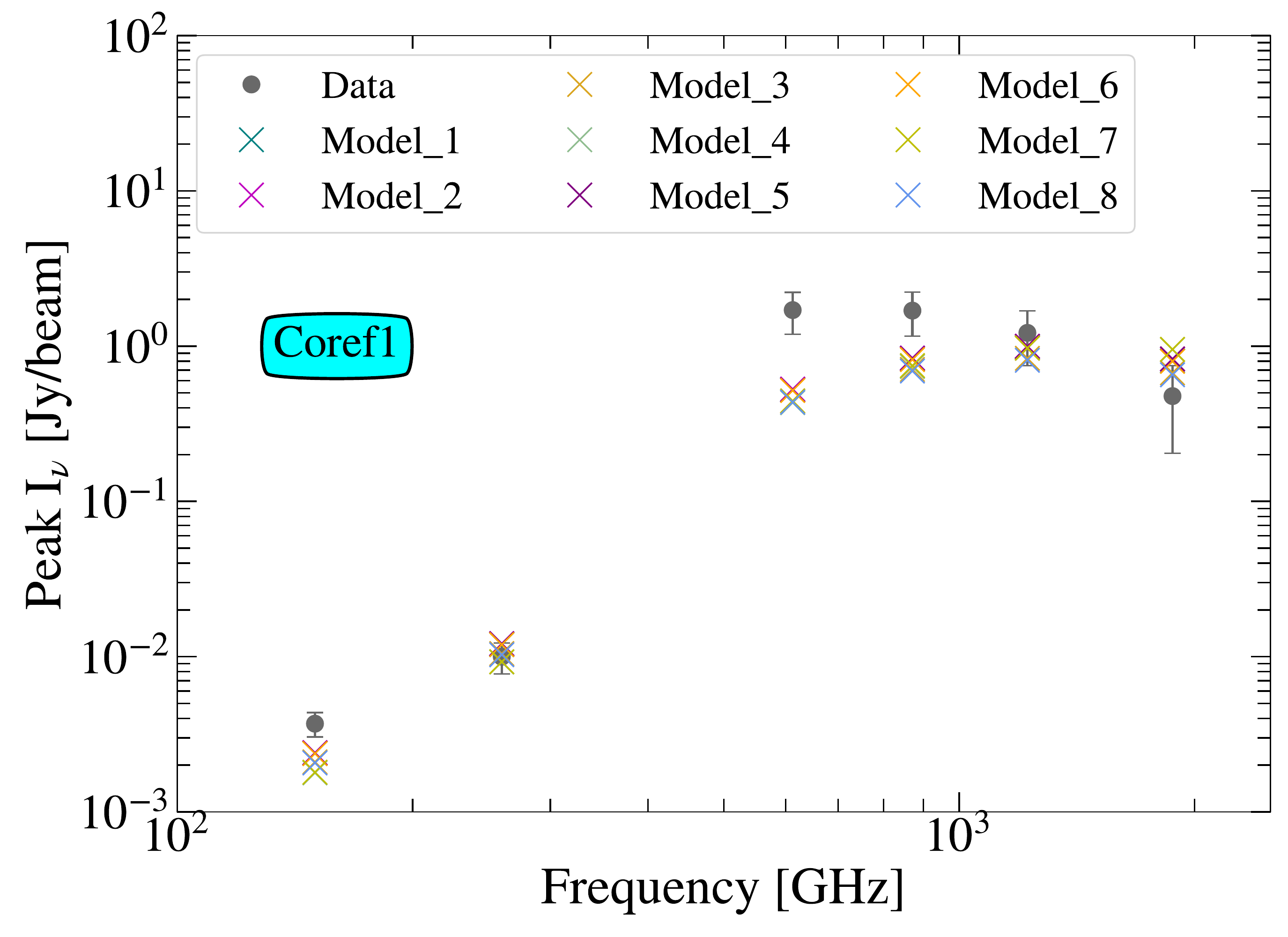} & \includegraphics[width=56mm]{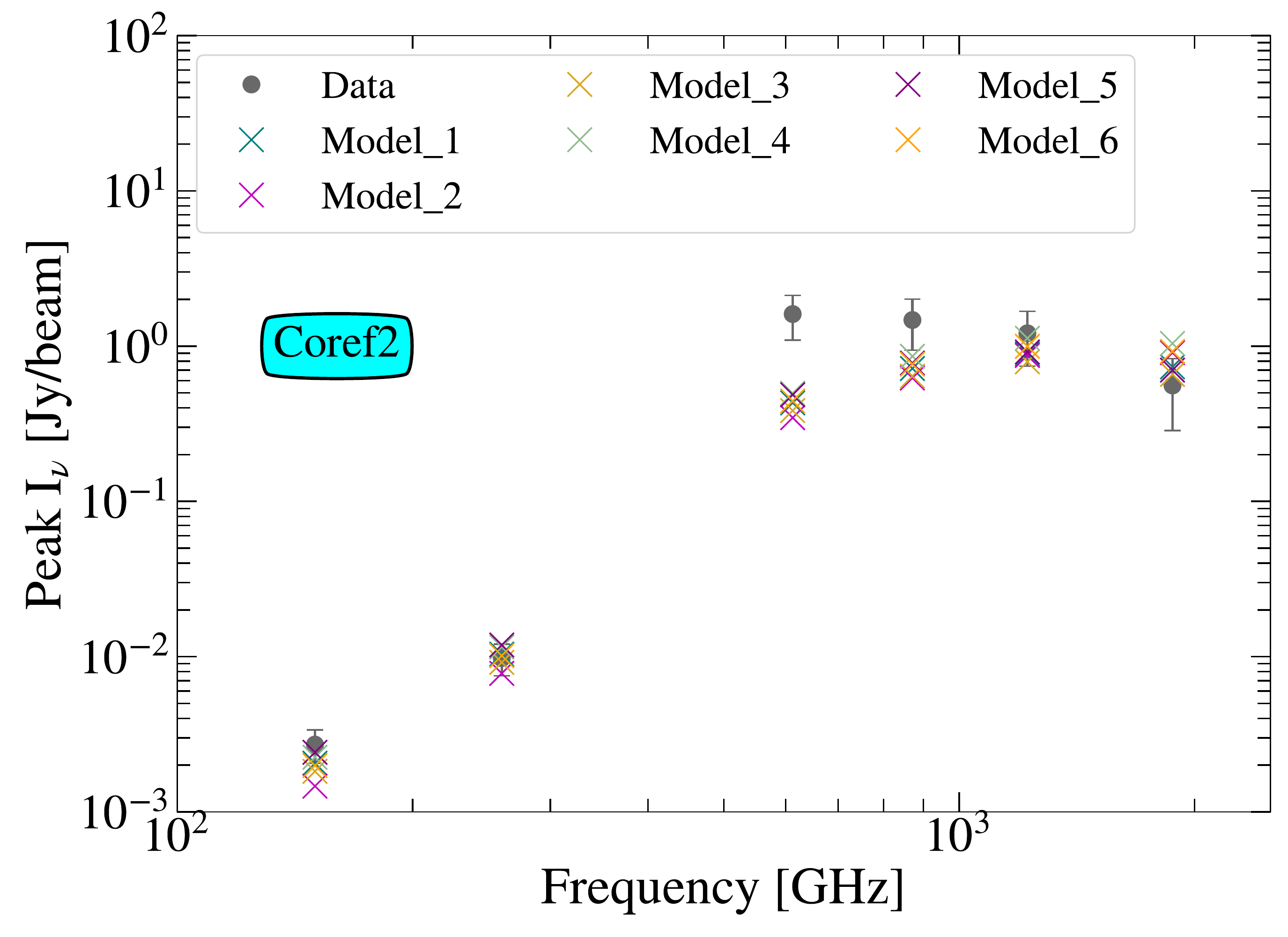} 
\end{array}$
\end{center}
\caption{\label{all_seds} Peak intensities and their errors (from left to right) for the data at 2.0mm, 1.2mm, 500$\mu$m, 350$\mu$m, 250$\mu$m, and 160$\mu$m are plotted as grey points. Each panel represents a different core and labeled via a cyan box to the left in each panel. In each panel the best-fit \textit{pandora} models are plotted as `X' symbols ordered starting from `Model\_1' for each core. Only models that have peak 1.2mm and 250$\mu$m intensities that lie within the errors are considered as `best-fits' to the SED. }
\end{figure*}

\begin{figure*}
\centering
\begin{center}$
\begin{array}{cccccccccccccc}
\includegraphics[width=56mm]{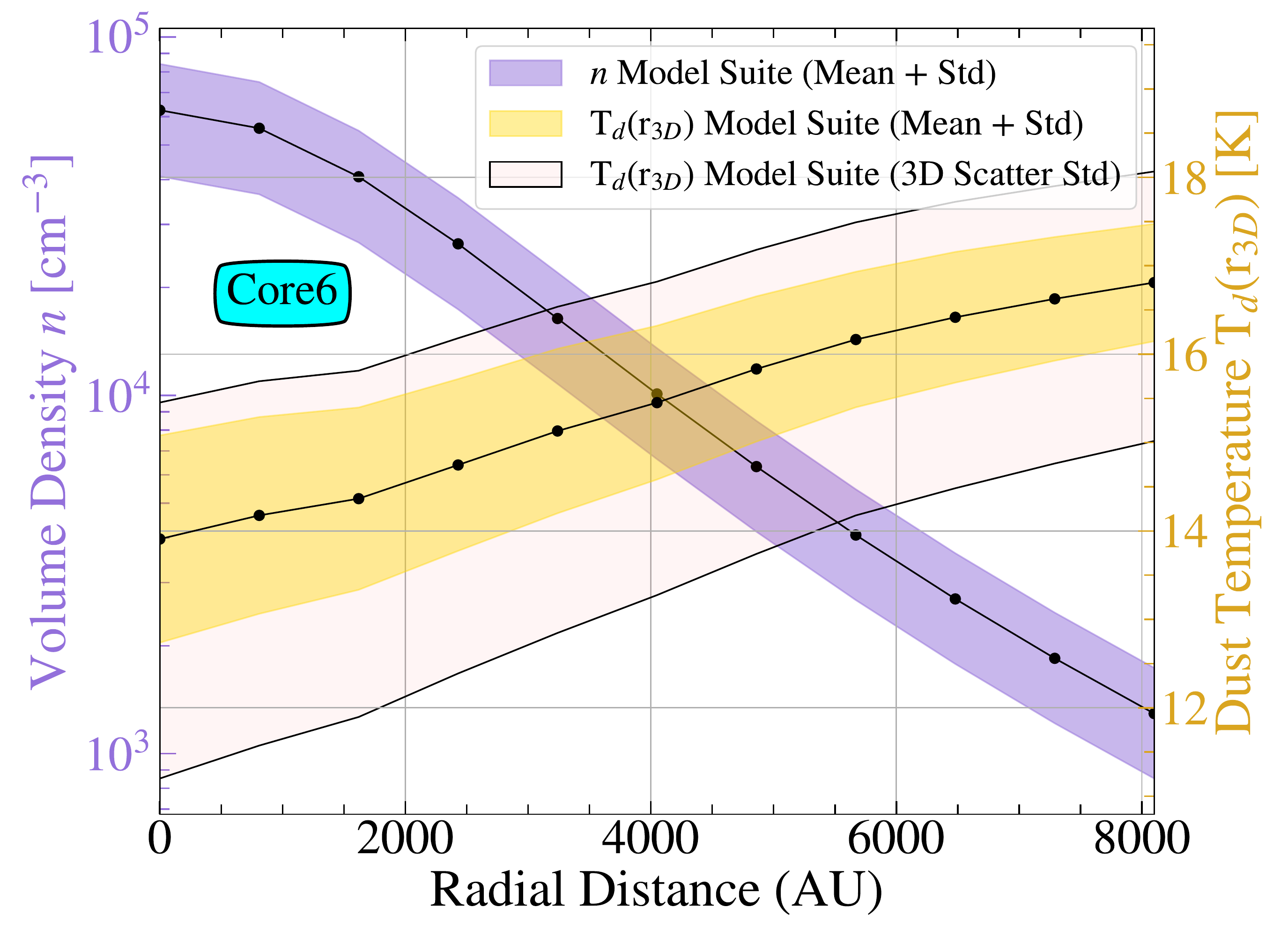} & \includegraphics[width=56mm]{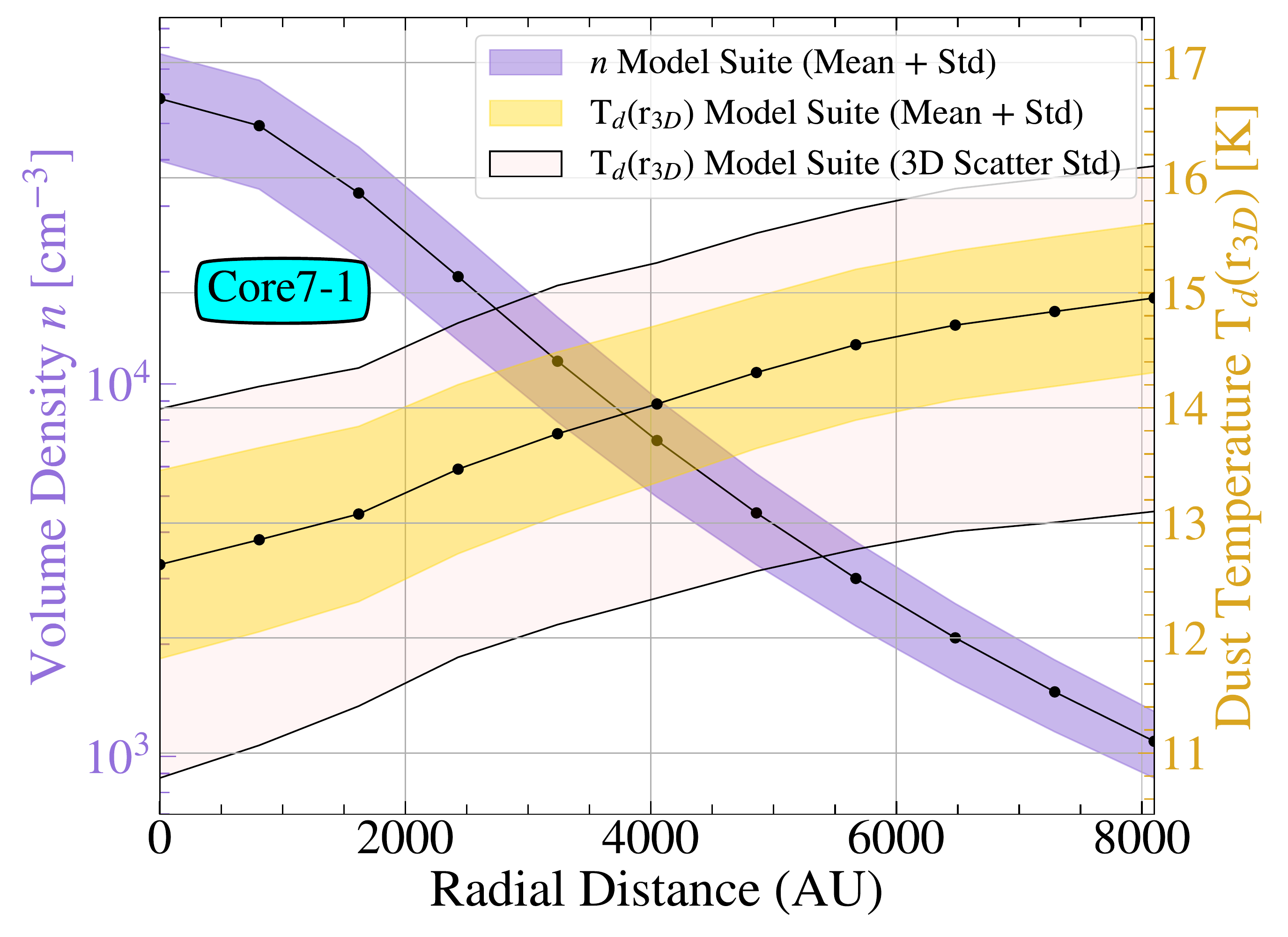} & \includegraphics[width=56mm]{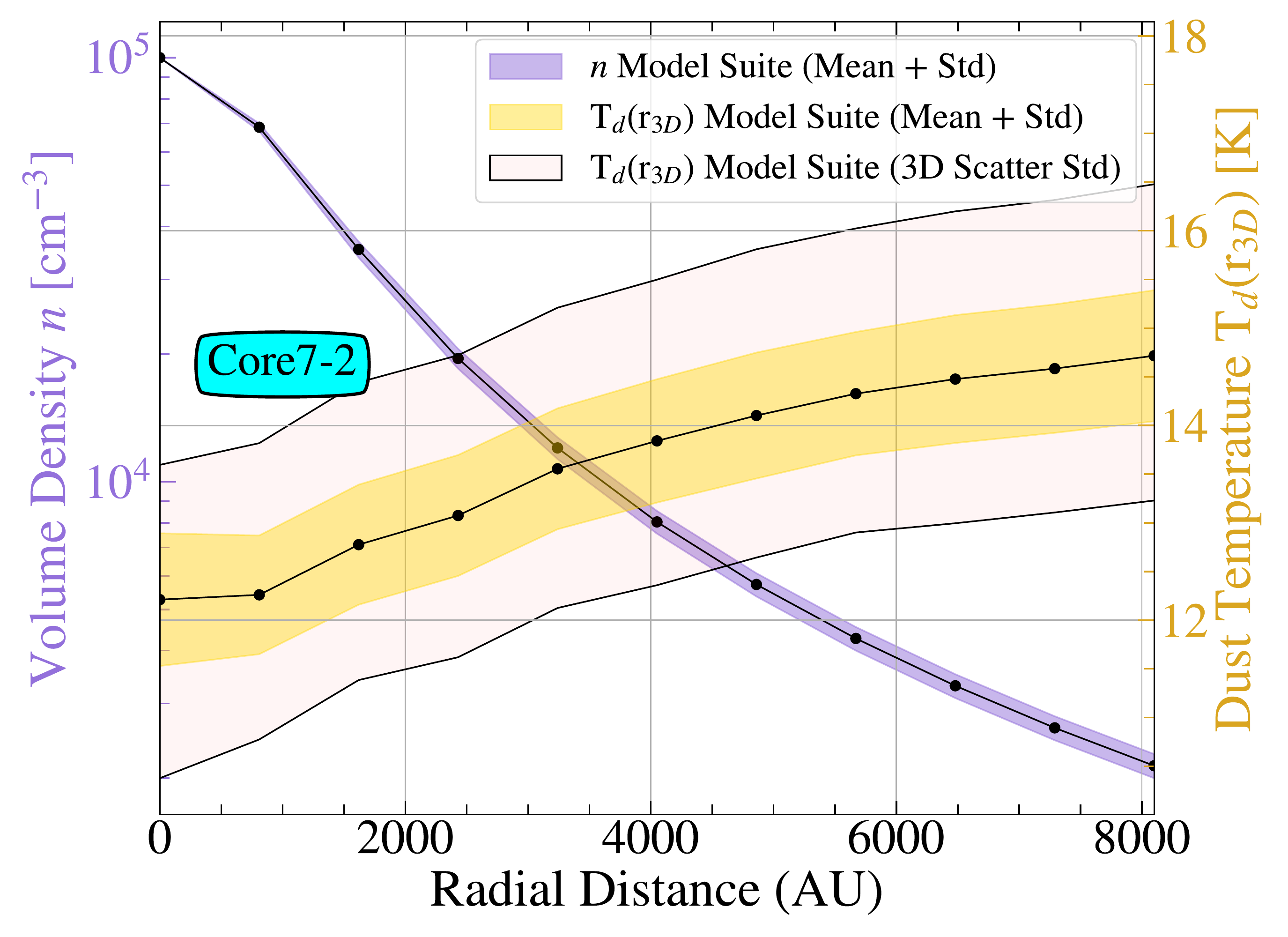} \\
\includegraphics[width=56mm]{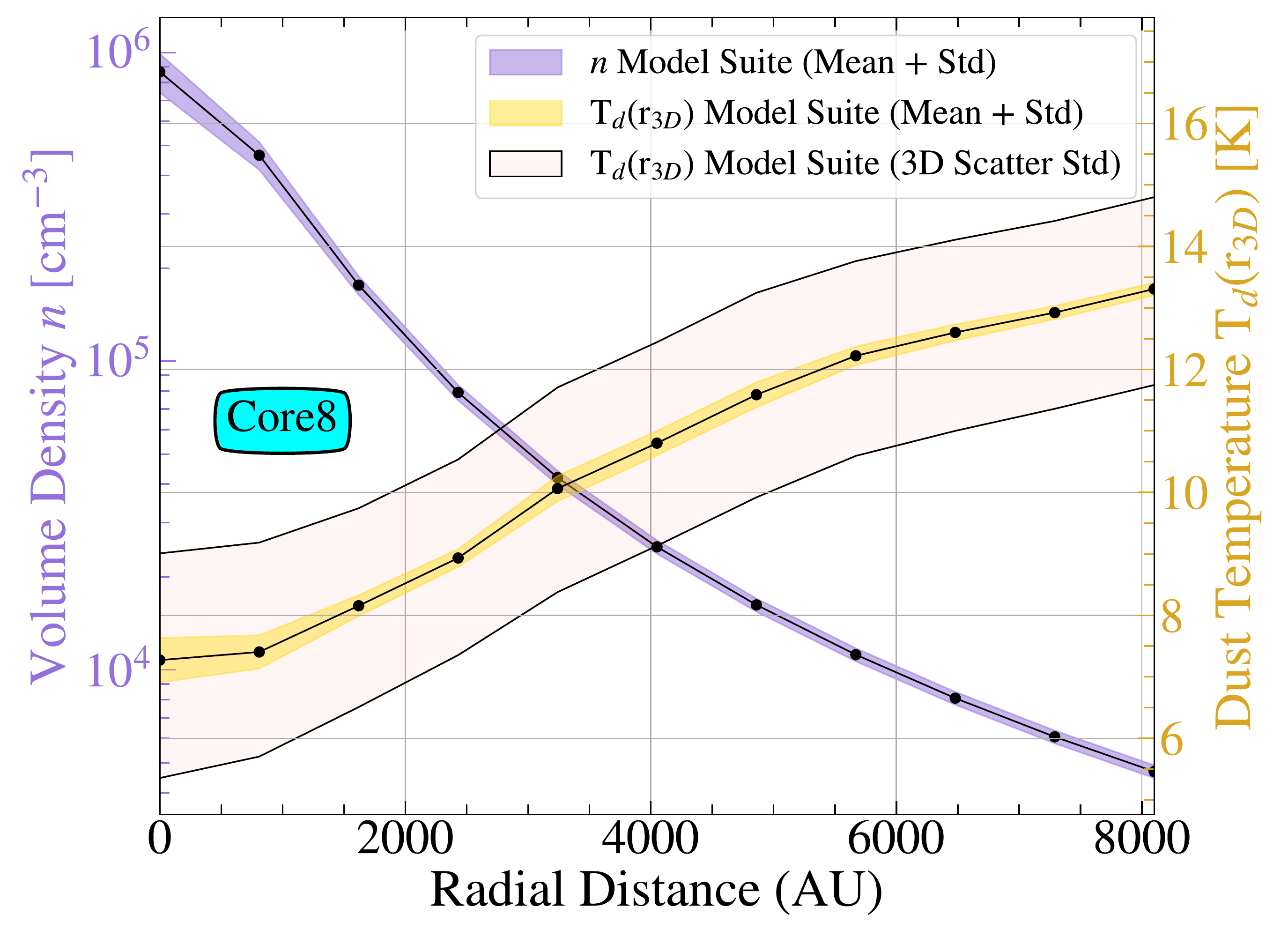} & \includegraphics[width=56mm]{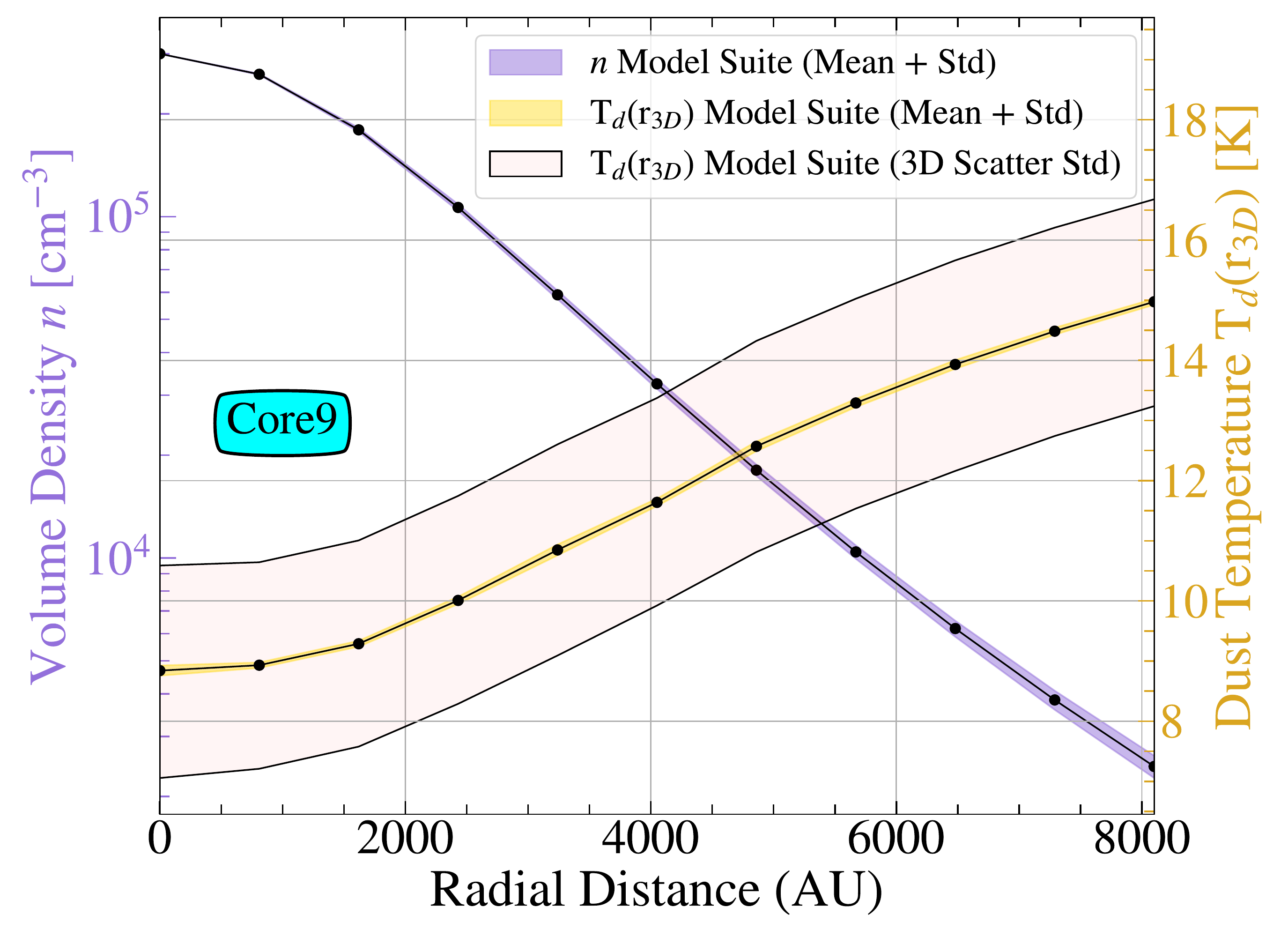} & \includegraphics[width=56mm]{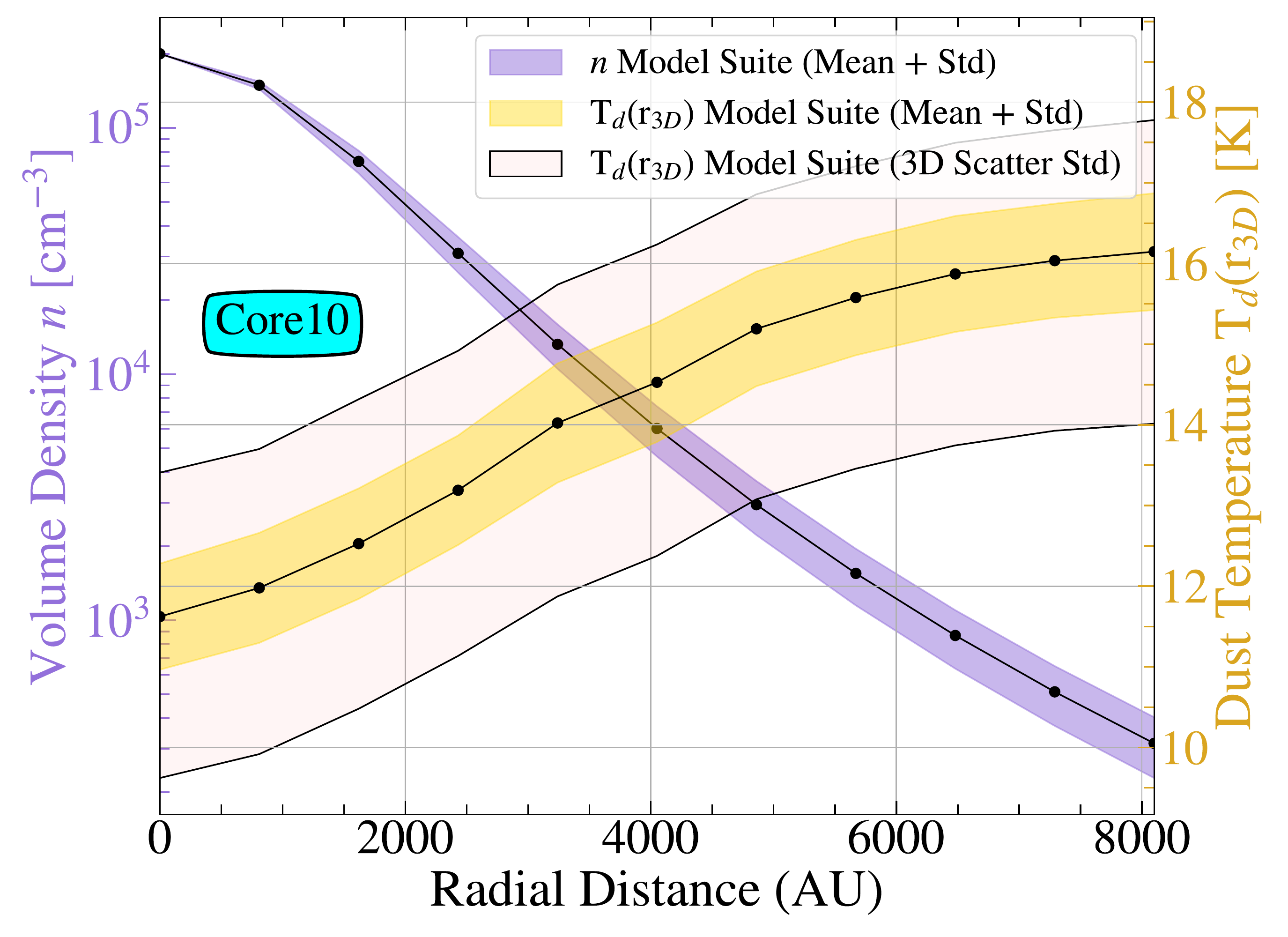} \\
\includegraphics[width=56mm]{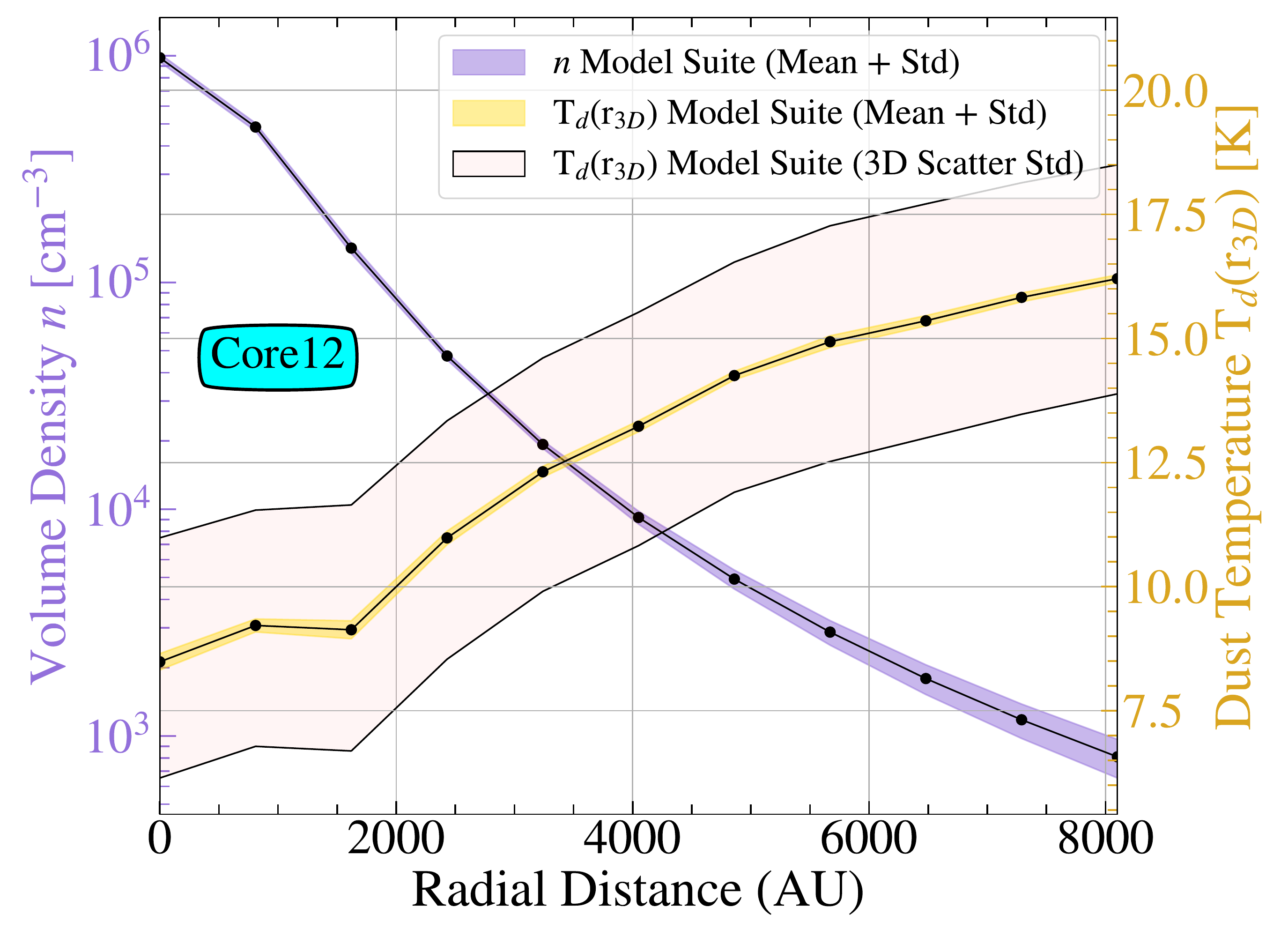} & \includegraphics[width=56mm]{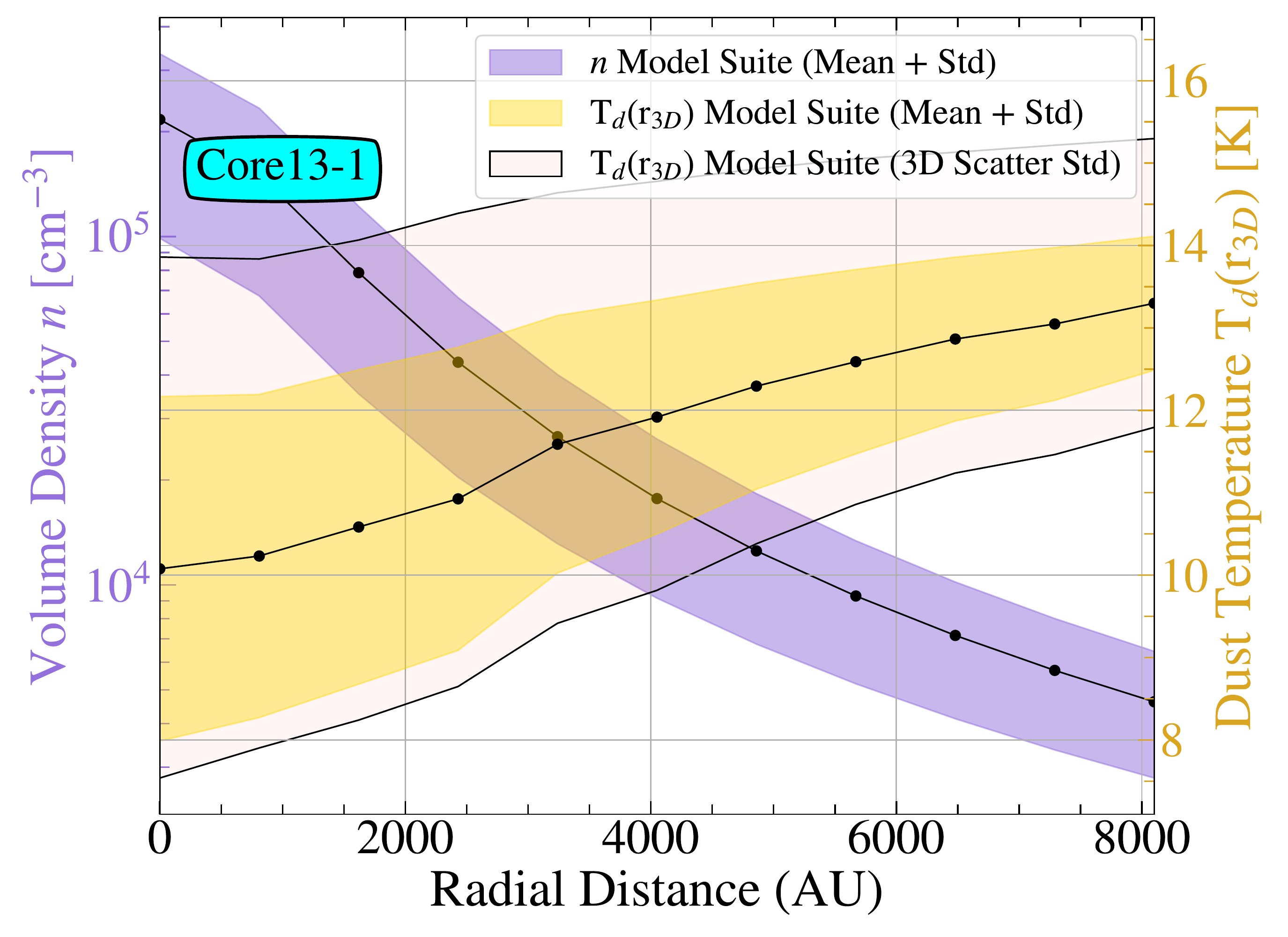} & \includegraphics[width=56mm]{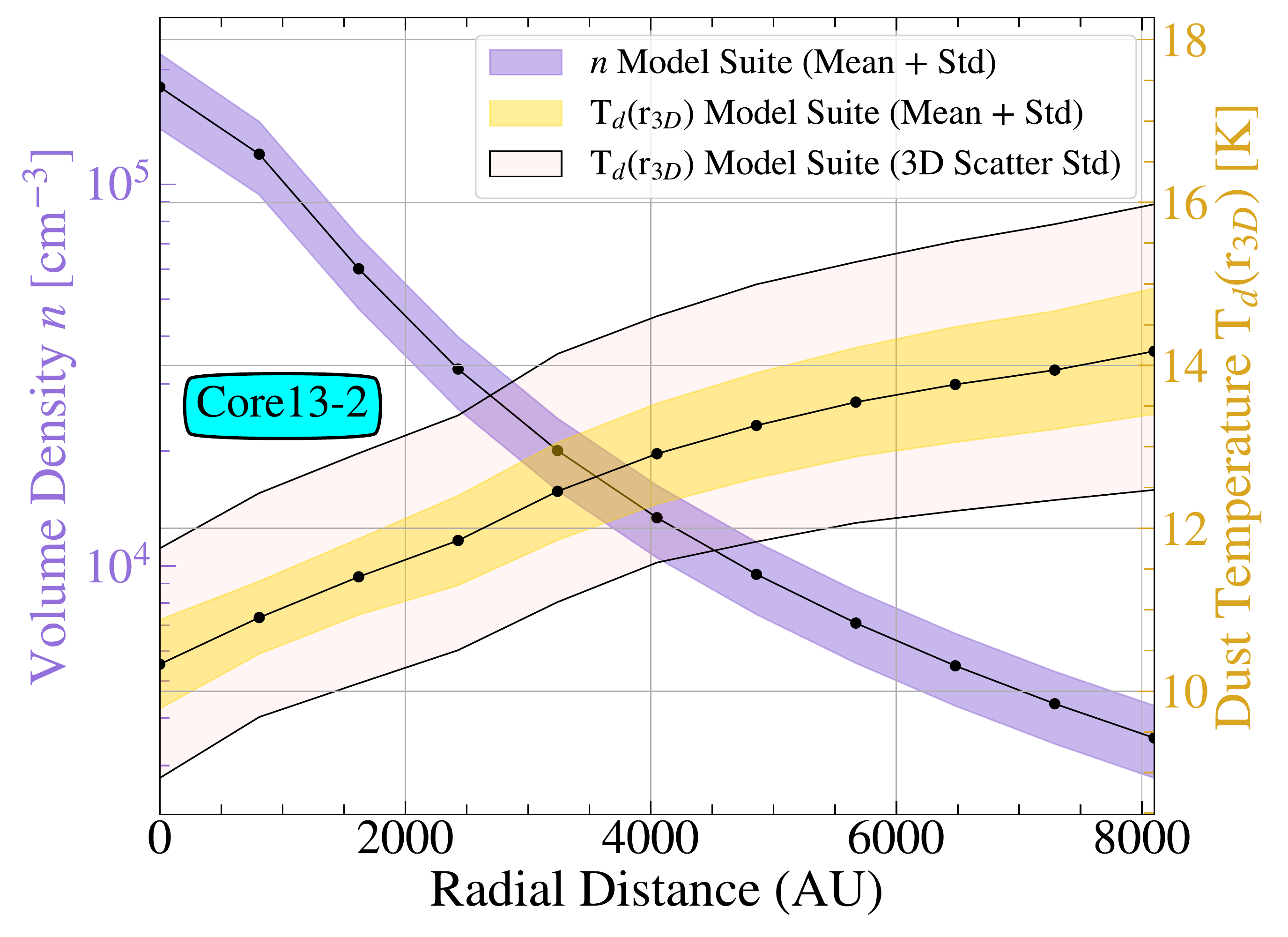} \\
\includegraphics[width=56mm]{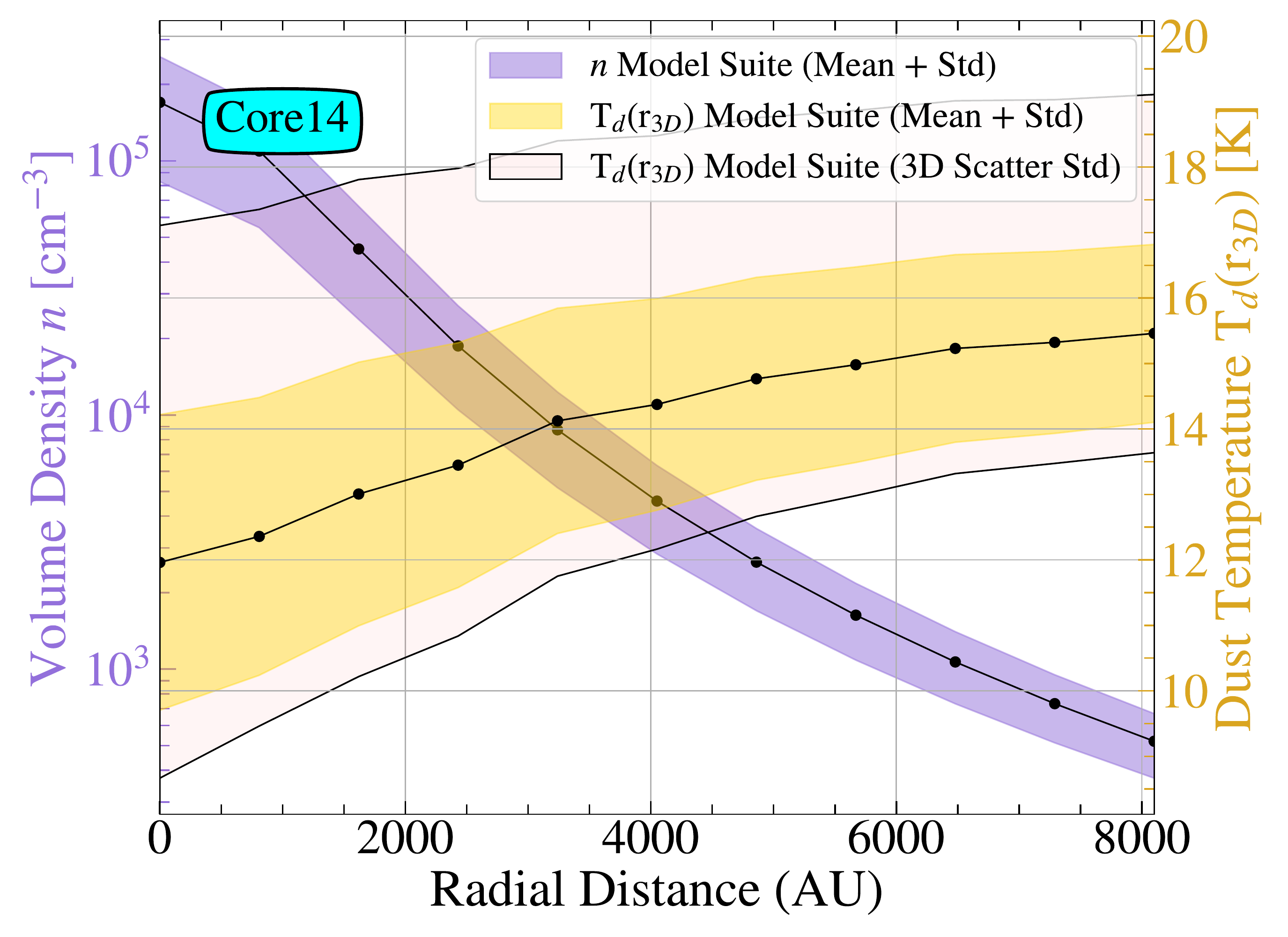} &
\includegraphics[width=56mm]{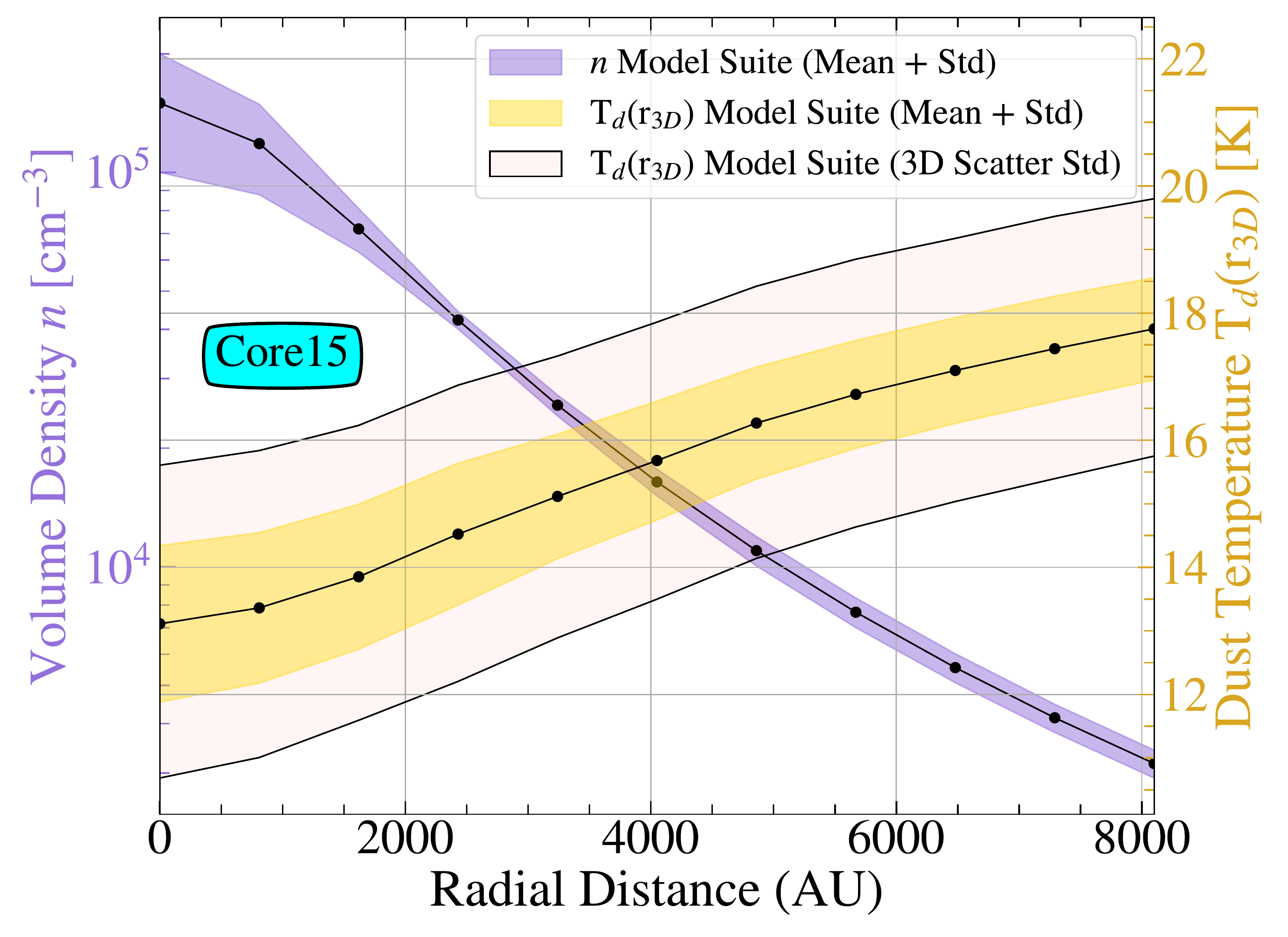} &
\includegraphics[width=56mm]{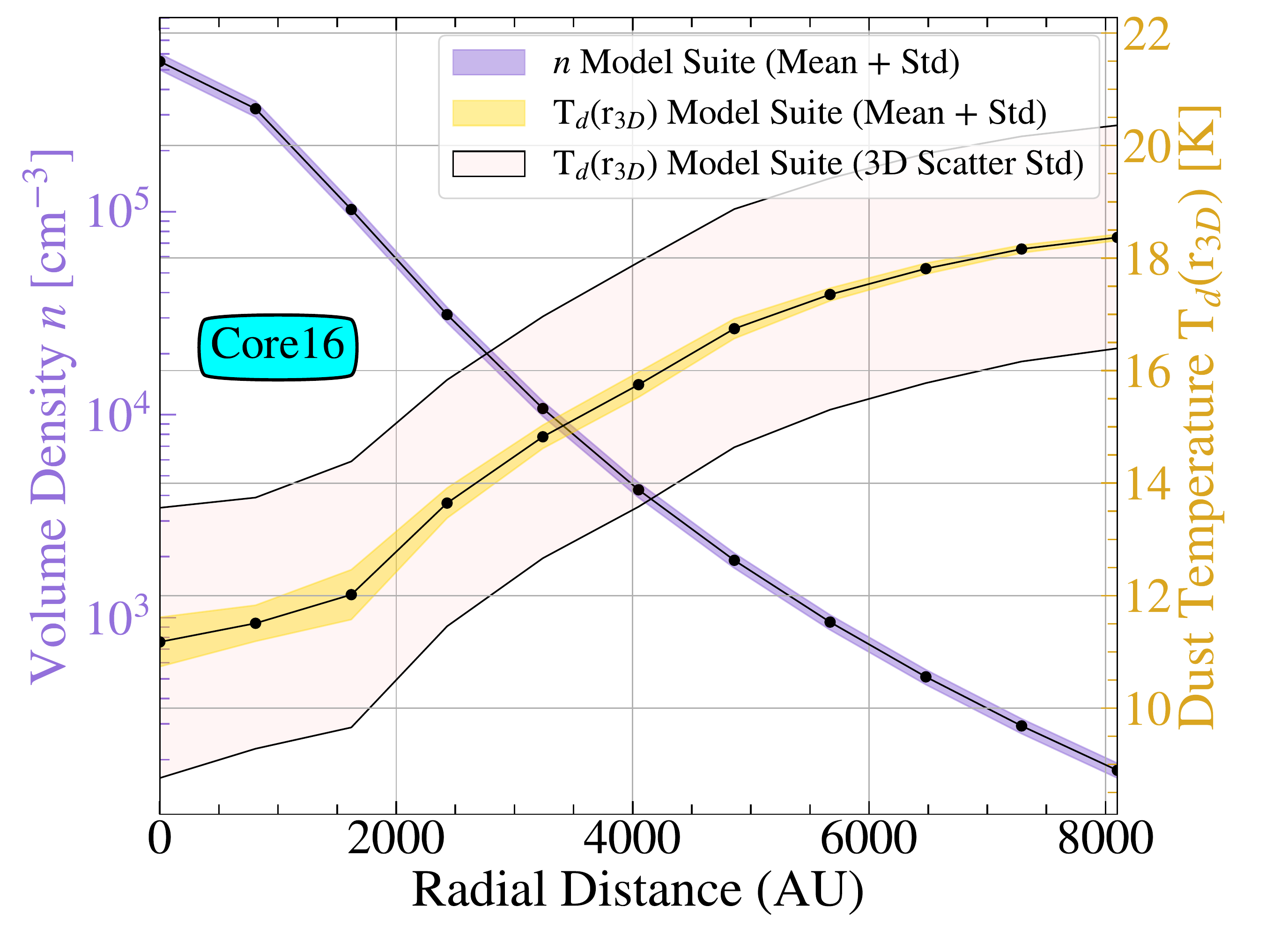} \\
\includegraphics[width=56mm]{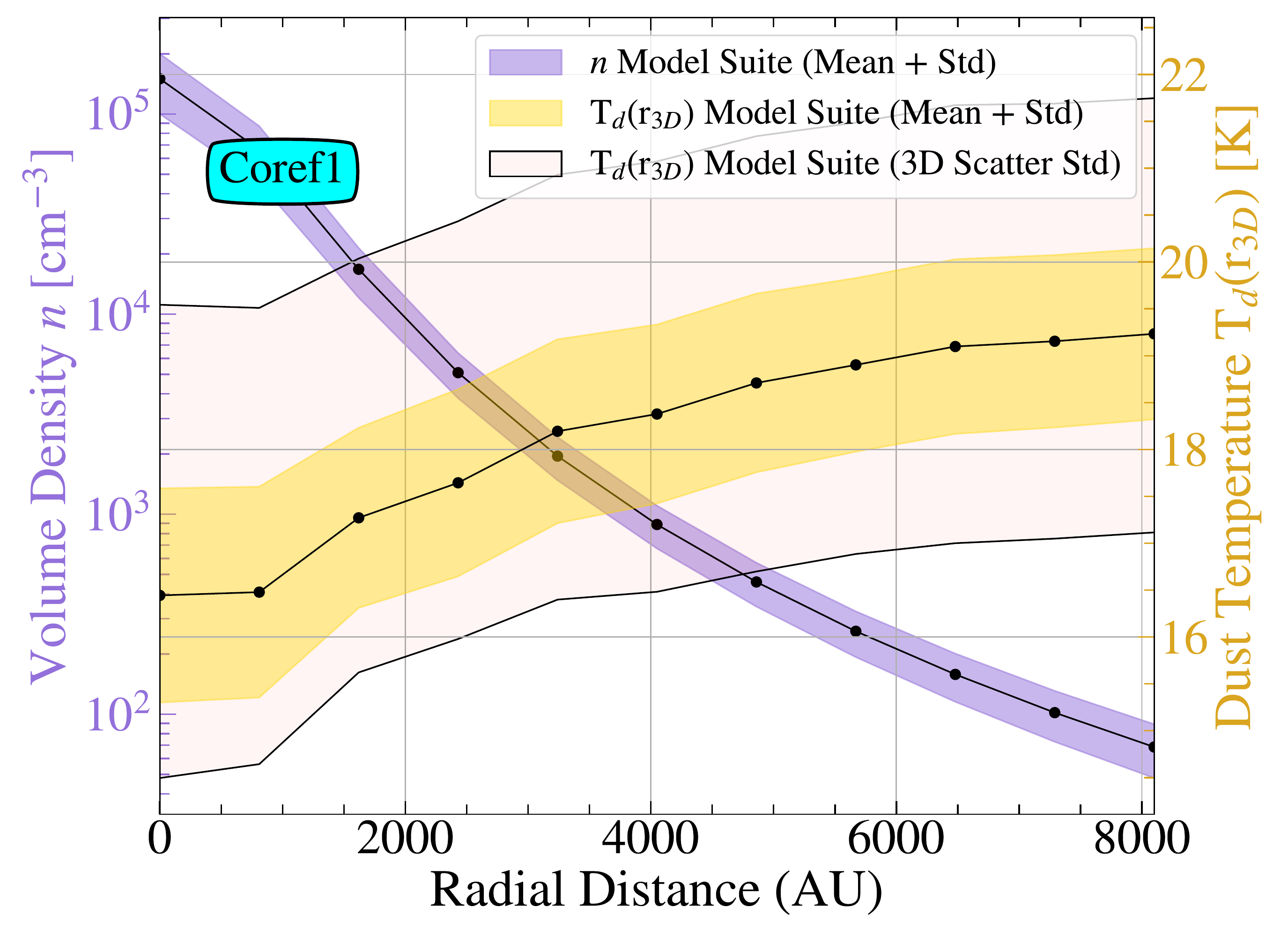} &
\includegraphics[width=56mm]{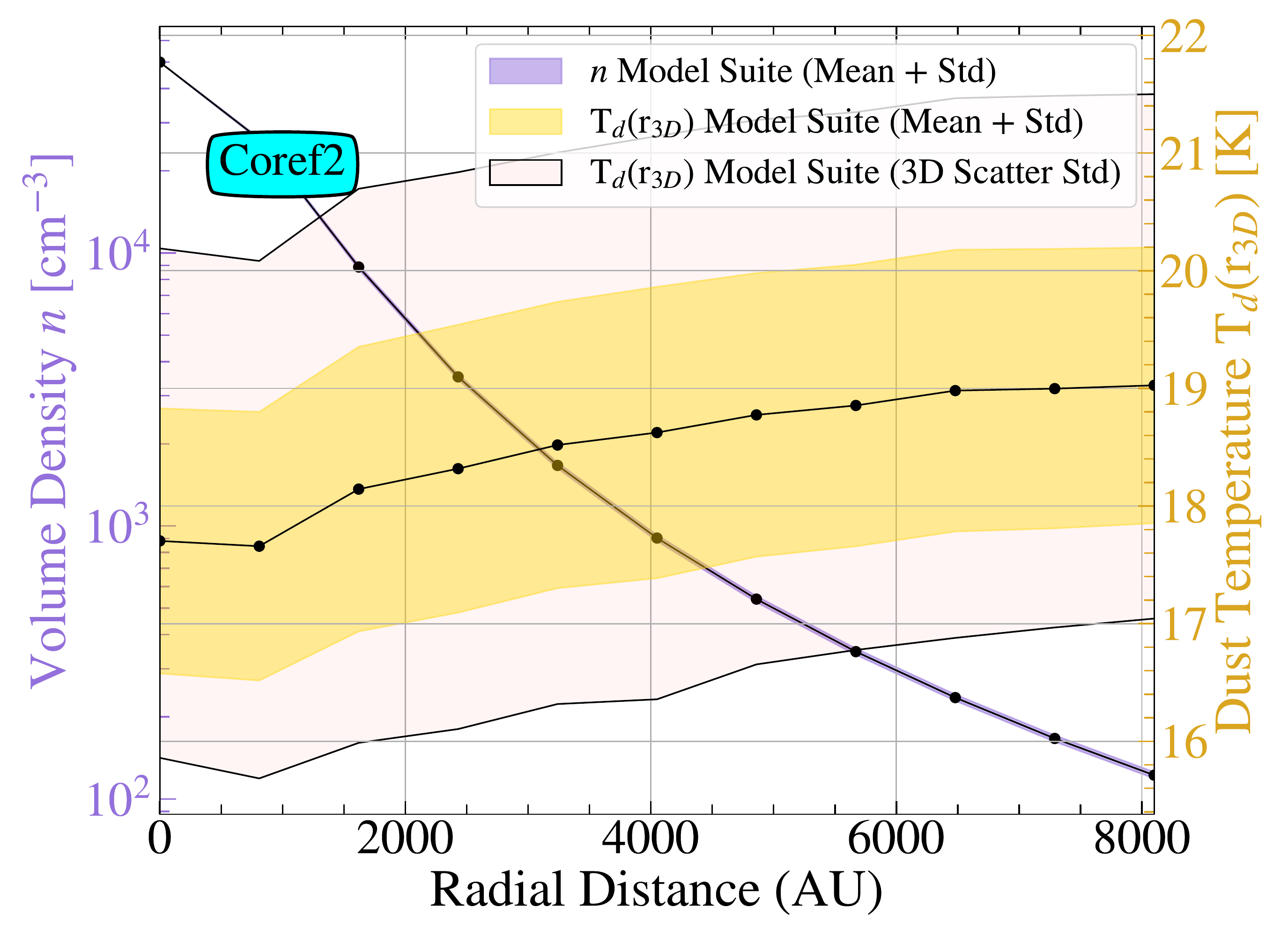} 
\end{array}$
\end{center}
\caption{\label{tdust_plots} Averaged model best-fit density $n$ and 3D dust temperature T$_d$({r$_{3D}$}) distributions for each of the cores with respect to radial distance (in AU) out from the center as black connected points. The standard deviation for all best-fit models are represented by filled in regions in purple (for $n$) and in yellow (for T$_d$({r$_{3D}$})). Averages can be unrepresentative of the full range in T$_d$({r$_{3D}$}), so an additional lighter shaded region outlined in black is shown that depicts the standard deviation in scatter within $r_{3D}<60''\,(8100\, \mathrm{AU})$ of the modeled dust temperature maps. The y-axis on the left side of each plot denotes the volume density $n$ in units of cm$^{-3}$ calculated from the modified Plummer profile and normalized radius. The y-axis on the right side of each plot denotes the dust temperature in units of K, extracted from each \textit{pandora} model's 3D temperature map. }
\end{figure*}

\subsubsection{Remaining Best-fits and Core Statistics}\label{compare}

For the remainder of the cores, as for core 12, the same analysis is done to determine best-fit models and the results can be found in Table\,\ref{tab:pandora_best_fit_params} and Figure\,\ref{sector_plot}. Additionally, the best fit SEDs are plotted in Figure\,\ref{all_seds} for all the cores. In general, the best-fit models slightly under-predict the 2.0mm and 500$\mu$m peaks and slightly over-predict the 160$\mu$m peak, but the overall fits do well to reproduce observed continuum peaks. Again, the best-fits are selected to be those that fit the 1.2mm and 250$\mu$m data the best, in order to probe the Rayleigh-Jeans side and peak of the SED. 

Compared to the very dense, dynamically evolved prestellar core L1544 with central densities $\geq 10^7$\,cm$^{-3}$ \citep{2007A&A...470..221C, 2010MNRAS.402.1625K, 2019ApJ...874...89C}, the cores in B10 are more `typical' with modeled central densities ranging from 5\,$\times$\,10$^4$\,-\,1\,$\times$\,10$^6$\,cm$^{-3}$, with a mean value of 2.6\,$\times$\,10$^5$\,cm$^{-3}$ and a median value of 2.0\,$\times$\,10$^5$\,cm$^{-3}$ for the modeled cores. For many of the cores the central densities are as we would expect. For example, core 12 has a central density of $\sim 10^6$ cm$^{-3}$, consistent with \cite{2019PASJ...71...73T}. Additionally, for core 14 we find best-fit models ranging from 5\,$\times$\,10$^4$\,-\,2\,$\times$\,10$^5$\,cm$^{-3}$, consistent with the \cite{2019PASJ...71...73T} estimate of $\sim 1-3 \times 10^5$ cm$^{-3}$. We are surprised that core 8 is fit by models with the high densities of $7-10 \times$\,10$^5$\, cm$^{-3}$. From looking at core 8's radial profile in Figure\,\ref{sector_plot}, it is clear that, unlike neighboring core 9, the shallower slope and smaller inner radius values raise the core's central density. The smaller cores, and thus those cores with larger error bars in their radial profiles, produce a larger number of best-fits (see, for example, core 13-1, core f1 and core f2). Still, given this range, for individual cores the best-fit density values do not differ by more than a factor of five. 

Modeled volume density and 3D dust temperature radial profiles, T$_d$({r$_{3D}$}), are plotted in Figure\,\ref{tdust_plots}. The volume density is calculated directly from the modified Plummer profile. The RADMC-3D generated dust temperature maps were created in the \textit{pandora} framework for each best-fit model at the same size and resolution as the 1.2mm map and thus radial profiles were extracted similarly, within a 6$''$ annulus out to 60$''$ (8100 AU). We find, for example, in the case of core 12 that while the average T$_d$({r$_{3D}=0$}) is at 8.5\,K, the lowest T$_d$({r$_{3D}$}) values scatter at values as low as 6.5\,K (see Figure\,\ref{tdust_plots}).

\begin{figure}
\begin{center}$
\begin{array}{ll}
\includegraphics[width=82mm]{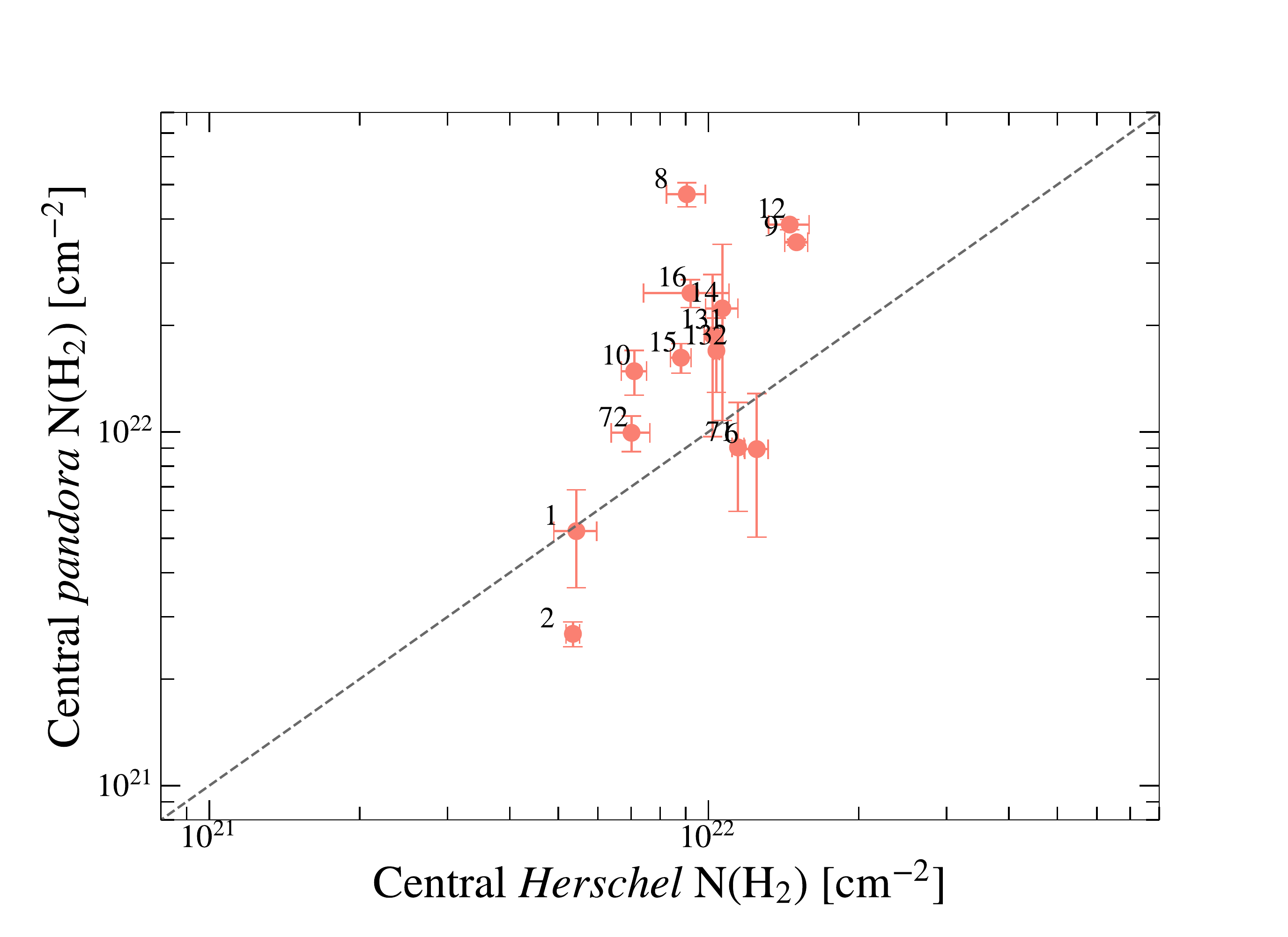} \\
\includegraphics[width=82mm]{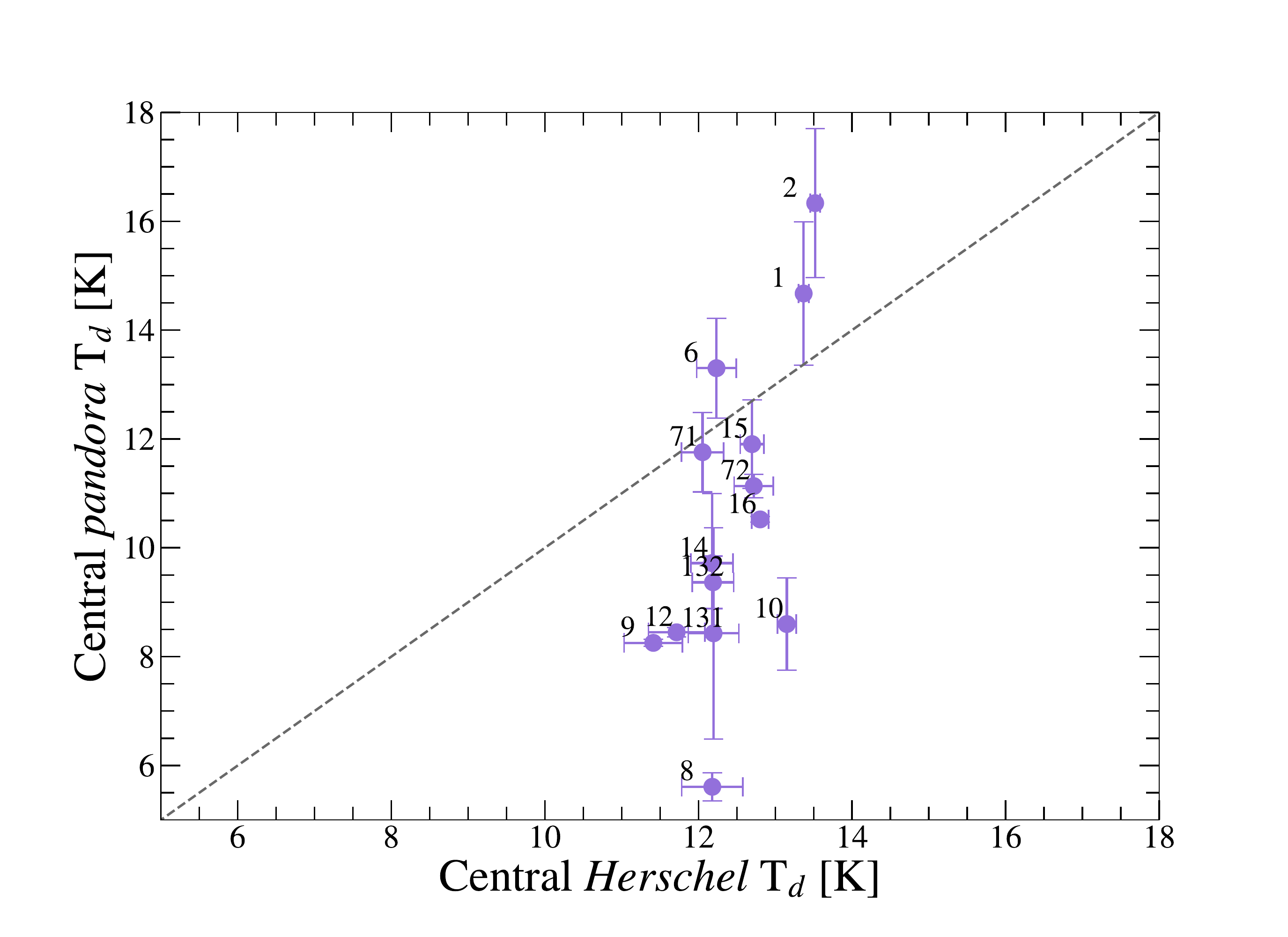} \\
\end{array}$
\end{center}
\caption{\label{hersch_compare} Comparison of (top) peak column density and (bottom) peak line-of-sight dust temperature for our modeled \textit{pandora} cores versus the \textit{Herschel} maps. In general the modeled cores have higher column density's and lower dust temperatures. The `1' and `2' labeled points refer to cores `f1' and `f2' respectively. We note that the \textit{Herschel} maps use an opacity law where $\beta$ is $\sim\,1.75$ in the B10 region.
}
\end{figure}

In Figure\,\ref{hersch_compare} we compare the average line-of-sight central dust temperature, T$_d$, as well as the central column density N(H$_2$) for all the cores modeled, finding in general the modeled T$_d$ is lower and N(H$_2$) is higher than \textit{Herschel} values. The N(H$_2$) values are extracted from the RADMC-3D generated column density map and the line-of-sight central dust temperature, T$_d$, from a RADMC-3D generated 2D dust temperature map collapsed along the z axis. For core 12, the best-fit column density peaks\,$\sim$2.6 times higher than the peak \textit{Herschel} value and the dust temperature  (T$_{d}$ = 11.80 K for \textit{Herschel}) drops to a lower value of 8.5 K in our models. Overall, the best-fit models behave as we expect, where the average distribution of T$_d$ drops by at least a few Kelvin toward the center of the core. In general the average difference in the modeled core outer dust temperature, $T_o$, vs $T_c$ is $\langle T_o - T_c \rangle = 3.8$ K, where $T_o$ is calculated at the core boundary (column 8 in Table\,\ref{Core_Parameters}).

The narrow range of line-of-sight T$_d$ that \textit{Herschel} probes compared to our \textit{pandora} models shows that \textit{Herschel} is very sensitive to the extended emission in the B10 region (Figure\,\ref{hersch_compare}). We note, however, that a fixed opacity law with $\beta \sim 1.75$ was used to construct the \textit{Herschel} maps (see section\,\ref{betasection}), thus the comparisons to our models should be used only as a general guide.

\begin{figure}
\begin{center}$
\begin{array}{l}
\includegraphics[width=82mm]{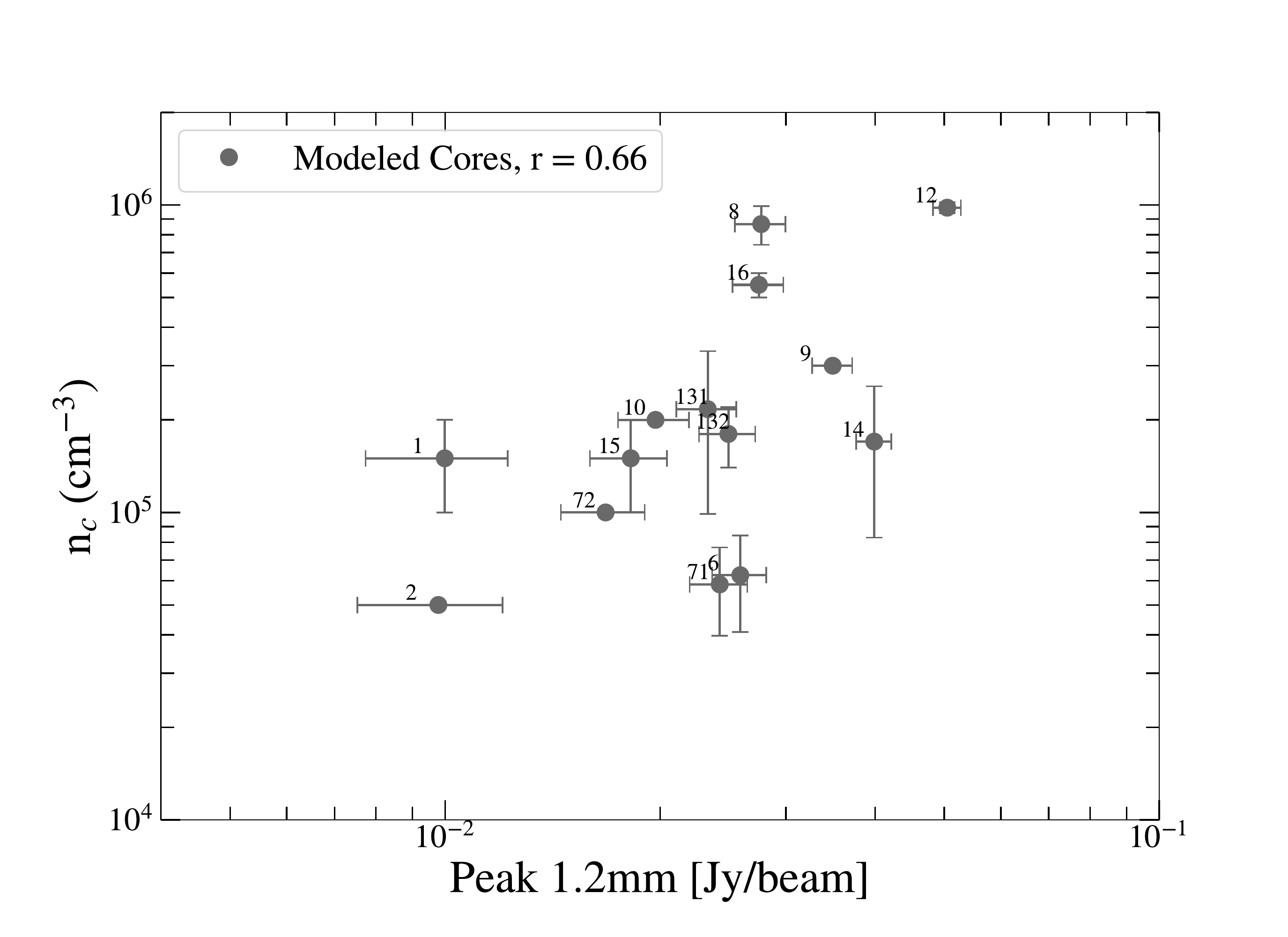} 
\end{array}$
\end{center}
\caption{\label{n_c_vs_peak} Modeled central volume density values versus observed peak 1.2mm emission values for each core, labeled as grey circles. There is a slight positive correlation (r = 0.66).
}
\end{figure}

\begin{figure}
\begin{center}$
\begin{array}{c}
\includegraphics[width=82mm]{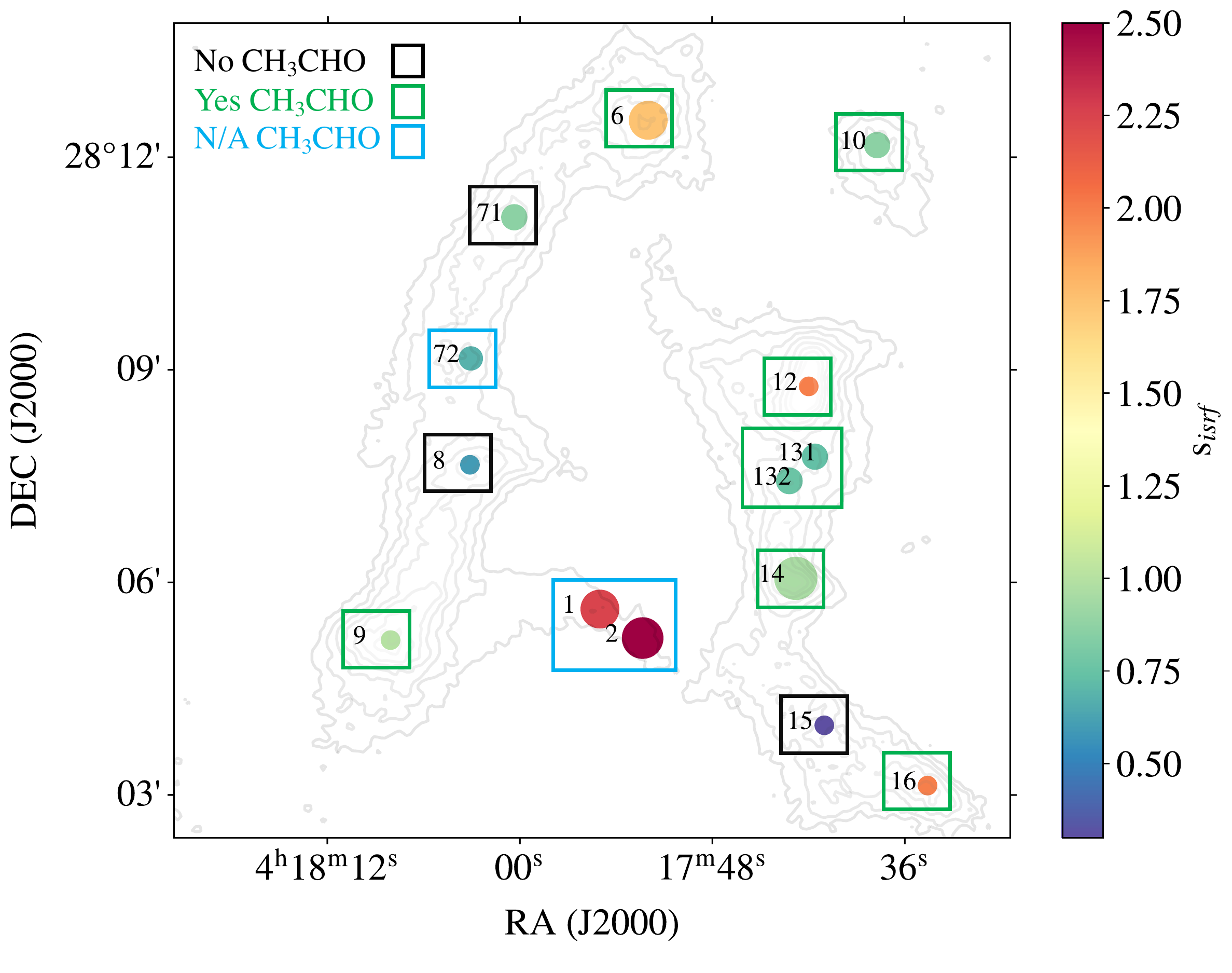} 
\end{array}$
\end{center}
\caption{\label{sisrf_corr} A map of B10 (grey contours as in Figure\,\ref{NIKA2map} for the 1.2mm NIKA2 map) where scatter points show where each of the 14 cores lie. The color of the points represent the average interstellar radiation field scale factor ($s_{isrf}$) value for the model suite and the size of the scatter point corresponds to increasing standard deviation error. The cores are boxed-in by squares color-matched to whether the complex molecule CH$_3$CHO was detected (green), not detected (black), or not targeted in that core (blue). The cores with low $s_{isrf}$ (i.e., core 8 and core 15) appear in more the shielded regions of B10 and show lower chemical complexity (no CH$_3$CHO was detected).
}
\end{figure}

We do find that there is a general positive correlation (Spearman's rank coefficient of r=0.66) with central density (n$_c$) and the observed 1.2mm peak emission (Figure\,\ref{n_c_vs_peak}). There is no trend, however, with n$_c$ versus the normed 2D aspect ratio $|r_{2D}|$, or with n$_c$ versus $\eta$, n$_c$ versus $s_{isrf}$, and n$_c$ versus opacity law. 

By averaging over the best-fits for each core, we find the overall median $s_{isrf}$ is 0.91 and mean $s_{isrf}$ is 1.2, with a standard deviation of 0.68. There are only a couple of cores that have models with consistent $s_{isrf}$ values at or below 0.6 (i.e., core 15 and core 8), which appear further embedded within the B10 filament (see Figure\,\ref{sisrf_corr}). Conversely, the few models with $s_{isrf}$ values well above or at 2.0 (i.e., core f1 and core f2), are in a more irradiated and less bright (in the 1.2mm map) portion of the B10 filament (Figure\,\ref{sisrf_corr}).

For the cores with low $s_{isrf}$ values (core 15 and core 8), they show low chemical complexity compared to the other cores in B10 (i.e., no CH$_3$CHO detected at the dust peak from \citealt{2020ApJ...891...73S}). These cores have likely been shielded from photodesorption by UV photons from the ISRF, reducing the gas-phase chemical inventory. Core 8 in particular is in the highest extinction region of the B10 region, at A$_\mathrm{V} \sim 30$\,mag, and, while the same A$_\mathrm{V}$ applies for the nearby core 9, core 8 is a smaller core that is fully surrounded by this high extinction, whereas core 9 is larger and encompassed by lower ($<$ 12\,mag) extinction values. Following from this, core 15 is also likely shielded by the higher extinction regions surrounding it (see maps in \citealt{2010ApJ...725.1327S}). This shielding and uneven illumination from the ISRF has been shown to affect how COMs (as well as carbon-chain molecules) are distributed spatially in and around starless cores \citep{2016A&A...592L..11S, 2020A&A...643A..60S}. Our results thus provide evidence to support the idea that gas-phase complex molecules, like CH$_3$CHO, as well as their reaction partners, are potentially still locked in the grains (or `frozen out') toward the dust peak where these observations were taken in \cite{2020ApJ...891...73S}, leading to an uneven distribution of complex chemistry in the B10 region.  

\section{Discussion} \label{discussion}

To better put in context the opacity laws used in our 3D modelling of starless cores in B10, we create a dust emissivity index, or $\beta$, map using our NIKA2 millimeter continuum and \textit{Herschel} dust temperature maps. 

We also perform an analysis of core stability through a virial analysis. Given that we have constraints on the physical parameters for the starless cores in B10 directly from our models, we take advantage of the fact that we know the 3D density distribution of our best-fits to calculate virial terms. Unlike in integrated line-of-sight observations, where disentangling emission from cores and larger-scale structures like filaments and parent clouds can be difficult, our models allow for self-consistent calculations for the starless cores. 

\subsection{Dust Emissivity Index ($\beta$) Analysis} \label{betasection}

The dust opacity, $\kappa_\nu$,  is dependent on $\beta$ by the relation, 

\begin{equation}\label{kappanu}
    \kappa_\nu  = \kappa_0 \left(\frac{\nu}{\nu_0}\right)^\beta,
\end{equation} where $\kappa_0$ is the emissivity cross-section per gram of dust and gas at the frequency $\nu_0$. In the optically thin limit ($\tau_\nu << 1$), the emission from each of the NIKA2 maps can be related to $\kappa_\nu$ by, 

\begin{equation}
    S_{\nu} = B_{\nu}[T_d] \Omega \tau_\nu
    \label{Snu},
\end{equation} where the opacity $\tau_\nu = \kappa_\nu \Sigma$ and $\Sigma$ is the mass surface density, $B_\nu$ is the Planck black body function at certain frequency, $\nu$, and average line-of-sight dust temperature, $T_d$, and $\Omega$ is the solid angle. 

\begin{figure}
\begin{center}$
\begin{array}{c}
\includegraphics[width=75mm]{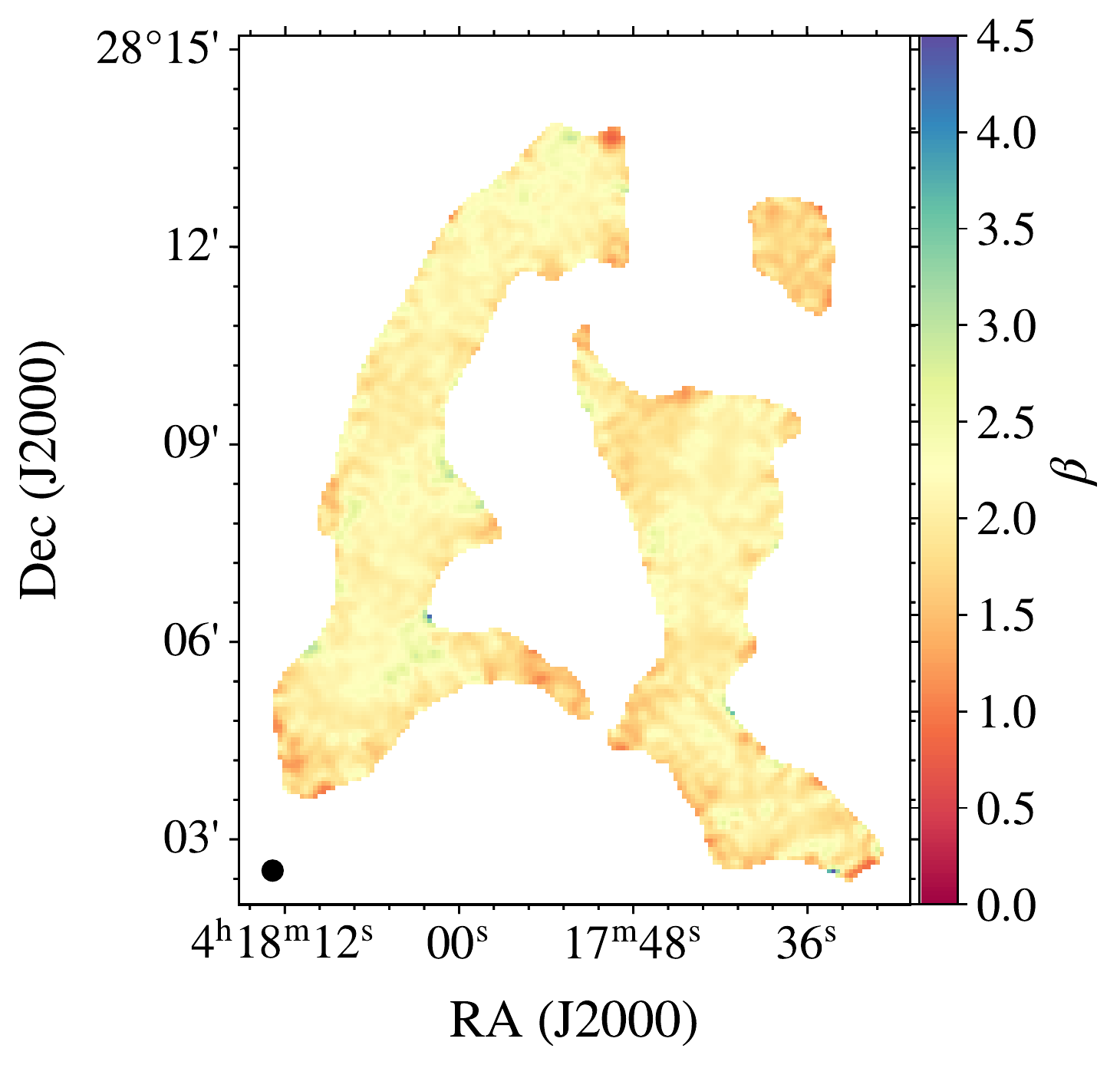} 
\end{array}$
\end{center}
\caption{\label{beta_map} Map of the dust emissivity index, $\beta$, constructed using the ratio of the NIKA2 1.2mm and 2.0mm maps as well as the \textit{Herschel} dust temperature map (see equation\,\ref{betaeqn}). The resolution is 18$''$ and only the pixels above 6$\sigma_{1.2mm}$ are included. 
}
\end{figure}

\begin{figure}
\begin{center}$
\begin{array}{c}
\includegraphics[width=80mm]{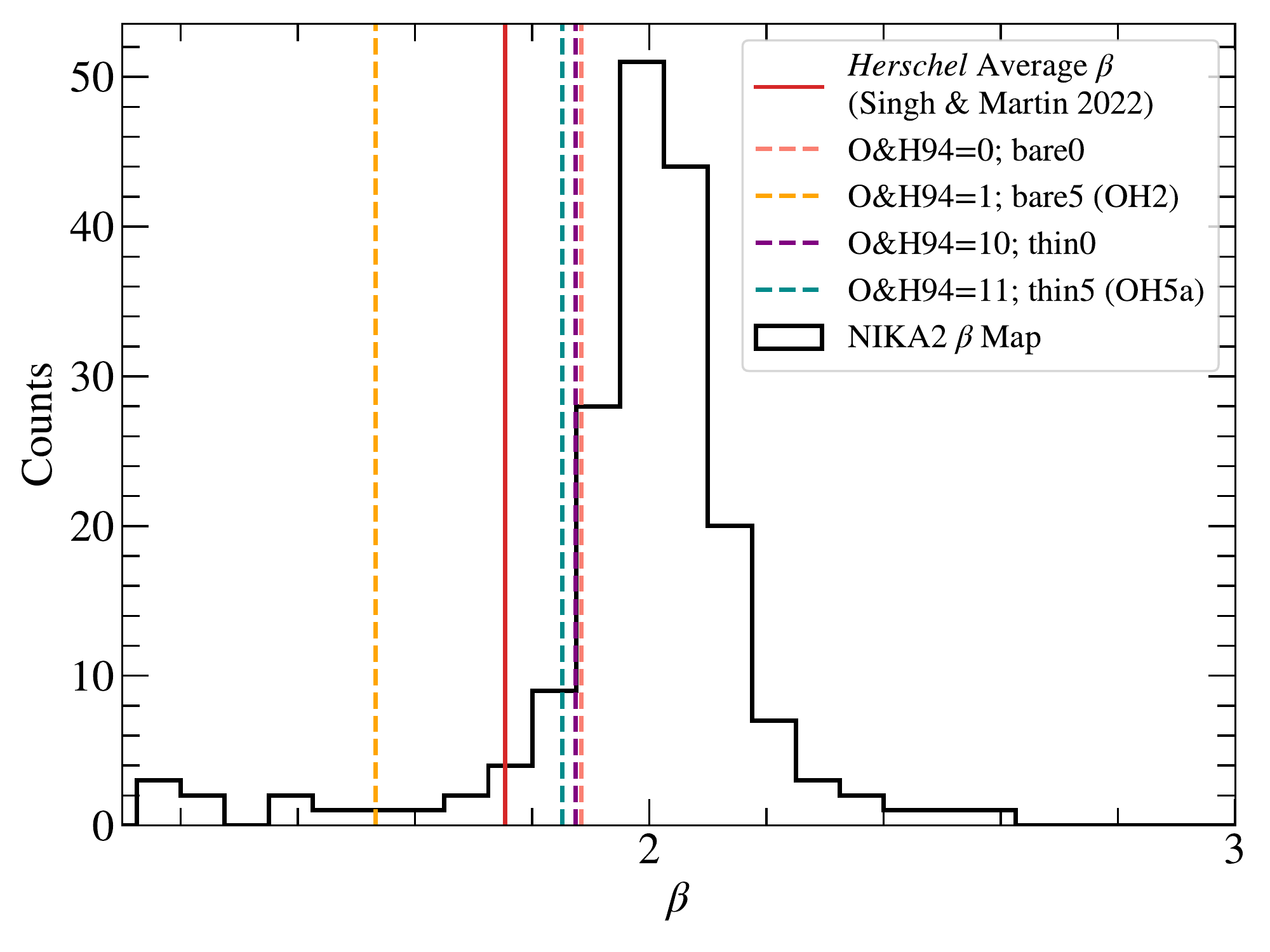} 
\end{array}$
\end{center}
\caption{\label{hist} Distribution of $\beta$'s from the map in  Figure\,\ref{beta_map} (black). In red as a solid line is the mean \textit{Herschel} derived $\beta$ from \citealt{2022ApJ...941..135S}. The $\beta$'s for the opacity laws used in our 3D modelling analysis are shown as dashed lines. 
}
\end{figure}

The NIKA2 1.2mm map was first deconvolved from it's 12$^{''}$ FWHM beam and then convolved to the 2.0mm resolution of 18$^{''}$ using a single Gaussian FWHM beam via the Python package \code{radio-beam}. The maps were then masked to include only pixels $>$ 6$\sigma_{1.2mm}$ and converted from mJy/beam to MJy/sr. The ratio of newly convolved emission maps, $R_{1,2}$, was then used to construct a $\beta$ map given that we can use equation\,\ref{Snu} to write the ratio as, 

\begin{equation}
    R_{1,2} = \frac{B_{\nu_{1.2mm}}[T_d]}{B_{\nu_{2.0mm}}[T_d]} \left(\frac{\nu_{1.2mm}}{\nu_{2.0mm}}\right)^\beta, 
    \label{R12_betaeqn_2}
\end{equation} where, since each map was convolved to the same resolution, $\Sigma$, $\Omega$, $\kappa_0$ and $\nu_0$ all dropped. Therefore $\beta$ is written as,

\begin{equation}
    \beta = \frac{\log(R_{1,2} \times B_{\nu_{2.0mm}}[T_d]]/B_{\nu_{1.2mm}}[T_d])}{\log(\nu_{1.2mm}/\nu_{2.0mm})},
    \label{betaeqn}
\end{equation} where $T_d$ comes from the \textit{Herschel} map \citep{2022ApJ...941..135S} re-gridded to match the NIKA2 ratio map. In Figure\,\ref{beta_map} we present the $\beta$ map for B10 constructed using equation\,\ref{betaeqn} above, which has a mean value of 2.01$\pm$0.48 for the entire region (where the error has been derived from the standard deviation). Comparatively, \cite{2017A&A...604A..52B} found that for starless cores in B213, $\beta = 2.4\pm0.3$ from NIKA data when corrected for possible line-of-sight temperature gradients through the Abel transform inversion technique.

The analysis presented in this work is not entirely self-consistent, since the $T_d$ map from \textit{Herschel} was fit using a different opacity law. Their maps show $\beta$ in the B10 region is on average lower, at $\sim$1.75. Given that the \textit{Herschel} opacity law sets $\kappa_0$\,=\,10\,cm$^2$/g at $\nu_0$ = 1000 GHz, extrapolating $\kappa_\nu$ at  260 GHz (1.2mm) gives $\sim 0.95$\,cm$^2$/g for this B10 region in the \textit{Herschel} maps \citep{2022ApJ...941..135S}. This value is within the range of our models, where the absorption cross section in cm$^2$ per gram of refractive material in the dust times the gas-to-dust ratio of 100 extrapolated 260 GHz is 0.39, 1.1, 0.64 and 0.89 cm$^2$/g for opacity laws O\&94 = 0, 1, 10 and 11 respectively (see Figure\,\ref{opacityfig}).

In Figure \,\ref{hist} we plot directly the distribution of $\beta$'s from the map in Figure\,\ref{beta_map} and compare to the mean \textit{Herschel} derived $\beta$ as well as the $\beta$'s from our opacity laws used in the radiative transfer modelling (section\,\ref{pandora}; where $\beta$ is determined by a linear regression over 350 - 1300$\mu$m). Each modeled opacity law $\beta$ falls on the lower end but within the NIKA2 $\beta$ map distribution. 

It is perhaps not surprising that only one best-fit model (Model\_5 for core 14) out of 68 was fit with the opacity law O\&H=1 which has a $\beta$ of 1.53, the farthest out from the mean distribution in Figure \,\ref{hist}. The remaining laws have $\beta$ values closer to the mean: 1.88, 1.87 and 1.85 for O\&H=0, 10, and 11, respectively. Still, each O\&H $\beta$ falls on the lower end of the distribution and explains why in our modelling we generally have a hard time fitting both sides of the SED, i.e., both the 2mm and 160$\mu$m intensity peaks (see Figure\,\ref{all_seds}). This suggests that the opacity laws we use in our radiative transfer analysis, which are also used elsewhere throughout the literature, do not accurately describe the observed $\beta$.

Our $\beta$ map provides a more realistic estimate for the dust emissivity index, since the NIKA2 1.2mm and 2.0mm observations probe the submillimeter regime far from the peak of the SED. However, as mentioned above, the analysis does rely on $T_d$ values from \textit{Herschel} and is therefore not self-consistent. We cannot use the $T_d$ values calculated from our 3D modelling since they assume a dust opacity law with a fixed $\beta$.

\subsection{Virial Analysis}

\begin{figure}
\begin{center}$
\begin{array}{c}
\includegraphics[width=82mm]{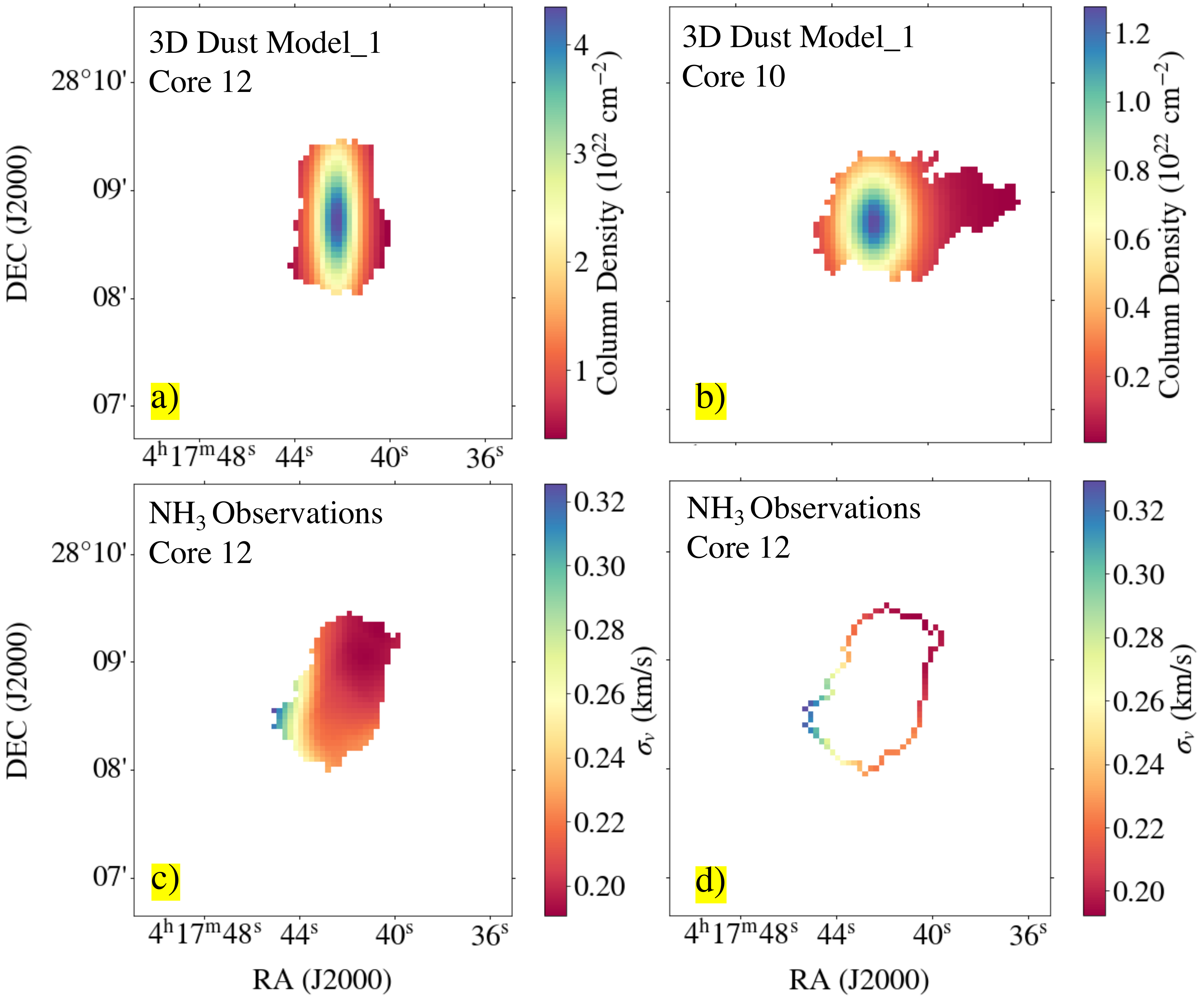} 
\end{array}$
\end{center}
\caption{\label{virial_extract} Example extractions within dendrogram regions. (a)  Modeled hydrogen column density, N(H$_2$), map for Model\_1 for Core 12 and for (b) Core 10. (c) Total velocity dispersion, $\sigma_\mathrm{tot}$, from the NH$_3$ observations of core 12 that has been re-gridded onto the modeled grid (3'' pixels). (d) The outer velocity dispersion, $\sigma_\mathrm{tot_\mathrm{out}}$, within a one pixel annulus around core 12. Note: all models are oriented so that the position angle for each core is at 0 degrees and the dendrogram region reflects this on the model grid, whereas the dendrogram region for cores in the observational maps match the core's position angle from the original analysis. 
}
\end{figure}

\begin{table*}
	\caption{Physical Properties \& Virial Terms}
	\setlength{\tabcolsep}{2.6pt}
	\label{virial_table}
	\begin{tabular}{lccccccccccc} 
        Core & $\sigma_\mathrm{tot}$ & $T_\mathrm{kin}$  & $\mathrm{v}_{z}$ & $N$ & $\Omega_K$ & n$_c$ & $M(R_C)$ & $\Omega_G$ & $n_\mathrm{out}$  & $\sigma_\mathrm{tot_{out}}$ & $\Omega_P$  \\ 
         & [km s$^{-1}$]  & [K] & [km s$^{-1}$] &  [10$^{22}$ cm$^{-2}$]  &[10$^{41}$ ergs] &  [10$^5$ cm$^{-3}$] & [M$_\odot$]& [10$^{40}$ ergs] &  [10$^3$ cm$^{-3}$] & [km s$^{-1}$] & [10$^{42}$ ergs]  \\
        \hline
        6   & 0.23$\pm$0.01  & 9.63$\pm$0.66  &  6.90$\pm$0.05 & 0.48$\pm$0.20  & 3.80$\pm$1.59 &  0.63$\pm$0.22  &  0.17$\pm$0.064 &  -6.99$\pm$5.34   &  3.22$\pm$2.03  &  0.22$\pm$0.01  &  -4.33$\pm$2.73  \\ 
        7-1 &  0.19$\pm$0.01  &  9.31$\pm$0.46 & 6.83$\pm$0.03  & 0.35$\pm$0.11 &  1.89$\pm$0.58 & 0.58$\pm$0.19 & 0.12$\pm$0.044 & -3.83$\pm$1.72 & 4.34$\pm$1.64  &  0.19$\pm$0.01  & -4.62$\pm$1.75    \\ 
        7-2 &  0.20$\pm<$0.01  &  9.39$\pm$0.65 & 6.79$\pm$0.01  & 0.54$\pm$0.06 &  0.80$\pm$0.09  & 1.00$\pm$0.00 & 0.04$\pm$0.003 & -0.63$\pm$0.09 & 3.82$\pm$0.89&  0.20$\pm$0.01  & -0.20$\pm$0.05 \\
        8 &  0.20$\pm$0.01  &  9.22$\pm$0.19 & 6.84$\pm$0.02  & 2.87$\pm$0.17 &  4.48$\pm$0.27 &  8.67$\pm$1.25 & 0.20$\pm$0.011 & -15.8$\pm$2.36& 1.85$\pm$0.39&  0.20$\pm$0.01  & -0.58$\pm$0.03\\
        9 &  0.20$\pm$0.01  &  8.66$\pm$0.40 & 6.97$\pm$0.06  & 1.40$\pm$0.03 &  8.52$\pm$0.19  & 3.00$\pm$0.00 & 0.62$\pm$0.019& -78.4$\pm$4.78 & 5.66$\pm$0.54&  0.20$\pm$0.01  & -7.70$\pm$0.73 \\
        10&  0.26$\pm$0.04  &  8.94$\pm$1.92 & 5.74$\pm$0.15  & 0.53$\pm$0.09 &  5.41$\pm$0.89 & 2.00$\pm$0.00  & 0.16$\pm$0.020& -5.26$\pm$1.35 &  0.85$\pm$0.46 &  0.29$\pm$0.04  & -0.69$\pm$0.37\\
        12 & 0.22$\pm$0.03 & 9.28$\pm$0.50 & 5.96$\pm$0.05  & 1.73$\pm$0.10 &  8.71$\pm$0.49  & 9.80$\pm$0.40   & 0.36$\pm$0.052& -40.0$\pm$3.00 & 8.05$\pm$0.57&  0.24$\pm$0.05  & -4.00$\pm$0.30\\
        13-1 &  0.22$\pm$0.01  &  8.70$\pm$0.75 & 5.97$\pm<$0.01  &  1.86$\pm$0.90 & 0.37$\pm$0.18  &  2.17$\pm$1.18 &0.02$\pm$0.103 & -0.44$\pm$0.31 & 7.24$\pm$2.82&  0.22$\pm$0.01  & -0.01$\pm<$0.01\\
        13-2&  0.24$\pm$0.01  &  10.63$\pm$0.77 & 6.02$\pm$0.02  &1.67$\pm$0.40  &  0.40$\pm$0.09  &  1.80$\pm$0.40 & 0.02$\pm$0.004& -0.32$\pm$0.11 & 6.22$\pm$1.72&  0.25$\pm$0.02  & -0.02$\pm<$0.01 \\ 
        14 &  0.23$\pm$0.01  &  9.80$\pm$0.62 & 6.05$\pm$0.03  & 1.34$\pm$0.68  &  2.96$\pm$1.50  &  1.70$\pm$0.87 & 0.11$\pm$0.060& -6.44$\pm$0.49 & 10.21$\pm$5.90 &  0.24$\pm$0.03  &  -1.21$\pm$0.70  \\
        15&  0.23$\pm$0.02  &  8.89$\pm$0.47 & 6.83$\pm$0.04  & 1.37$\pm$0.12  &  1.31$\pm$0.12   & 1.50$\pm$0.50  & 0.07$\pm$0.010 & -2.27$\pm$0.71 &5.69$\pm$2.40 &  0.22$\pm$0.02  & -0.21$\pm$0.09 \\ 
        16 &  0.20$\pm$0.01  &  9.62$\pm$0.48 & 6.81$\pm$0.03  &  0.96$\pm$0.08 &  4.21$\pm$0.33 & 5.50$\pm$0.50  & 0.24$\pm$0.017 & -13.8$\pm$2.02 & 0.21$\pm$0.02 &  0.21$\pm$0.02  &  -0.17$\pm$0.02\\ 
        
        f1 &  0.29$\pm$0.01  &  11.59$\pm$1.05 & 7.15$\pm$0.03  &  0.22$\pm$0.07 &  0.47$\pm$0.14   & 1.50$\pm$0.50   & 0.02$\pm$0.001& -0.36$\pm$0.22 & 0.07$\pm$0.02&  0.28$\pm$0.02  &  -0.001$\pm<$0.001 \\ 
        
        f2 &  0.32$\pm$0.01  &  9.27$\pm$1.46 & 7.01$\pm$0.11  &  0.11$\pm$0.01 &  0.32$\pm$0.04  &  0.50$\pm$0.00 & 0.01$\pm<$0.001& -0.03$\pm<$0.01 & 0.27$\pm$0.05 &  0.32$\pm$0.01  &  -0.003$\pm<$0.001\\
		\hline
	\end{tabular}
      \small
      \begin{description} 
        \item Mean values extracted from dendrogram region, or directly from the 3D modelling (i.e., $n_c$ and $n_\mathrm{out}$), for each core (errors are standard deviation values). 
      \end{description}
\end{table*}

In this work, we find 14 starless cores within the B10 region that we can perform a stability analysis on via virial parameters. We consider the contributions from the kinetic ($\Omega_K$), gravitational ($\Omega_G$) and external pressure ($\Omega_P$) energy terms in our virial analysis. A magnetic ($\Omega_B$) term should be added for a full treatment, but unfortunately it is difficult to get accurate estimates of the magnetic field strengths at core scales. We can estimate, however, what effective magnetic field strength, $\Delta B_\mathrm{eff}$, the cores would need to bring them back to a stable equilibrium (see below). We start first by calculating an average $\Omega_K$ over all viewing angles, which follows from \cite{2021ApJ...922...87S}, 

\begin{equation}
    \Omega_K = \Omega_{K_\mathrm{bulk}} + \frac{3}{2} \int \sigma_\mathrm{tot}(x,y)^2 \Sigma_{c}(x,y) dx dy,
\end{equation} where $x$ and $y$ are sky coordinates and $\Sigma_{c}$ is the core column density. The $\sigma_\mathrm{tot}$ is the total velocity dispersion and we included both the thermal and non-thermal (turbulent) support as follows,

\begin{equation}
\sigma_\mathrm{tot}^2 = \sigma_{NT}^2 + \sigma_{T}^2 = \left(\sigma_{\mathrm{NH}_3}^2 - \frac{k_B T_\mathrm{kin}}{\mu_{\mathrm{NH}_3}m_{H}}\right) + \frac{k_B T_\mathrm{kin}}{\mu_\mathrm{avg}m_{H}},
\label{sigmatot}
\end{equation} where $\sigma_{\mathrm{NH}_3}$ is the measured velocity dispersion and $T_\mathrm{kin}$ the measured kinetic temperature from the ammonia data described in \cite{2015ApJ...805..185S}, with $\mu_{\mathrm{NH}_3}$\,=\,17 and $\mu_\mathrm{avg}$\,=\,2.34. We write, 

\begin{equation}
    \Omega_{K_\mathrm{bulk}} = \frac{3}{2} \int [\mathrm{v}_{z}{(x,y)} - \mathrm{v}_\mathrm{CM,z}]^2 \Sigma_{c}(x,y) dx dy, 
\end{equation} as the `bulk' kinetic energy, which is defined in \cite{2021ApJ...922...87S} as the contribution from resolved variations in the line-of-sight velocity $\mathrm{v}_z$ across the core. The center-of-mass velocity is mass-averaged and thus is calculated following,

\begin{equation}
    \mathrm{v}_\mathrm{CM,z} =\frac{ \int \Sigma_{c}(x,y) \mathrm{v}{_z}(x,y) dx dy }{\int \Sigma_{c}(x,y) dx dy }.
\end{equation}

\begin{figure}
\centering
\begin{center}$
\begin{array}{c}
\includegraphics[width=82mm]{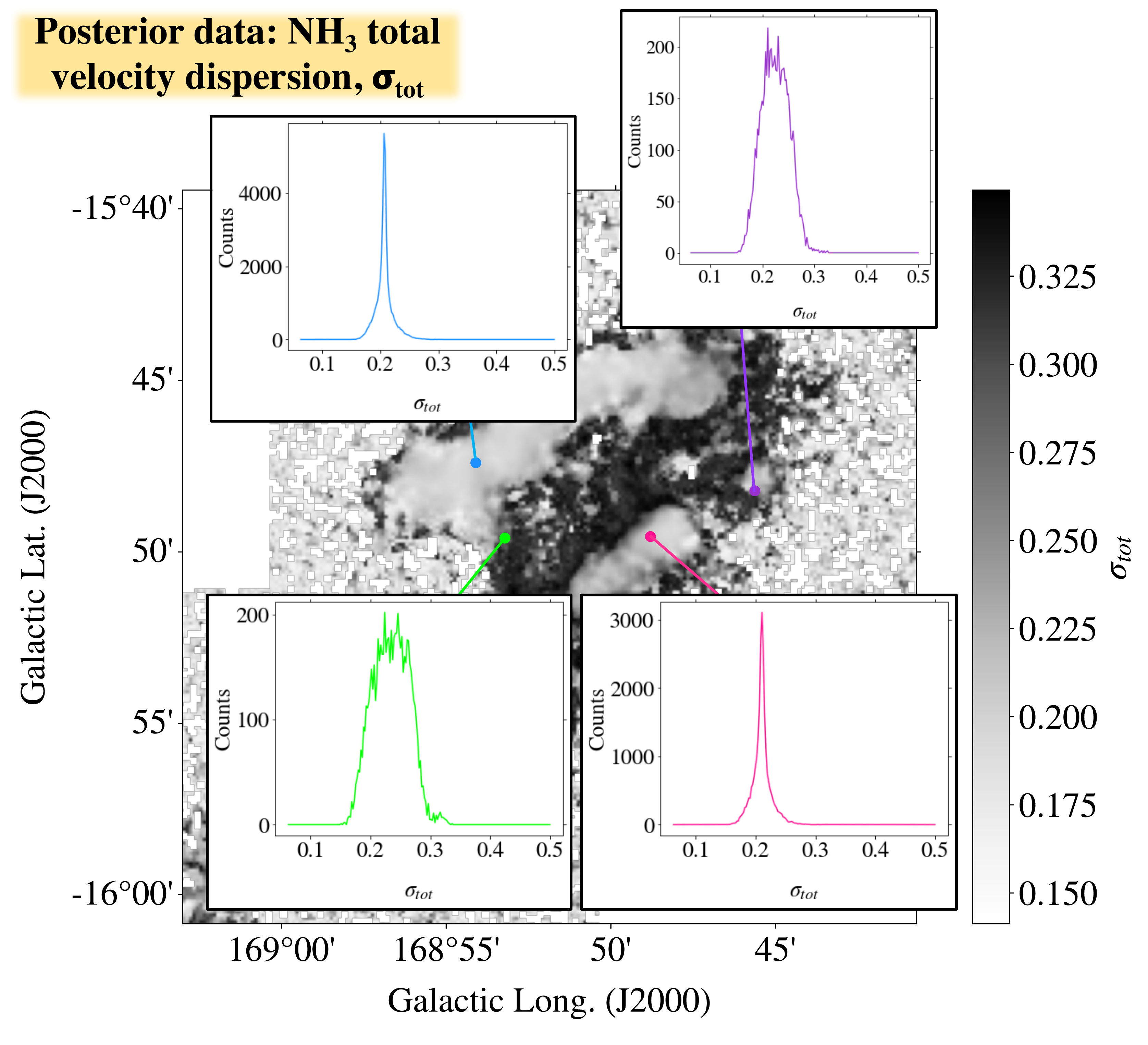} 
\end{array}$
\end{center}
\caption{\label{pdf_sigmav} The posteriors for the total velocity dispersion, $\sigma_\mathrm{tot}$, in the B10 region of Taurus derived from NH$_3$ maps and created using the \code{nestfit} code (Svoboda in prep.). The 2D MAP image for B10 is shown in greyscale, where the darker grey represents a higher $\sigma_{tot}$ value. Over-plotted are examples of extracted PDFs (used to create the MAP image) at four center pixel positions corresponding to core 8 (blue), core 10 (purple), core 12 (pink) and core f1 (green). } 
\end{figure}

\begin{figure}
\centering
\begin{center}$
\begin{array}{c}
\includegraphics[width=82mm]{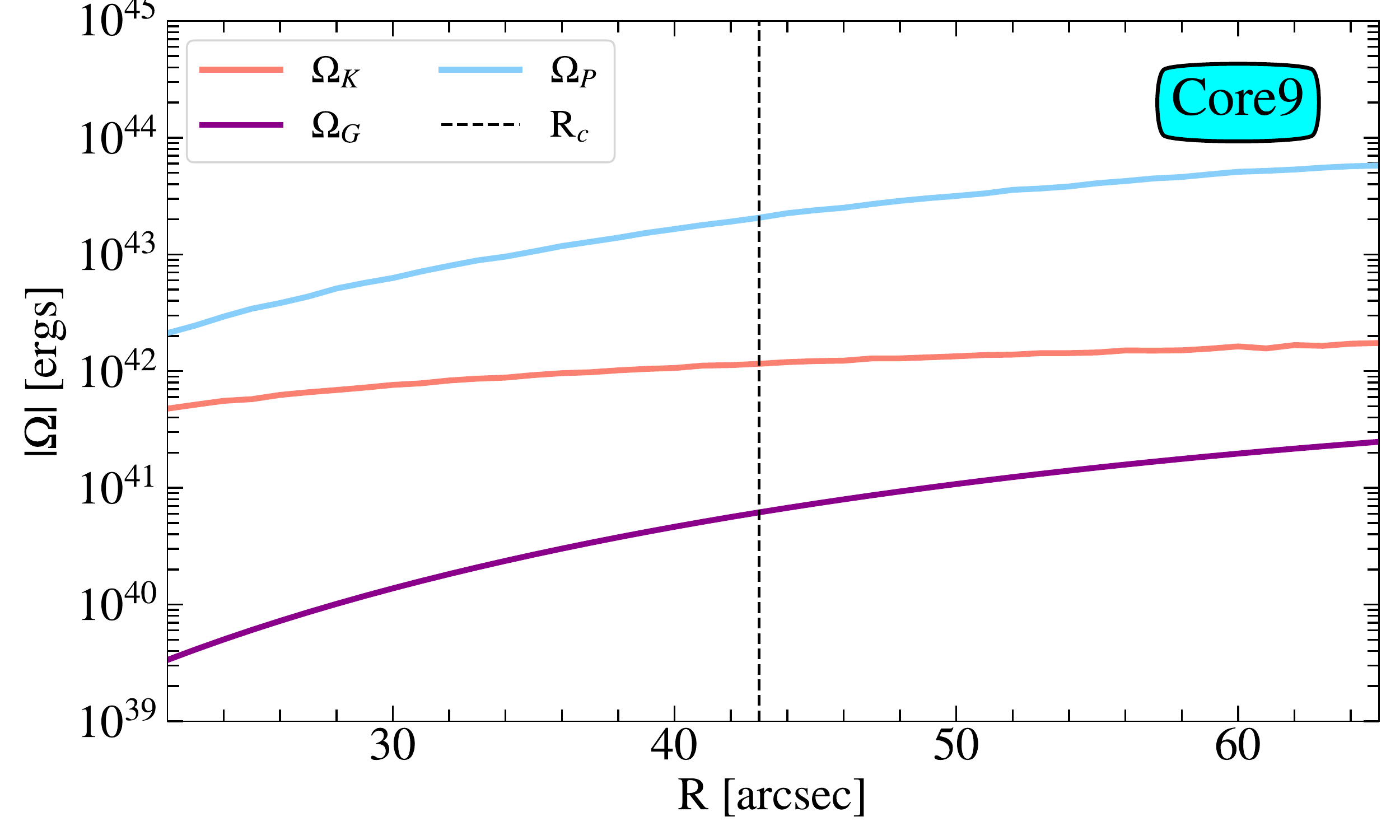} 
\end{array}$
\end{center}
\caption{\label{virial_rc_core9} Energy terms versus changing spherical radius, R, for the more isolated core 9. As a black dashed line is the true R$_c$ value used from the dendrogram analysis. The highest contributing energy term, $\Omega_P$, in blue, followed by $\Omega_K$ in red and $\Omega_G$ in purple.} 
\end{figure}

The \textit{pandora} framework produces 2D column density maps at the resolution of high resolution 1.2mm NIKA2 data (12'' resolution and 3'' pixel scale) that we use for $\Sigma_{c}$ in our calculations. We sum over every pixel in the core and our full equation for $ \Omega_K $ becomes, 

\begin{equation}
\begin{split}
    \Omega_K = \frac{3}{2} \Sigma_\mathrm{pix} [\mathrm{v}_{z\mathrm{(x,y)}} - \mathrm{v}_\mathrm{CM,z}]^2 \mu m_{H} N(x,y) A_\mathrm{pix} \\
    + \frac{3}{2} \Sigma_\mathrm{pix} \sigma_\mathrm{tot}(x,y)^2 \mu m_{H} N(x,y) A_\mathrm{pix},
\end{split}
\end{equation} where $N(x,y)$ is the modeled column density of $H_2$ through the model at each pixel position, $A_\mathrm{pix}$ is the area of that pixel, $\mu = 2.8$ and m$_H = 1.6733 \times 10^{-24}$ grams. We extract out modeled column density values within an individual core's dendrogrammed region (i.e., Figure\,\ref{virial_extract}) for each of our best-fit models (listed in Table\,\ref{tab:pandora_best_fit_params}). 

\begin{figure*}
\centering
\begin{center}$
\begin{array}{c}
\includegraphics[width=175mm]{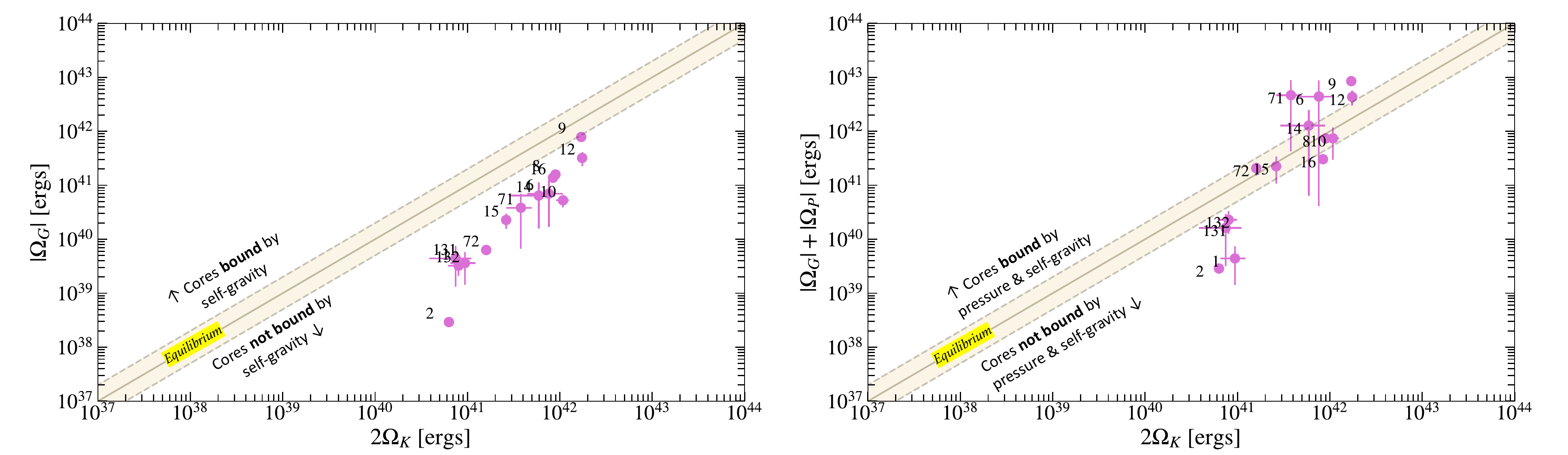}
\end{array}$
\end{center}
\caption{\label{virial_plots}   We plot mean virial parameters with standard deviation errors for each of our core's given their suites of best-fit models. (left) We plot the absolute value $|\Omega_G|$ vs. $2\Omega_K$, where cores above the `equilibrium' shaded region are considered to be bound by self-gravity and cores below this region are not bound by self-gravity. (right) We plot the absolute value $|\Omega_G|$ + $|\Omega_P|$ vs. $2\Omega_K$, where cores above the `equilibrium' shaded region are considered to be bound by both external pressure and by self-gravity. The `1' and `2' labeled points refer to cores `f1' and `f2' respectively.} 
\end{figure*}

The observational data are from NH$_3$ maps presented in \cite{2015ApJ...805..185S} and provide the measurements for the velocity and temperature estimates in this analysis. The $\sigma_\mathrm{tot}$, T$_\mathrm{kin}$ and $\mathrm{v}_z$ have been re-calculated using the bayesian fitting code, \code{nestfit} (Svoboda in prep.)\footnote{\url{https://github.com/autocorr/nestfit}}. The code uses MultiNest \citep{2009MNRAS.398.1601F} to perform Nested Sampling Monte Carlo to do a one component fit on the NH$_3$ data cubes. The NH$_3$ model assumes constant excitation temperature and uses the rotational partition function to calculate level populations (see also \citealt{2008ApJS..175..509R, 2017ApJ...843...63F}).

The maxiumum a posteriori (MAP) value is calculated for each pixel in the B10 region and the posteriors are also saved as a data cube (see Figure\,\ref{pdf_sigmav}). It is important to stress that the extracted MAP values for cores in areas of the map with low signal-to-noise, which are cores 10, f1 and f2, are more uncertain than for the remaining cores. 
As shown in Figure\,\ref{virial_extract}, the pixels in the MAP
NH$_3$ images ($\sigma_\mathrm{tot}$, T$_\mathrm{kin}$, $\mathrm{v}_z$) are re-gridded onto the model grid that are spaced at 3$''$ per pixel (the same as the NIKA2 1.2mm map).

The gravitational term, $\Omega_G$, follows from \cite{1992ApJ...395..140B}, 

\begin{equation}\label{OmegaG}
    \Omega_G = - a \frac{3G M(R_c) ^2}{5R_c},
\end{equation} where $G$ is the gravitational constant, $M(R_c)$ is the mass of the core and $R_c$ is the radius of the core (column 8 in Table\,\ref{Core_Parameters}). In observational studies, the mass is equal to $M_\mathrm{obs} = \Sigma_\mathrm{pix} \mu m_H N(x, y) A_\mathrm{pix}$. In this analysis, since we have the actual density profile from our \textit{pandora} models (equation\,\ref{nr_pandora}), the mass can be calculated more accurately via a numerical integration given our modeled density profile,

\begin{equation} \label{MRC}
    M(R_c) = \mu m_H n_c \int_{r=0}^{r=R_c} \frac{4 \pi r^2 }{ \left[1 + (\frac{r}{r_0})^2 \right]^{\eta/2}} dr, 
\end{equation} where $n_c$ is the modeled central density and $r_0$ is given by a 3D geometric mean, $\sqrt[3]{r_{0,x}  r_{0,y}r_{0,z}}$, of the best-fit aspect ratios for that model (Table\,\ref{tab:pandora_best_fit_params}). Again, we define $R_c$ as the spherically averaged radius of the exact dendrogram area (column 8 in Table\,\ref{Core_Parameters}).

Additionally, the $a$ in equation\,\ref{OmegaG} is a dimensionless parameter that measures the effects of a nonuniform or nonspherical mass distribution on the gravitational energy and is on order unity. Following the derivation in the appendix of \citep{1992ApJ...395..140B}, the parameter $a$ can be calculated explicitly for a Plummer sphere density profile where,

\begin{equation}
    a = \frac{5 R_c\int_0^{R_c/r_0}\frac{y dy}{(1+y^2)^{\eta/2}} \int_0^y \frac{x^2 dx}{(1+x^2)^{\eta/2}}}{3r_0 \left[ \int_0^{R_c/r_0}\frac{x^2dx}{(1+x^2)^{\eta/2}}\right]^2}
\end{equation} 

We find our Plummer spheres span the range from $a = 1.02 - 1.25$ with a typical value of 1.13. Thus, a standard value of $a = 1.13$ is used in our calculations. We note that \cite{2021ApJ...922...87S} independently find an average of $a \simeq 1.13$ for their lower mass sample of Gould Belt clumps.

For our suite of models the $M_\mathrm{obs}$ method (that utilizes the modeled N(H$_2$) column density maps) compared to the $M(R_c)$ method in equation\,\ref{MRC}, differing by factors of 0.3 to 6, with a median value of 1.3. 
Similarly, if we use instead the \textit{observed} column density maps from \textit{Herschel} (re-gridded to our model grid), we find $\Omega_\mathrm{G_{obs}}$ differ from $\Omega_\mathrm{G_{3D}}$ by at most a factor of 14 and median factor of 2. We find that line-of-sight mass estimates more often overestimate $\Omega_G$ by up to an order of magnitude for cores in the B10 region.

Lastly, a treatment of the external pressure, 
\begin{equation}
    \Omega_P = - 4 \pi P_\mathrm{out} R_\mathrm{out}^3,
\end{equation} can be considered following from the ideal gas law where $\Omega_P$ in our models is calculated by finding the outer density, $n_\mathrm{out}$, at each `outer' radius (i.e., beyond the effective radius) around the core, $R_\mathrm{out}$. This density, $n_\mathrm{out}$, is then used in the following equation for the outer pressure, 

\begin{equation}
    P_\mathrm{out} = \rho_\mathrm{out} \sigma^2_\mathrm {tot_\mathrm{out}} = \mu m_H  n_\mathrm{out} \sigma^2_\mathrm {tot_\mathrm{out}}.
\end{equation}

We calculate the outer velocity dispersion by extracting $\sigma^2_{\mathrm{tot_\mathrm{out}}}$ within an annulus one pixel (3'') outside the dendrogram area (see d) in Figure\,\ref{virial_extract}). We then calculate the corresponding average n$_\mathrm{out}$ using the \textit{pandora} density profiles for each radius point ($R_\mathrm{out}$) along the outer dendrogram perimeter. We find for the modeled best-fit cores that the averaged $n_\mathrm{out}$ ranges from $0.07 - 10.21$\,$\times$10$^3$\,cm$^{-3}$ and the mean value is  4.45$\,\times$10$^3$\,cm$^{-3}$ (Table\,\ref{virial_table}).

Now that we have defined each energy term, we also quantify how $\Omega$'s may change with varying R$_c$ (Figure\,\ref{virial_rc_core9}). We do this for core 9, which is in a rather isolated portion of the B10 map that does not overlap with other cores (see Figure\,\ref{dendofig}). Instead of using the dendrogram region, we use circular apertures at different R$_c$ values from 22 to 65 arcseconds (on either side of the true R$_c$ value of 43 arcseconds), or $\sim$ 14 pixels. We find, perhaps not surprisingly given that R$_c$ itself is not used in this energy term, that $\Omega_K$ does not vary by more than a factor of 4. $\Omega_P$ varies by an order of magnitude and $\Omega_G$ varies by roughly two orders of magnitude. However, within $\pm$6 arcseconds (2 pixels) of R$_c$, the variation in spherical radius drops down to less than factors of 3 for each energy term (Figure\,\ref{virial_rc_core9}).

We perform our full stability analysis for each modeled core. Ignoring $\Omega_B$ for now, in general, if $-\Omega_G$/$\Omega_K$ $>$ 2, then gravity is sufficient to bind the core, while $\Omega_G$/$\Omega_P$ $<$ 1 simply indicates the cloud external pressure dominates over gravity. A dense core is in virial equilibrium when $-(\Omega_G + \Omega_P) = 2\Omega_K$. We find, from averaging the values for suites of models for each core and calculating the standard deviation as our error, that all (100\%) of the B10 cores fall towards the `unbounded' side of the $\Omega_G$ vs. $2\Omega_K$ plot (Figure\,\ref{virial_plots}). 

When external pressure is added to the gravity term, we find that most cores (64\%) are either in virial equlibrium (36\%) or on the `bound' side (28\%) of the -($\Omega_G + \Omega_P$) vs. $2\Omega_K$ plot (see Figure\,\ref{virial_plots}). We mention that the B10 cores picked out by \cite{2015ApJ...805..185S} were also unbound by gravity from the analysis in \cite{2020ApJ...891...73S}, which was done via global measurements of physical parameters from projected \textit{Herschel} maps.

The small cores f1 and f2 lie outside of the equilibrium region and are unbound by both gravity and pressure, which is perhaps not so surprising since they were not picked out by the original NH$_3$ fitting. These cores are also in a low signal-to-noise region of the map and thus the variation in the pdf in the observed NH$_3$ parameters is higher, meaning the pdf resembles more the priors in the fitting routine. Interestingly, across f1 and f2 polarisation was recently detected in the BISTRO POL2 survey, and they also conclude that may be the youngest region in B10 \citep{2023arXiv230212058W}. The small cores 13-1 and 13-2 (one core `13' in the original NH$_3$ fitting) are also unbound by gravity and external pressure  but closer to equilibrium given their errors (Figure\,\ref{virial_plots}). 

The magnetic energy term, $\Omega_B$ can be estimated from,

\begin{equation}
    \Omega_B = \frac{(B^2 - B_0^2) R^3}{6},
\end{equation} where $B$ and $B_0$ are the uniform magnetic field interior and exterior to the core such that this term is the difference in magnetic energy inside and outside the core \citep{2017stfo.book.....K}. For cores that are bound by external gravity and pressure in the right panel of Figure\,\ref{virial_plots} (i.e., core 6, 7-1, 9, 12 and 14) we calculate how large the effective magnetic field difference, 

\begin{equation}
\Delta B_\mathrm{eff} = \sqrt{(B^2 - B_0^2)},
\end{equation} would have to be to bring these cores back to virial equilibrium when $-(\Omega_G + \Omega_P) = 2\Omega_K$ + $\Omega_B$ (see \citealt{2022MNRAS.515.5219G}). We find $\Delta B_\mathrm{eff}$ needs to be 13, 15, 16, 14 and 14\,$\mu$G for cores 6, 7-1, 9, 12 and 14, respectively. 
At much larger scales (200$''$ resolution) in the material surrounding dense cores in the Taurus Molecular Cloud, magnetic field values range from $5-82\mu$G \citep{2011ApJ...741...21C}. Higher resolution dust polarisation measurements toward B10 estimate an average magnetic field strength of $\sim46\mu$\,G \citep{2023arXiv230212058W}. In this study they also estimate an upper limit of $\approx 70\mu$\,G for the magnetic field strength in core 12 (their core 1), leading to a $B_\mathrm{eff} \lesssim 24\mu$\,G for the core. Thus, it is quite plausible that the effective magnetic field difference could be on the order of $15\mu$G for the B10 cores. 

Overall, from our unique virial analysis that uses best-fit densities and mass distributions from our 3D high resolution models, we find the majority (64\%) of our B10 cores are either in virial equilibrium, or are bound by external pressure and self-gravity where a relatively small effective magnetic field difference of $\sim 15 \mu$G would be needed to push the five bounded cores back to equilibrium.

Starless cores take longer than the free-fall timescale ($t_\mathrm{ff}$) to collapse, around $\sim 5\times t_\mathrm{ff}$ for average density cores in our B10 sample \citep{2014prpl.conf...27A}. A small contribution from  magnetic fields could be supporting the cores and keeping them in equilibrium. Since we find only a small $\Delta B_\mathrm{eff}$ is needed for support, magnetic fields should not be ignored when considering a core's evolutionary state through a virial analysis.

Traditional definitions of whether a starless core is `prestellar' have relied on virial analyses that ignore terms such as external pressure or magnetic fields. It is clear from our analysis that gravity (as well as external pressure) should not be the sole identifiers when considering a core as `starless' or `prestellar.' Magnetic fields appear to play an important role. A more restrictive definition of prestellar should also consider dynamical motions of a core, such as their observed infall properties \citep{2005ApJ...619..379C, 2007ApJ...664..928S, 2019ApJ...871..134S}. More recent magnetohydrodynamic simulation work, that has categorized cores based on coherence, also warn readers that one should not rely on `a conventional virial analysis' to predict a core's evolutionary state, go on to predict that that $\geq 20\%$ of observed starless cores (spanning a density range similar to our B10 sample of cores) will not go on to form stars \citep{2022MNRAS.517..885O}. 

Now that robust physical models have been created, subsequent papers detailing the depletion and deuteration chemistry, as well as the kinematic structure of the B10 starless cores will be done to further constrain their exact evolutionary states.

\section{Conclusions} \label{conclusions}

A detailed analysis of the physical properties and stability conditions for 14 starless cores in the B10 region of the Taurus Molecular Cloud has been carried out. A grid of over one million core models was run within the sophisticated 3D radiative transfer framework \textit{pandora} that utilizes the RADMC-3D code. We list below our major findings:
\begin{itemize}
    \item  Through radial profile and SED diagnostics, best-fit models  constraining central densities, density distributions, aspect ratios, opacity laws and strengths of the interstellar radiation field were found for each core. The 14 `typical' cores in B10 span central densities from 5\,$\times$\,10$^4$\,-\,1\,$\times$\,10$^6$\,cm$^{-3}$, with a mean value of 2.6\,$\times$\,10$^5$\,cm$^{-3}$. The central line-of-sight dust temperature, $T_c$, in general is lower than what is extracted from the \textit{Herschel} map where $T_c$ averages to $10.57\pm0.56$ in our models. The average difference in the modeled core line-of-sight outer dust temperature, $T_o$, vs $T_c$ is $\langle T_o - T_c \rangle = 3.8$ K. We stress, however, that 3D dust temperatures T$_d$({r$_{3D}$}) do reach lower values $\sim\,6$\,K at the center in some higher density models.
    Additionally, we find for the cores in B10 the overall median scale factor to the interstellar radiation field, $s_{isrf}$, is 0.91, and the mean $s_{isrf}$ is 1.2$\pm$0.68. 

    \item A dust emissivity index ($\beta$) map was constructed for the full B10 region using the NIKA2 1.2mm and 2.0mm emission maps, as well as the \textit{Herschel} dust temperature map. We find a mean $\beta = 2.01\pm0.48$ in the B10 region, which is $16\%$ larger than \textit{Herschel} estimates and $9-29\%$ larger than the $\beta$ from the opacity laws we use in our 3D modelling. This result suggests that the opacity laws used in this analysis and throughout the literature do not accurately describe the observed $\beta$. 
    \item Self-consistent calculations for a virial analysis, from 3D density structures, were performed on the 14 modeled cores. We find that the majority of the cores (9 out of  14) are either in virial equilibirum or are bound by external pressure self-gravity. Additionally, we find that a small effective magnetic field difference of $\sim15\mu$G would be needed to push the bounded cores (6, 7-1, 9, 12 and 14) back to equilibrium. Our results support that external pressure and magnetic fields should not be ignored when considering if a starless core will go on to form a star. 
    
\end{itemize}

The constraints on the physical properties from our high resolution models of these more `typical' cores in Taurus will allow for detailed follow-up studies regarding core chemistry and velocity structure. Because starless cores are the earliest stage of low-mass star formation, the constraints on their evolution put forward in our analysis give insight into the initial conditions needed for later-stage star and planet formation. 

\section*{Acknowledgements}

The authors would like to thank Bilal Ladjelate for his guidance regarding the NIKA2 observations and data reduction. We also thank the anonymous reviewer for the helpful comments. Samantha Scibelli has been supported by a National Science Foundation Graduate Research Fellowship (NSF GRF) Grant DGE-1143953. Samantha Scibelli (PI) and Yancy Shirley (co-I) have also been supported by the Universities Space Research Association (USRA) Stratospheric Observatory of Infrared Astronomy (SOFIA) grant 09-0155. 

These results are based on observations carried out under project number 028-19 with the IRAM 30-meter telescope. IRAM is supported by INSU/CNRS (France), MPG (Germany) and IGN (Spain). This research has also made use of data from the Herschel Gould Belt survey (HGBS) project (http://gouldbelt-herschel.cea.fr). The HGBS is a Herschel Key Programme jointly carried out by SPIRE Specialist Astronomy Group 3 (SAG 3), scientists of several institutes in the PACS Consortium (CEA Saclay, INAF-IFSI Rome and INAF-Arcetri, KU Leuven, MPIA Heidelberg), and scientists of the Herschel Science Center (HSC).
Part of the analysis for this paper relied on \code{astrodendro}, a Python package to compute dendrograms of Astronomical data (http://www.dendrograms.org/). This work made extensive use of Astropy:\footnote{http://www.astropy.org} a community-developed core Python package and an ecosystem of tools and resources for astronomy. The modelling was carried out using High Performance Computing (HPC) resources supported by the University of Arizona TRIF, UITS, and Research, Innovation, and Impact (RII) and maintained by the UArizona Research Technologies department.

\section*{Data Availability}

The data underlying this article will be shared on reasonable request to the corresponding author.




\bibliographystyle{mnras}
\bibliography{B10_dust} 







\bsp	
\label{lastpage}
\end{document}